\documentclass[a4paper,twocolumn, superscriptaddress, aps, pra, 10pt]{revtex4-2}
\pdfoutput=1
\usepackage{cmap}
\usepackage{preamble}

\usepackage{times}

\usepackage{tikz}

\usepackage{eczoo}
\usepackage{soul}

\renewcommand{\braket}[1]{\left\langle #1 \right\rangle}

\def\llangle{\langle\!\langle}
\def\rrangle{\rangle\!\rangle}
\def\des{a}

\def\sketbra#1{\mathinner{|{#1}\rrangle}\!\mathinner{\llangle{#1}|}}

\usepackage[OT2,T1]{fontenc}
\DeclareSymbolFont{cyrletters}{OT2}{wncyr}{m}{n}
\DeclareMathSymbol{\Sha}{\mathalpha}{cyrletters}{"58}

\newtcolorbox{mybox}{colback=gray!10!white, colframe=gray}

\begin{document}
\title{Continuous-variable designs and design-based shadow tomography from random lattices}
\author{Jonathan Conrad}
\thanks{\url{jonathan.conrad@epfl.ch}}
\affiliation{Institute of Computer and Communication Sciences, École Polytechnique Fédérale de Lausanne (EPFL), Lausanne CH-1015, Switzerland}
%\affiliation{Dahlem Center for Complex Quantum Systems, Physics Department, Freie
 %   Universit{\"a}t Berlin, Arnimallee 14, 14195 Berlin, Germany}
%\affiliation{Helmholtz-Zentrum Berlin f{\"u}r Materialien und Energie, Hahn-Meitner-Platz 1, 14109
 %   Berlin, Germany}
\author{Joseph~T.~Iosue}
\affiliation{Joint~Quantum~Institute, NIST/University of Maryland, College Park, Maryland 20742, USA}
\affiliation{Joint Center for Quantum Information and Computer Science, NIST/University of Maryland, College Park, Maryland 20742, USA}
\author{Ansgar~G.~Burchards}
\affiliation{Dahlem Center for Complex Quantum Systems, Physics Department, Freie
    Universit{\"a}t Berlin, Arnimallee 14, 14195 Berlin, Germany}
\author{Victor~V.~Albert}
\affiliation{Joint Center for Quantum Information and Computer Science, NIST/University of Maryland, College Park, Maryland 20742, USA}
\date{\today}

\begin{abstract}
    We investigate state designs for continuous-variable quantum systems using the aid of lattice-like quantum states. These are code states of Gottesman-Kitaev-Preskill (GKP) codes. 
    We show that the set of all GKP states forms a rigged continuous-variable state $2$-design for an $n$-mode system. 
    We use these lattice state designs to construct a continuous variable shadow tomography protocol, derive sample complexity bounds for both global and local GKP shadows under reasonable physical assumptions, and provide the physical gadgets needed to implement this protocol.
\end{abstract}
\maketitle

\notoc

\section{Introduction \& summary}

Designs~\footnote{a.k.a. averaging sets, quadratures, or cubatures}  simplify averages over a space $X$ by providing an approximate version, or ``skeleton'', of the space which can be used instead of $X$ to simplify calculations.
Designs are ubiquitous in numerical integration \cite{stroud1971approximate} and coding theory \cite{Delsarte1977,ConwaySloane,Colbourn_designs}, and designs over the space of quantum states and the unitary group are valuable tools in quantum information processing and even black-hole physics (e.g., see Refs.~\cite{renes2004symmetric,huang2019predicting, Huang_2020, Elben_2022, Haferkamp_2022, Gross_2007, Haferkamp:2022jqk, Jain_2024}).

Designs exist on finite-dimensional spaces, but recent work \cite{Iosue_2024} heralded an extension of designs to infinite-dimensional spaces of normalizable functions. The prominence of such spaces in quantum information processing, signal processing, machine learning, and quantum field theory warrants further study of such designs.

In quantum mechanics, such function spaces parameterize quantum state spaces of continuous-variable (CV) systems, e.g., photonic or phononic degrees of freedom.
As modern technological developments allow for precise control of CV degrees of freedom in engineered quantum systems --- and as physics is continuous at its fundamental level --- the development of quantum computers embedded in natively CV systems is increasingly recognized as a promising strategy \cite{Albert_2018, Terhal_2020, Bourassa_2021_blue}. 
Design-based protocols for CV systems are a well-behaved and quantifiable way to extract information which may further aid in their classification and simulation, analogous to their qubit-based counterparts.

A \eczoo[design]{t-designs} of strength $t$ --- a $t$\textit{-design} --- is a subset of points $\CE\subseteq X$ such that a uniform average over the design of every degree$\leq t$ polynomial equals its average over $X$. Reference~\cite{Iosue_2024} introduced a family of appropriately defined (and necessarily non-Gaussian \cite{Blume_Kohout_2014}) CV $2$-designs for a space of functions on the real line, $\R^{d=1}$, leaving open the question of existence of natural designs for real-space functions of higher dimension $d$.

We answer this question in the affirmative using quantum lattice states --- infinite quantum superpositions of coherent states
at all points in a symplectic lattice in the CV phase space.
We show that the set of such states for all displaced versions of all possible $2n$-dimensional symplectic lattices form a family of CV $2$-designs \textit{for all} $n$.

Quantum lattice states form codewords of the  \eczoohref[Gottesman-Kitaev-Preskill code]{multimodegkp} \cite{GKP} (see also Refs.~\cite{Albert_2018, sivak2023real}) and the Zak basis \cite{PhysRevLett.19.1385,PhysRevX.11.011032}, and are relevant to quantum optics \cite{Perelomov1986, Aharonov1969}, post-quantum cryptography \cite{Regev_2009,conrad2023good}, signal processing \cite{feichtinger1998gabor}, quantum metrology \cite{Duivenvoorden_Sensor}, and condensed-matter physics \cite{PhysRevLett.19.1385,PhysRevX.11.011032}.
Henceforth, we colloquially refer to quantum lattice states as GKP states.

Infinite families of finite-dimensional $2$-designs feature prominently in a popular qubit-based tomography protocol called shadow tomography \cite{huang2019predicting,Huang_2020, acharya2021informationallycompletepovmbasedshadow}, and our results readily yield similarly powerful CV variants of both the ``global'' and ``local'' variants of this protocol. 
This result shows that GKP quantum error correction --- realized in several experiments \cite{Fluehmann_2019,Campagne_Ibarcq_2020,de2022error,sivak2023real,lachance2024autonomous,valahu2024quantum,brock2024quantum} --- is a powerful tomographic resource for CV systems, on par with Clifford circuits and Pauli measurements for qubit systems.

\section{GKP states form rigged $2$-designs}

A \textit{rigged $t$-design} is an ensemble of pure states $\CE$ together with a measure $\mu_{\CE}$ such that its constituents resolve permutation-symmetric subspaces~\cite{Iosue_2024}, satisfying
    \begin{equation}\label{eq:t-design}
        \int_{\CE} d\mu_{\CE}\lr{X} \ketbra{X}^{\otimes t'} = \des_{t'} \Pi_{t'} \; \forall t' \leq t~,
    \end{equation}
with non-negative \textit{design coefficients} $\des_{t^{\prime}}$, and with $a_1$ set to one. The adjective ``\textit{rigged}'' indicates that our ensemble contains non-normalizable vectors, which are required to satisfy the above equation.
% improper states, i.e., it is an ensemble over tempered distributions for a physical Hilbert space.}
Here, $\Pi_{t'}$ is a projector onto the permutation-symmetric subspace of the tensor-product space of $t$ factors, with each factor consisting of $n$ modes.

We denote the ensemble of all $n$-mode GKP (stabilizer) states for all lattices by
\begin{equation}
    \CX_n = \lrc{\ket{\Lambda; \bs{\alpha}}=D\lr{\bs{\alpha}}\ket{\Lambda},\; \Lambda \in Y_n,\; \bs{\alpha}\in \CP\lr{\Lambda}}. \label{eq:GKP_ensemble}
\end{equation}
Here, $\ket{\Lambda}$ is an equal superposition of coherent states at all points in the symplectically self-dual lattice $\Lambda=\Lambda^{\perp}\subset \R^{2n}$. %
This state is displaced by all possible vectors $\bs{\alpha}$ in the lattice's fundamental domain $\CP\lr{\Lambda}$ --- the unit cell centered at the origin --- 
via displacement operators $D(\bs{\alpha})$.
Each lattice is defined by a symplectic matrix, and the space $Y_n = \Sp_{2n}\lr{\Z}\backslash \Sp_{2n}\lr{\R}$ of all lattices is parameterized by all such matrices, up to basis transformations done by integer-valued symplectic matrices. This space happens to be measurable and comes with its own Haar measure, $\mu\lr{\Lambda}$~\cite{Sarnak1994,conrad2023good}.

The core contribution of this manuscript is the following.

\begin{them}[GKP states form a rigged $2$-design]\label{them:2design_summary}
The ensemble $\CX_n$ forms a rigged $2$-design with respect to the Haar measure over the space of symplectic lattices $Y_n$ and the uniform measure over the fundamental domain $\CP\lr{\Lambda}$.
\end{them}

We prove this Theorem in three different ways. The key step is afforded by the following formula, which converts certain integrals over lattices to integrals over phase space.
\begin{lem}[Symplectic mean value formula {\cite{Sarnak1994}\citep[theorem 2]{Moskowitz2010}}]
    \label{lem:mean-value_main}
    Let $f:\, \R^{2n}\rightarrow\R$ be a (Riemann) integrable function such that the following quantities converge absolutely, and let $\Lambda \in Y_n$ denote a symplectic lattice. For
    \begin{equation}
    \label{eq:F_lambda}
        F\lr{\Lambda}=\sum_{\bs{\lambda}\in \Lambda-\lrc{0}} f\lr{\bs{\lambda}}~,
    \end{equation}
    it holds that 
    \begin{equation}
        \int_{Y_n} d\mu\lr{\Lambda} F\lr{\Lambda} = \int_{\mathbb{R}^{2n}} d\bs{x}\, f\lr{\bs{x}} \label{eq:mean_val}~.
    \end{equation}
    % relative to the Haar measure over the space of symplectic lattices $Y_n$.
\end{lem}

We now summarize our proofs. 
From the decomposition of the CV Hilbert space into orthogonal sectors via displaced GKP states $D(\bs{\alpha})\ket{\Lambda}$ for each lattice $\Lambda$, we obtain a complete covering of phase space.
This means that each sub-ensemble $(\Lambda,\bs{\alpha})$ with $\Lambda$ fixed forms a $1$-design.
For the $2$-design condition, known formulas about GKP states~\cite{Conrad_2022} reduce the left-hand side of Eq.~\eqref{eq:t-design} to a lattice integral over $\Lambda \in Y_n$. 
The mean-value formula allows us to substitute that integral with one over the entire phase space, 
and solving that integral completes the proof.
% implying that the ensemble $\CX_n$ forms a $2$-design.

By replacing each GKP state in $\mathcal{X}_n$ with a regularized version of the state, this ensemble can further be turned into a set of \textit{physical} (read:~square-integrable and finite-energy) quantum states that conforms to the notion of \textit{approximate} rigged designs introduced in Ref.~\cite{Iosue_2024}. 
Without this modification, however, the rigged design is still useful as it allows for the definition of a POVM useful for state tomography.

%%%%%%%%%%%%%%%%%%%%%%%%%%%%%%%%%%%%%%%%%
\section{Design-based shadow tomography}
%%%%%%%%%%%%%%%%%%%%%%%%%%%%%%%%%%%%%%%%%

Our design construction can be used to implement CV classical-shadow tomography protocols using access to GKP states and an appropriately chosen Gaussian circuit.
We formulate two protocols --- a global and a local GKP shadow protocol --- which are CV cousins of global~\cite{huang2019predicting} (a.k.a.~Clifford) and local~\cite{huang2019predicting,Huang_2020} (a.k.a.~Pauli) qubit shadow protocols, respectively. 
We first place our protocols in the context of their qubit counterparts, and then proceed with a summary of our technical results.

Any ensemble $\CE$ satisfying the $2$-design condition, either conventional or rigged, yields the following map,
\begin{align}
  \CM_{\CE}\lr{\rho}&=\int_{\CE}d\mu\lr{X}\,\braket{X|\rho|X}\ketbra{X}\nonumber \\
  &=\frac{\des_{2}}{2}(I+\rho)~.\label{eq:MX_1}
\end{align}
This map shows that a POVM over the set of \textit{pointer ``states''} $|X\rangle$ with probability densities $\langle X|\rho|X\rangle$ contains complete information about $\rho$.
Plugging the map into the expectation value of an observable $O$ with repsect to $\rho$ expresses this value in terms of a statistical average,
\begin{align}
    \langle O\rangle=\frac{2}{\des_{2}}\mathop{\mathbb{E}}_{X|\rho}\langle X|O|X\rangle-\Tr[O]\equiv\mathop{\mathbb{E}}_{X|\rho}o(X)~,\label{eq:estimator}
\end{align}
over estimators $o(X)$ stemming from pointers $\ket{X}$. The subscript $X|\rho$ indicates that the average is relative to the distribution over the pointers $X$ inherited from the state $\rho$.%. with probability distribution $\braket{X|\rho|X}$.

The key to classical shadows is choosing an ensemble that yields a favorable scaling of the number of samples needed for a good estimate.
This sample complexity can be obtained by examining the estimator variance $\text{Var} ~o(X)$.

\paragraph*{Global shadows}
In the case of global qubit shadows, the ensemble is the set of all $n$-qubit stabilizer states, $\CE = \CS_n$, and the POVM reduces to a depolarizing channel with $\des_2 = 2/(2^n+1)$.
Stabilizer states form a quantum state $3$-design with $\des_3 = 6/[(2^n+1)(2^n+2)]$, and the variance can be bounded in terms of the trace of $O^2$ and a ratio of its design coefficients (see supplemental material~\cite{supplement}),
\begin{equation}
    \mathop{\text{Var}}_{X|\rho}o(X)\leq\frac{2\des_{3}}{\des_{2}^{2}}\Tr[O^{2}]~.
\end{equation}
The design-coefficient ratio approaches $3$ for large $n$, implying that the sample complexity depends only on the trace term.

The GKP analogue of this protocol --- \textit{global GKP shadows} --- uses the set of all GKP states, $\CE = \CX_n$. 
While we do not know if GKP states form a $3$-design, we show that the ensemble possesses just enough structure to yield a useful characterization of the variance.
Using favorable properties of GKP states and lattice moment formulas, we obtain in Thm.~\ref{them:CV_shadow_summary} a similar overall scaling with $\Tr[O^2]$ and an additional dependence on the $L^1$ norm of the characteristic function of $O$.

\paragraph*{Local shadows}
The pointers of the local qubit shadow protocol are $n$-fold tensor products of single-qubit stabilizer states, $\CE = \CS_1^{\otimes n}$.
For a traceless $k$-qubit tensor-product observable, the variance is the $k$th power of the above global shadow variance for the case of a single qubit --- upper bounded by $3^k \Tr[O^2]$.

Analogously, our \textit{local GKP shadow} protocol utilizes $n$-fold tensor products of single-mode GKP states, $\CE = \CX_1^{\otimes n}$.
As with its qubit counterpart, each subsystem is measured independently, and no entangled measurements are necessary.

We once again cannot use the $3$-design condition to bound the variance, and instead assume a conjecture~\eqref{eq:conj} about finite-energy bosonic states.
Equipped with this conjecture, we obtain a similar dependence of the variance on the observable as in the qubit case (see Thm.~\ref{them:CV_shadow_loc_summary}).

\subsection{Global GKP shadows}
We employ medians of means (MoM) estimation \cite{Huang_2020} to implement the classical post-processing of the protocol, where the $N=KB$ samples are divided into $K$ batches of size $B$, and the estimator $\rm MoM$ returns the median of the arithmetic means over each batch.
We establish a general upper bound using a recently derived second-moment bound for functions over the space of symplectic lattices \cite{Kelmer_2019}, which is parametrized by a bounded sequence $\lrc{f_n}_{n=1}^{\infty}$, defined as
\begin{equation}
    f_1=1,\quad 10\geq f_n=\frac{4\zeta\lr{n}^2}{\zeta\lr{2n}},\, n>1,
\end{equation}
where $\zeta$ is the Riemann-zeta function.
Combining this bound with the standard approach established in Ref.~\cite{Huang_2020}, we show the following theorem.

\begin{them}[GKP shadow tomography]\label{them:CV_shadow_summary}
    Let $\epsilon, \delta >0$, and let $O_i=O_i^{\dagger}, i=1,2,\cdots, M$ be a set of $M$ observables on an $n$-mode CV quantum system with integrable characteristic functions $c_{O_i}$. Let
    \begin{equation}
    \begin{aligned}\label{eq:global-variance}
        \tilde{V}_{O_i} &=\lr{\|c_{O_i}\|_1+\abs{\Tr\lrq{O_i}}}^2 + f_n\Tr\lrq{O_i^2} \\ \vspace{.5cm}
        \tilde{V}_{O}   & =\max_i \tilde{V}_{O_i},
    \end{aligned}
    \end{equation}
    and  $N=KB$, with
    \begin{equation}
        B= 34 \tilde{V}_O / \epsilon^2 \quad\text{and} \quad K=2\log\lr{2M/\delta}.
    \end{equation}
    Then, $N$ samples from the state relative to the 
    POVMs defined by the ensemble of GKP states $\CX_n$ suffice to approximate the expectation values of each observable $O_i$ with probability
\begin{equation}
        {\rm Pr}\lr{\max_{i=1,\dots,M}\left\lvert {\rm MoM}\lrq{\tilde{o}_i} -  \Tr\lrq{O_i\rho}\right\rvert \geq \epsilon} \leq \delta.
    \end{equation}
\end{them}

The trace term $\Tr\lrq{O^2}$ in Eq.~\eqref{eq:global-variance} stems from the $L^2$ norm of an observable's characteristic function, $\|c_O\|_2^2=\Tr\lrq{O^2}\leq \|O\|^2 {\rm rank}\lr{O}$, and is also present in the qubit shadow bound. 
However, rather than capturing the \textit{rank} of the observable as in the qubit case, the CV domain (where the rank need not be defined) renders this into a measure of how quickly the operator $O$ \textit{decays} in phase space.

The $L^1$ norm of the characteristic function in Eq.~\eqref{eq:global-variance} --- not present in the qubit bound --- takes a dominant role in the derived bound. 
For qubits, the $l^1$ norm of the Pauli coefficient vector of a state (the analogue of the characteristic function) serves as a quantifier for the ``non-stabilizerness'' of the state \cite{Campbell_2011}. 
In the CV case, the $L^1$ norm of a characteristic function of a convex combination of Gaussian pure states $\rho_i=G\lr{V_i, \bs{\overline{x_i}}}$ yields a value $\|c_{\rho}\|_1\leq 2^n$ \cite{GaussianQuantumInfo}, 
such that any state whose $L^1$ norm exceeds this value must necessarily contain non-Gaussian components.
If the $3$-design property of GKP states could be shown, more stringent bounds on the estimator variance and sample complexity would follow that do not depend on 
the observables' characteristic functions~\cite{Iosue_2024}.

\subsection{Local GKP shadows under physicality conjecture}

In order to bound the local GKP shadow sample complexity, we impose two additional ``physicality'' assumptions on the input state $\rho$.
We assume that the state's average photon number is upper bounded by a constant,  $\Tr[{\rho\hat{N}}]\leq\overline{N}$.
and that the state decays exponentially in phase space \footnote{This is not a very stringent assumption, given that physical states are purported to decay super-polynomially~\cite{albert2022bosoniccodingintroductionuse}, and that all Gaussian states decay exponentially.}.
Using these assumptions, we can bound the pointer expectation values $\langle X|\rho|X\rangle$ using an effective squeezing parameter~\cite{Duivenvoorden_Sensor}, 
\begin{equation}
\Delta^2_{\bs\lambda}\lr{\rho}= -\frac{2}{\pi}\log\lr{\abs{c_{\rho}\lr{\bs\lambda}}}~.
\end{equation}
We conjecture that this squeezing parameter is lower bounded by the inverse average photon number of the system as 
\begin{align}
\tilde{\Delta}^{2}\lr{\rho}&\coloneqq\min_{M\in\Sp_{2n}\lr{\R}}\min_{i}\Delta_{\bs\xi_{i}}^{2}\lr{\rho}\\&\geq4/\overline{N}\hspace{3.5cm}\text{(conjecture)},\label{eq:conj}
\end{align}
only depending on $\overline{N}$, and where $M=(\bs{\xi}_1,..,\bs{\xi}_{2n})^T$. This conjecture generalizes a previous observation~\cite{Duivenvoorden_Sensor} using information-theoretic bounds and corresponds to the minimal squeezing parameter achievable if all the photons of an $n$-mode system with average photon number $\overline{N}$ are concentrated in an individual mode. %
For large $\overline{N}\gg 1$, this yields the bound
\begin{equation}\label{eq:bound_phys}
    \langle X|\rho|X\rangle\leq\left(\overline{N}/\pi\right)^{2n}.
\end{equation}
We use this bound to show that the relevant variance parameter $\tilde{V}_{O_i}$ in Thm.~\ref{them:CV_shadow_summary} can be replaced by \begin{equation}
\tilde{V}_{O_i}=\lr{\overline{N}/\pi}^{2n}\lr{\Tr\lrq{O_i^2}+\Tr\lrq{O_i}^2},\label{eq:VOi_summary}
\end{equation}
which, up to the physicality factor $\lr{\overline{N}/\pi}^{2n}$, is equivalent to that obtained from rigged CV $3$-designs in Ref.~\cite{Iosue_2024} and takes the comparable functional form as that of global qubit shadows \cite{Huang_2020}. 

Following a similar calculation that leads to the variance scaling in Eq.~\eqref{eq:VOi_summary}, we find that the performance of the local shadow tomography protocol derived here is not limited by the \textit{locality} of the observable, but rather by its \textit{in-separability}.

\begin{them}[CV shadow tomography, local and physical]\label{them:CV_shadow_loc_summary}
    Let $\epsilon, \delta >0$, and let $O_i=O_i^{\dagger}, i=1,2,\cdots, M$ be a set of $M$ observables on a $n$-mode CV quantum system with integrable characteristic functions $c_{O_i}$. Assume the conjecture in Eq.~\eqref{eq:conj} and that the input state has average photon number at most $\overline{N}\gg 1$ with exponentially decaying characteristic function. Let
    \begin{align}
        \tilde{V}_{O_i}^{\rm loc} & = \lr{\overline{N}/\pi}^{2n}\sum_{\bs{k} \subseteq \bs{n}} \Tr_{\bs{n}-\bs{k}}\lrq{\Tr_{\bs{k}}\lrq{O_i}^2} \\
        \tilde{V}_{O}^{\rm loc}   & =\max_i \tilde{V}_{O_i}^{\rm loc},
    \end{align}
    and  $N=BK$, with
    \begin{equation}
        B= 34 \tilde{V}_O^{\rm loc} / \epsilon^2 \quad\text{and} \quad K=2\log\lr{2M/\delta}.
    \end{equation}
    
    Then, $N$ samples from the states relative to the POVMs defined by the ensemble of GKP states $\CX_1^{n}=\bigotimes_{i=k}^n D\lr{\bs{\alpha}_k}\ket{\Lambda_k}$, where $\ket{\Lambda_k}$ is the unique GKP state corresponding to $\Lambda_k=\Lambda_k^{\perp}\subset \R^{2}$ and where $\bs{\alpha}_k\in \CP\lr{\Lambda_k}$, suffice to approximate the expectation values of each observable $O_i$ with
   \begin{equation}
        {\rm Pr}\lr{\max_{i=1,\dots,M}\left\lvert {\rm MoM}\lrq{\tilde{o}_i} -  \Tr\lrq{O_i\rho}\right\rvert \geq \epsilon} \leq \delta.
    \end{equation}
    % \textcolor{red}{Joe: This should be 
    % \begin{equation}
    %     {\rm Pr}\lr{\max_{i=1,\dots,M}\left\lvert {\rm MoM}\lrq{\tilde{o}_i} -  \Tr\lrq{O_i\rho}\right\rvert \geq \epsilon} \leq \delta.
    % \end{equation}
    % Or alternatively, we could phrase it as: 
    % With probability at least $1-\delta$, $\left\lvert {\rm MoM}\lrq{\tilde{o}_i} -  \Tr\lrq{O_i\rho}\right\rvert < \epsilon$ for all $i$.
    % This is how they phrase it in Eq (S14) of \url{https://arxiv.org/pdf/2002.08953}
    % }
\end{them}

In words, the native estimator variance is given by the \textit{total purity} of the observable, measured over every bi-partition $\bs{k}\subseteq \bs{n}$ given by a subset of mode labels $\bs{n}$ and its complement $\bs{n}-\bs{k}$. 
As long as the total purity of the observable across any bi-partition remains small, the local shadow protocol will be efficient. 
This is a notable difference from the sample complexity bound on local qubit shadows, which depends heavily on the tensor-product locality of the observable~\cite{Huang_2020}.

We test how our protocols fare on estimating an observable proportional to the thermal state of the total-energy Hamiltonian, $H = \overline{N}$, defined on both qubits and modes.
We obtain similar exponential scaling with qubit/mode number $n$ for all four protocol combinations --- local and global, qubit and GKP --- summarized in Table~\ref{tab:table_thermal_state}. In the supplemental material~\cite{supplement}, we analyze Hamiltonians with a pertubative polynomial part $\hat{H}=\hat{N}+\epsilon P\lr{\bs{\hat{x}}}$, with $P\lr{\bs{\hat{x}}}$ a polynomial expression in the quadrature operators, and show how the corresponding estimator values $|\braket{X|O|X}|$ are bounded by functions related to so-called weighted theta series~\cite{Elkies_weighted}. 
The general techniques outlined there show how these and other non-trivial observables can be examined with our protocol.

\begin{table}
    \centering
    \begin{tabular}{l c}
        Method &  Sample complexity ($\beta \ll 1$)\\ \hline\hline
         Global qubit shadows \cite{Huang_2020}& $\|O'\|_{\rm shadow}\sim \Tr\lrq{O'^2} = O\lr{2^n}$ \\
         Global GKP shadows & $O\lr{\lr{\overline{N}/\pi \beta}^{2n}}$\\
         \hline
          Local qubit shadows \cite{Huang_2020}& $\|O'\|_{\rm shadow} \sim 4^n\|O'\|_{\infty}=O\lr{4^n}$ \\
         Local GKP shadows & $ O\lr{\lr{\overline{N}/\pi \beta}^{2n}}$\\
    \end{tabular}
    \caption{Sample complexity scaling of qubit- and GKP based shadow tomography protocols with $\beta \ll 1$ for the thermal state observables $O=e^{-\beta \hat{N}}$ and $O'=e^{-\beta \sum_{i=1}^n \sigma_{z,i}}$.
    Here, $\overline{N}$ is the average energy of the underlying bosonic state.
    }
    \label{tab:table_thermal_state}
\end{table}

\section{Implementing shadow tomography}

Given an input state $\rho$, our global shadow protocol proceeds as follows, with the local version substituting $\CX_n \to \CX_1^{\otimes n}$.
To obtain samples from the effective state $\CM_{\CX_n}\lr{\rho}$~\eqref{eq:MX_1},
\begin{enumerate}
    \item sample a lattice $\Lambda \in Y_n$ specified by a generator matrix $M=\lr{\bs{\xi}_1,\hdots, \bs{\xi}_{2n}}^T$ \cite{Conrad_2022};
    \item measure each displacement operator $\lrc{D\lr{\bs{\xi}_i}}_{i=1}^{2n}$ by performing GKP syndrome extraction.
\end{enumerate}
In this strategy, for each sampled lattice $\Lambda$, the displacement operator measurement returns an outcome determined by a vector $\bs{\alpha}\in \CP\lr{\Lambda}$ with probability
\begin{equation}
    P_{\Lambda}\lr{\bs{\alpha}} = \braket{\Lambda |D^{\dagger}\lr{\bs{\alpha}} \rho D\lr{\bs{\alpha}}|\Lambda},\label{eq:P_alpha_sum}
\end{equation}
which is precisely the expression $\braket{X|\rho|X}$ that determines the distribution over pointers $X=(\Lambda,\bs{\alpha})$ that we are seeking. To understand how this expression arises, note that by symplectic self-duality of $\Lambda \in Y_n$, the POVM element $D\lr{\bs{\alpha}}\ketbra{\Lambda}D^{\dagger}\lr{\bs{\alpha}}$ describes the projector onto the $\bs{\alpha}$-``error sector'' of the GKP code specified by $\Lambda$. This sector is labeled by syndrome $MJ\bs{\alpha}$ for $M=(\bs{\xi}_1,..\bs{\xi}_{2n})^T$. Sampling over uniformly random $\Lambda \in Y_n$, we find that the shadow can be obtained via standard syndrome extraction for the GKP code. This strategy is analogous to the strategy employed in Ref.~\cite{Huang_2020} and can be implemented with standard qubit-assisted or GKP-state-assisted techniques to implement GKP stabilizer measurements.
We depict the corresponding circuits for these in Fig.~\ref{fig:Disp_circ} and provide a detailed analysis in the supplemental material~\cite{supplement}.

The qubit-assisted implementation is an experimentally widely used protocol, based on dominant-bit phase estimation \cite{Terhal_2016,Fluehmann_2019, Campagne_Ibarcq_2020, sivak2023real, lachance2024autonomous}, and we show how the probabilities $P_{\Lambda}\lr{\bs{\alpha}}$ can be directly estimated from Pauli expectation values on the qubit.
This strategy leads to an effective compression of the shadow tomography protocol, and we show how statistical techniques can be used to bound the sample overhead to estimate these probabilities.

Qubit-assisted protocols only yield binary outcomes in each execution of the circuit. An alternative strategy with higher information gain is to execute a continuous-valued phase estimation routine using an auxiliary GKP mode. Here we introduce new circuits that use a single GKP auxiliary mode to measure multimode GKP stabilizers, akin to the displacement sensor routine developed in Ref.~\cite{Duivenvoorden_Sensor}. This strategy is, to our knowledge, novel, and we show how the Fourier coefficients of the measurement statistics encode the relevant displacement expectation values. We expect the circuits presented here and their analysis to be of general use also outside of GKP-based shadow tomography. %

Finally, we note that, given access to a resource of GKP states $\ket{X}\in \CX_n$, one can also directly estimate the probabilities $\braket{X|\rho |X}$ by performing a SWAP test \cite{Buhrman_2001, Milburn_Fradkin, ding2024quantumcontroloscillatorkerrcat} between the resource state $\ket{X}$ and the input $\rho$. This further strenghtens the argument that GKP states form powerful resources for quantum tomography.
\begin{figure}
    \centering
    \includegraphics[width=.7\linewidth]{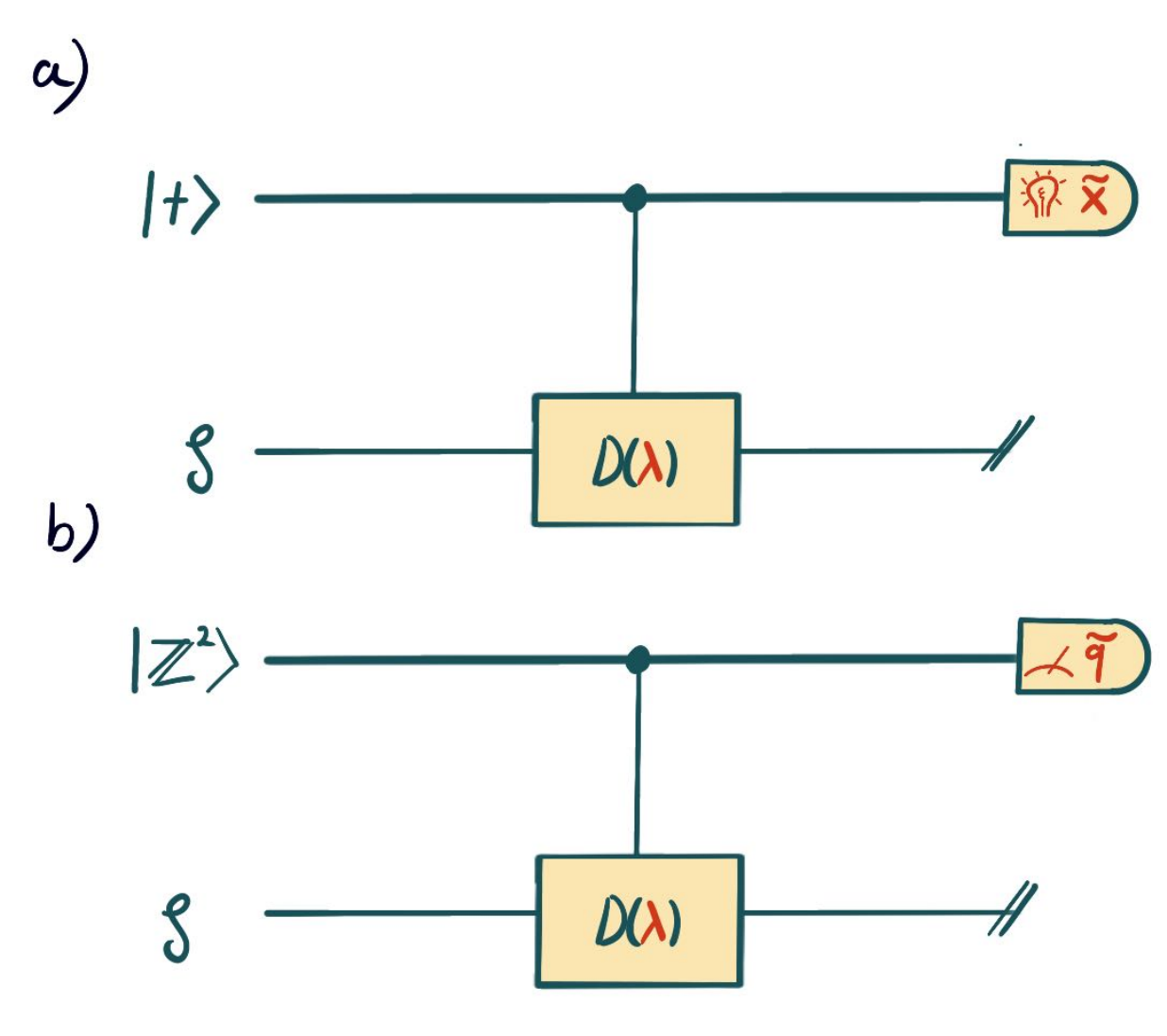}%
    \caption{
    Qubit-assisted and GKP-state-assisted stabilizer measurement circuits that implement the shadow tomography protocol. 
    $a)$ An implementation using a qubit-controlled displacement, where the qubit expectation value, $\langle \tilde X \rangle = P_{+}-P_{-}=\Re\lrc{\Tr\lrq{D\lr{\bs{\lambda}}\rho}}$, encodes the sampling probabilities.  
    $b)$ An auxiliary square-lattice GKP state $|\mathbb{Z}^2\rangle$ together with a mode-controlled displacement operation (a Gaussian unitary gate) is used to sample the probability distribution $P\lr{\tilde{p}}$ of momenta on the auxiliary mode, whose Fourier series coefficients encode all values $\Tr\lrq{D\lr{\ell\bs{\lambda}}\rho}$, $\ell \in \Z$.
    }
    \label{fig:Disp_circ}
\end{figure}

\section{Discussion \& Outlook}

We show that quantum lattice states (code words of all GKP codes~\cite{GKP}) form a continuous-variable $2$-design (in the sense of Ref.~\cite{Iosue_2024}).
This puts them on a similar footing with stabilizer states of qudits of dimension $q >2$, which also form a $2$-design.
We use this result to introduce global \textit{and} local GKP-based versions of qubit shadow tomography, present recipes on how to implement them, and derive bounds to understand their performance.

Our key technical tools are two formulas from lattice theory --- a mean-value formula and a second-moment bound, both applied over the space of symplectic lattices.
Ensembles of random lattices and more general quantum states that mimic lattice structure are relevant objects in classical (post-quantum) cryptography \cite{Regev_2009, conrad2023good}.  
We believe that the presented applications of these powerful tools are only the beginning of their use in quantum information processing. 

For example, it would be interesting to build a better understanding of average-case hardness of the learning with errors (LWE) problem~\cite{Regev_2009}, and how exactly average case hardness of the underlying problem relates to the ability of the corresponding lattice ensemble to realize the mean value formula.
Any such investigation may also shed light on which discrete subsets of the space of lattices constitute practical sample sets.
Provided that a lattice-based cryptosystem was used in Ref.~\cite{conrad2023good} to construct a suitable distribution of lattices, we 
can speculate whether average-case hardness of typical lattice problems over such distributions could be such a sufficient requirement.

For another example, bounds similar to the utilized second-moment formula have previously been derived for higher moments of euclidean lattices~\cite{Rogers1955}. 
While they have yet to be derived for the symplectic case, it would be interesting to understand any information theoretic implications. 

Our global GKP shadow sample complexity bound features the $L^1$ norm, a measure of non-Gaussianity of the state~\cite{Weedbrook_2012}. This is conceptually interesting, as the shadow reflects an ensemble of ``classical snapshots'' of a quantum state. 
A similar measure for observables on qubit systems is provided by the stabilizer entropy~\cite{chen2024nonstabilizernessenhancesthriftyshadow}, and it would be interesting if the $L^1$ norm of the characteristic function of observables could be endowed with similar operational meaning as their qubit analogues~\cite{Leone_2024}. 

The shadow bounds, derived under physicality assumptions and imposed locality, are similar to the bounds derived for qubit-based protocols in terms of their dependency on the trace of the squared observable. In stark contrast to such bounds, however, our GKP shadow bound does not directly depend on an observable's \textit{locality}, but merely on its native variance across bi-partitions, permitting the observable to maintain spatially very non-local support as long as it is of sufficiently fast phase-space decay. In these derivations, we have conjectured a lower bound on the effective squeezing measure for arbitrary states scaling with the inverse average photon number. We note that, through their operational interpretation as variances in a displacement-estimation problem~\cite{Duivenvoorden_Sensor}, the proof of our conjecture would establish lower bounds for the tomography of displacements, including entangled strategies. 
It may be valuable to base a resource theory of non-Gaussian quantum states directly off such measures.

It is interesting to note that the sample complexity scaling derived for GKP shadows, albeit assuming a conjecture, exhibits the same scaling in the properties of the observable as the expected scaling for a shadow tomography protocol derived for rigged $3$-designs in Ref.~\cite{Iosue_2024}.
While we have been unable to answer the question whether our ensemble of GKP states forms a $3$-design, this is an encouraging factor. 
Even if the GKP ensemble were to fail to provide a $3$-design property, it would be interesting to further investigate the extent to which this failure occurs \cite{zhu2016cliffordgroupfailsgracefully}.

%%The shadow bounds, derived under physicality assumptions and imposed locality, are similar to the bounds derived for qubit-based protocols in terms of their dependency on the trace of the squared observable. In stark contrast to such bounds, however, our GKP shadow bound does not directly depend on an observable's \textit{locality}, but merely on its native variance across bi-partitions, permitting the observable to maintain spatially very non-local support as long as it is of sufficiently fast phase-space decay. It is interesting to note that the sample complexity scaling derived for GKP shadows, albeit assuming a conjecture, exhibits the same scaling in the properties of the observable as the expected scaling for a shadow tomography protocol derived for rigged $3$-designs in Ref.~\cite{Iosue_2024}. While we have been unable to answer the question whether our ensemble of GKP states forms a $3$-design, this is an encouraging factor. 

\acknowledgments
J.~C. and A.~B. thank the University of Maryland and its members for their hospitality, where the essential steps presented in this work were developed and J.~C. further thanks Oma Eichler for providing the environment where the manuscript was initially developed. 
We thank 
R.~Alexander,
L.~Bittel,
A.~J.~Brady, 
J.~Eisert,
S.~T.~Flammia,
M.~J.~Gullans, 
J.~Haferkamp, 
L.~Leone,
A.~A.~Mele,
J.~P.~Seifert,
and N.~Walk
for helpful discussions as well as F.~A. Mele for pointing us to an error in an earlier version of this manuscript.
J.~T.~I. thanks the Joint Quantum Institute at UMD for support through a JQI fellowship.
We acknowledge support from NSF grants OMA-2120757 (QLCI) and CIF-2330909, the BMBF (RealistiQ, MUNIQC-Atoms, PhoQuant, QPIC-1, and QSolid),  the DFG (CRC 183, project B04, on entangled states of matter), the Munich Quantum Valley (K-8), the ERC (DebuQC), Quantum Berlin as well as the ERU on quantum devices.

\bibliography{Shadow_bib}

\setcounter{section}{0}
\setcounter{lem}{0}
\setcounter{them}{0}

\onecolumngrid
\newpage
\appendix 

\renewcommand{\tocname}{appendix table of contents}
\tableofcontents

% Everything here and below is included in the table of contents 

%\tableofcontents

\section*{Outline of appendix}
The following material is structured as follows. In section~\ref{sec:toolbox}, we review the technical tools necessary to follow the mathematical developments in this manuscript. In this section we cover the relevant tools from physics, briefly review GKP coding- and lattice theory and explain the structure of CV designs as defined in Ref.~\cite{Iosue_2024}. We close this section by constructing a ``warm up'' example of a CV $1$-design based on random lattice states. Section~\ref{sec:designs} presents the main result. We show that uniformly randomly displaced GKP states constructed from a random symplectic lattice yield a $2$-design as per the definition provided in Ref.~\cite{Iosue_2024} and we discuss how random ensembles of lattices can be obtained in practice. Sections~\ref{sec:CVshadows} and \ref{sec:GKPShadows} apply the core result by developing a CV shadow tomography protocol based on the design property. We investigate the example of estimating a thermal state observable as well as general bounds on the sample complexity without and with additional physical assumptions. In section~\ref{sec:Implementation} we discuss various recipes to physically implement the protocol. While we do not show whether our ensembles also attain the $3$-design property, we explore the ramifications of $3$-design based classical shadow tomography in sec.~\ref{sec:design_based_shadows}. Finally, we outline an application of our protocol in the variational preparation of a bosonic thermal state in section~\ref{sec:thermal}.

%\renewcommand{\tocname}{appendix table of contents}

% Everything here and below is included in the table of contents
\toc  

\section{The toolbox}\label{sec:toolbox}
\subsection{Continuous-variable physics}
An $n$-mode continuous-variable system is naturally described using a $2n$-dimensional phase space $\R^{2n}$ which supports the bosonic quadratures $\bs{\hat{x}}=(\hat{q}_1,\hdots \hat{q}_n,\hat{p}_1,\hdots, \hat{p}_n)$, which we assumed to be normalized such that $\lrq{\hat{x}_i, \hat{x}_j}=iJ_{ij}$ with $J=-J^T=-J^{-1}$ the symplectic form. An important class of operators is given by the displacement operators
\begin{equation}
    D\lr{\bs{\xi}}=e^{-i\sqrt{2\pi}\bs{\xi}^TJ\bs{\hat{x}}},
\end{equation}
which satisfy
\begin{equation}
    D\lr{\bs{\xi}}D\lr{\bs{\eta}}=e^{-i2\pi \bs{\xi}^TJ\bs{\eta}}D\lr{\bs{\eta}}D\lr{\bs{\xi}} = e^{-i\pi \bs{\xi}^TJ\bs{\eta}}D\lr{\bs{\xi}+\bs{\eta}}\label{eq:disp}
\end{equation}
and form a basis, orthonormal in the Dirac sense, for the space of trace-class operators on the CV system. 
That is, every trace-class operator $O$ can be expressed as
\begin{equation}
    O=\int_{\R^{2n}} d\bs{\xi}\, c_O\lr{\bs{\xi}}D\lr{\bs{\xi}},\label{eq:op_disp_basis}
\end{equation}
where $c_O\lr{\bs{\xi}}=\Tr[D^{\dagger}\lr{\bs{\xi}}O]$ is called the characteristic function of the operator $O$. 

A relevant set of operators for the CV system is given by the Fock operators $\hat{n}_i=\hat{a}_i^{\dagger}\hat{a}_i$ and the annihilation operators $\hat{a}_i=\lr{\hat{q}_i+i\hat{p}_i}/\sqrt{2}$. The Fock operators count the \textit{photon number} of the individual modes, and the total photon number is given by $\hat{N}=\sum_{i=1}^n \hat{n}_i$. 

The \textit{total photon parity operator} is given by $\hat{\pi}=e^{i\pi \hat{N}}$ and has a constant characteristic function $c_{\pi}\lr{\bs{\xi}}=2^{-n}$ (see e.g. Ref.~\cite{ChasingShadows}) such that the symplectic Fourier transform of the characteristic function of an operator $c_{\rho}\lr{\bs{\xi}}$ can be obtained via the expectation value of displaced parity operators (a.k.a. phase-space point operators),
\begin{equation}
    \Omega\lr{\bs{\xi}} = D\lr{\bs{\xi}} \hat{\pi}D^{\dagger}\lr{\bs{\xi}},
\end{equation}
which satisfy
\begin{equation}
    \Tr\lrq{\Omega\lr{\bs{\xi}} \rho}=2^{-n} \int_{\R^{2n}} d\bs{\eta}\, e^{-i2\pi \bs{\xi}^T J \bs{\eta}} \Tr\lrq{D\lr{\bs{\eta}} \rho}
    = 2^{-n} W_{\rho}\lr{\bs{\xi}}.
\end{equation}
The function $W_{\rho}\lr{\bs{x}}$ is the \textit{Wigner function} of the state (or general operator) $\rho$. For quantum states, it represents a quasi-probability distribution over phase space and the formulas above imply that traces of operators and their Wigner functions can be obtained as
\begin{align}
    \Tr\lrq{A}=\int_{\R^{2n}} d\bs{\xi}\,W_A\lr{\bs{\xi}}, \quad
    \Tr\lrq{AB}=\int_{\R^{2n}} d\bs{\xi}\,W_A\lr{\bs{\xi}}W_B\lr{\bs{\xi}}.
\end{align}

The Wigner function of a state $\rho=\rho^{\dagger}$ always takes real values, and we have the bound $|W_A|\leq 2^n \|A\|$.
Throughout this manuscript we will make an important distinction between \textit{formal} and \textit{physical} operators, where we say an operator is physical if the corresponding Wigner function is a Schwartz function -- i.e., a function that admits finite expectation values of any power of Fock operators \cite{albert2022bosoniccodingintroductionuse}. In terms of the characteristic function of a state this requirement may be phrased as follows. As the Fourier transform is an automorphism of Schwartz space, the requirement of the Wigner function of any physical state to be Schwartz implies this attribute also to the characteristic function, which in particular means that $c_{\rho}\lr{\bs{\xi}}$ decays superpolynomially as $\|\xi\|\to \infty$ and we can restrict $c_{\rho}\lr{\bs{\xi}}$ to have compact essential support below a radius $R_{\rho}$. We call $R_{\rho}$ the \textit{bandwidth} of a state (or operator, respectively). As it quantifies the length of the longest relevant displacement appearing in the decomposition of Eq.~\eqref{eq:op_disp_basis}. As $D\lr{\bs{x}}D\lr{\bs{\xi}}D^{\dagger}\lr{\bs{x}}=e^{-i2\pi\bs{x}^TJ\bs{\xi}}$, this length-scale also determines the length $\|\bs{x}\|\sim \|\bs{\xi}\|^{-1}$ of the \textit{shortest} displacement $D\lr{\bs{x}}$ under which the corresponding state-component is displacement-invariant. One may hence think of the displacement vectors appearing in the decomposition of Eq.~\eqref{eq:op_disp_basis} as generalized \textit{frequencies} of the corresponding operator.

Furthermore it holds that, for $\beta \in (0, \infty)$  \cite{CahillGlauber, surfGKP},
\begin{equation}
    e^{-\beta \hat{N}}=\lr{1-e^{-\beta}}^{-n}\int_{\R^{2n}} d\bs{\gamma}\, e^{-\frac{\pi \|\bs{\gamma}\|^2}{\Delta^2}}D\lr{\bs{\gamma}},
\end{equation}
where $\Delta^2=2\tanh\lr{\beta /2}=\beta+O\lr{\beta^3}\in (0,2)$. We say that a state is \textit{physical}, if 
\begin{equation}
    N\lrq{\rho}\lr{\beta}=\Tr\lrq{\rho e^{-\beta \hat{N}}}=\lr{1-e^{-\beta}}^{-n}\int_{\R^{2n}} d\bs{\gamma}\,e^{-\frac{\pi \|\bs{\gamma}\|^2}{\Delta^2}} c_{\rho}\lr{\bs{\gamma}}.
\end{equation}
is an \textit{analytical} function of $\beta$ at $\beta=\Delta^2=0 $. In particular this means that each moment
\begin{equation}
    \lim_{\beta \to 0} \lr{-\frac{\partial}{\partial\beta}}^k N\lrq{\rho}\lr{\beta} = \Tr\lrq{\rho \hat{N}^k}
\end{equation}
is finite \cite{albert2022bosoniccodingintroductionuse, mele2024learningquantumstatescontinuous}.
A  direct route to understand this characterization is to recognize the characteristic function as the \textit{moment generating function} of a state, i.e. as 
\begin{equation}
    c_{\rho}\lr{\bs{\gamma}}=\Tr\lrq{D^{\dagger}\lr{\bs{\gamma}}\rho},
\end{equation}
the moments 
\begin{equation}
    \Tr\lrq{P\lr{\bs{\hat{x}}}\big\vert_{W} \rho}= \lim_{\bs{\gamma}\to 0} P\lr{-\frac{iJ\bs{\nabla}}{\sqrt{2\pi}}}c_{\rho}\lr{\bs{\gamma}},
\end{equation}
for any multivariate~\cite{Gerry2004} polynomial $P:\R^{2n}\to \R$ are encoded in the fluctuating behaviour of $c_{\rho}$ around $\bs{\gamma}=0$. The index ``$W$'' indicates that we assume the polynomial evaluated in the quadrature operators to be Weyl-odered~\cite{CahillGlauber, Gerry2004}, i.e. symmetrized amongst all permutations of ordering the quadrature operators  $(\hat{x}_i, \hat{x}_{i+n}),\; i =1,\hdots, n$.

A more general expression an be obtained by computing the partial derivatives of the displacement operators (see also ref.~\cite{bittel2024optimalestimatestracedistance} for an alternative form). Using the derivative of an exponential map~\cite{RossmanLie}
\begin{equation}
    \frac{d}{dt}e^{A(t)}=\int_{0}^1 dy\, e^{y A(t)}A'(t)e^{(1-y) A(t)}
\end{equation}
we obtain the expressions
\begin{align}
\bs{\nabla}D\lr{\bs{\gamma}} &= -i\sqrt{2\pi}J\lr{\bs{\hat{x}}-\sqrt{\frac{\pi}{2}}\bs{\gamma}} D\lr{\bs{\gamma}} \\
    \Delta D\lr{\bs{\gamma}} &=  
Q\lr{\bs{\gamma}}D\lr{\bs{\gamma}}, \hspace{3cm}\text{with} \\
    Q\lr{\bs{\gamma}}&=-2\pi \lr{\bs{\hat{x}}-\sqrt{\frac{\pi}{2}}\bs{\gamma}}^T\lr{\bs{\hat{x}}-\sqrt{\frac{\pi}{2}}\bs{\gamma}}=-2\pi D\lr{\frac{\bs{\gamma}}{2}}\bs{\hat{x}}^T\bs{\hat{x}}D^{\dagger}\lr{\frac{\bs{\gamma}}{2}}.
\end{align}
We recognize that 
\begin{equation}
    Q\lr{0}=-4\pi \lr{\hat{N}+n/2},
\end{equation}
as well as $\Delta Q\lr{\bs{\gamma}} = -\lr{2\pi}^2n$.
Now using the product rule for Laplaciancs
\begin{equation}
    \Delta\lr{f g} = \lr{\Delta f} g + 2\lr{\bs{\nabla}f}^T\lr{\bs{\nabla}g} + f\lr{\Delta g}
\end{equation}
it can be shown via induction that for every $k, n \in \N$ there exists a sequence of constants $c_m^{k,n}$ such that
\begin{equation}
    \Delta^k D\lr{\bs{\gamma}} = \sum_{m=0}^k c_m^{k,n} Q^m\lr{\bs{\gamma}} D\lr{\bs{\gamma}}.
\end{equation}
In particular, this also implies that

\begin{equation}
    \Delta^k c_{\rho}\lr{0} = \sum_{m=0}^k c_m^{k,n} (-4\pi)^m \Tr\lrq{\lr{\hat{N}+\frac{n}{2}}^m \rho}.
\end{equation}
That is, the fluctuations of the characteristic function $c_{\rho}(\bs{\gamma})$ around $\bs{\gamma}=0$ encode the moments of the photon number operator (see also ref.~\cite{Becker_2021}).

\subsection{Lattices \& GKP states}
Gottesman-Kitaev-Preskill (GKP) codes \cite{GKP, Conrad_2022} are stabilizer codes whose stabilizer group is generated by a full-rank set of displacement operators. The properties in Eq.~\eqref{eq:disp} imply that the stabilizer group becomes isomorphic to a (weakly) symplectically self-dual lattice $\CL\subseteq \CL^{\perp} \subset \R^{2n}$, i.e. the integer span of a full rank matrix $\CL=\Z^{2n}M$ with integral symplectic Gram matrix $A=MJM^T$, whose points label the displacements occuring in the stabilizer group.
Here duality is implied by the symplectic inner product represented by matrix $J$. The subspace symmetric under this stabilizer group has dimension $d=\sqrt{|\CL/\CL^{\perp}|}=\sqrt{|\det\lr{\CL}/\det\lr{\CL^{\perp}}|}=|\det\lr{\CL}|$ so that in particular symplectic (symplectic self-dual) lattices $\Lambda=\Lambda^{\perp}$ fix a one-dimensional subspace of the CV Hilbert space spanned by a formal state described by the (formal) group projector \cite{Conrad_2022}
\begin{equation}
    \Pi_{\Lambda}=\ketbra{\Lambda}=\sum_{\bs{\lambda}\in \Lambda} e^{i\Phi\lr{\bs{\lambda}}} D\lr{\bs{\lambda}}.
\end{equation}

Note, that throughout this manuscript we explicitly include the phases $e^{i\Phi\lr{\bs{\lambda}}}$, which result from the fact that -- in our slight abuse of notation -- the states $\ket{\Lambda}$ are in fact not fully determined by the lattice $\Lambda$ as a geometric object, but their specification typically involves a preferred choice of basis \cite{Conrad_2022, Royer_2022, burchards2024fiberbundlefaulttolerance}. Note, however, that in all subsequent discussions these phases will turn out to become irrelevant.

The phase $e^{i\Phi\lr{\bs{\lambda}}}$ takes values $\pm 1$ depending on the decomposition of $\bs{\lambda}$ into integer linear combinations of the minimal basis chosen to generate the corresponding stabilizer group \cite{Conrad_2022, burchards2024fiberbundlefaulttolerance}. Let $\ket{\bs{\lambda}}=D\lr{\bs{\lambda}}\ket{0}$ be the coherent state at $\bs{\lambda}\in \R^{2n}$. Coherent states are such that
\begin{equation}
\braket{\bs{\beta}|\bs{\alpha}}=e^{-\frac{\pi}{2}\lr{\|\bs{\alpha}-\bs{\beta}\|^2+i2\bs{\alpha}^TJ\bs{\beta}}}
\end{equation}
and form an overcomplete basis \cite{serafini2017quantum-continu}
\begin{equation}
    \int_{\R^{2n}} d\bs{\alpha} \ketbra{\bs{\alpha}}=I.
\end{equation}

\subsection{The mean value formula}
The space of symplectic lattices $Y_n=\Sp_{2n}\lr{\Z}\backslash \Sp_{2n}\lr{\R}$-- parametrized by symplectic matrices $\Sp_{2n}\lr{\R}$ \textit{up to basis transformations} $\Sp_{2n}\lr{\Z}$ acting from the left admits a natural normalized Haar measure. This insight had been extended from the theory of classical lattices $X_n=\SL_{2n}\lr{\Z}\backslash \SL_{2n}\lr{\Z}$, originally examined by Siegel \cite{Siegel_meanvalue}, to the case of symplectic lattices \cite{Sarnak1994, conrad2023good, garret_volumes}. Over this measure, an elegant formula holds for the mean value of a lattice sum when averaged over the space of respective lattices \footnote{In the literature the mean value formula is typically stated for compact functions. Note that any (Riemann) integrable function can be partitioned into a sum of functions over compactly restricted domains, so that this statement naturally generalizes. This observation is also implicitly present in Ref.~\cite{Rogers1955}}.

\begin{lem}[Symplectic mean value formula {\cite{Sarnak1994}~\citep[theorem 2]{Moskowitz2010}}]
    \label{lem:mean-value}
    Let $f:\, \R^{2n}\rightarrow\R$ be a (Riemann) integrable function following quantities converge absolutely and let $\Lambda \in Y_n$ denote a symplectic lattice. For
    \begin{equation}
        F\lr{\Lambda}=\sum_{\bs{\lambda}\in \Lambda-\lrc{0}} f\lr{\bs{\lambda}}
    \end{equation}
    it holds that 
    \begin{equation}
        \int_{Y_n} d\mu\lr{\Lambda} F\lr{\Lambda} = \int_{\mathbb{R}^{2n}} d\bs{x}\, f\lr{\bs{x}} \label{eq:mean_val_app}
    \end{equation}
    relative to the Haar measure over the space of symplectic lattices $Y_n=\Sp_{2n}\lr{\Z} \backslash \Sp_{2n}\lr{\R}$.
\end{lem}

In what follows, this formula will be the main ingredient for our construction of CV state designs via lattice-states.

\subsection{Continuous variable designs}
We briefly recall the relevant definition of continuous-variable state designs for $n$-mode quantum systems as stated in Ref.~\cite{Iosue_2024}. 
This requires a slightly extended definition of states.
Rather than requiring that states are representable as square integrable wave functions in $\CH=L^2\lr{\R^{n}}$, we also allow for \textit{improper} states which can be represented within the space of tempered distributions $S'\lr{\R^{n}}$, i.e. the %
continuous %
dual of the space of Schwartz functions $S\lr{\R^n}$ under the standard inner product. 
This sequence of spaces forms the well-known Gelfand triple $S\lr{\R^n}\subset L^2\lr{\R^{n}}\subset S'\lr{\R^{n}}$, with the first two spaces in the triple referred to as a \textit{rigged Hilbert space}. 
The point of this relaxation is, as we will see in the course of the manuscript, that whenever we encounter such improper states, they can still be made sense of as a measurement process that acts on a physical state.

Formally, a continuous-variable, or rigged, $t$-design is a measure space $\lr{\CX_n, \Sigma, \mu}$, where $\CX_n\subset S'\lr{\R^{n}}$ is a subset of tempered distributions, the $\Sigma$-algebra defines a means to take unions and complements of elements in $\CX_n$, and where $\mu_{\CX_n}$ is a measure on $\CX_n$.

\begin{definition}[Rigged designs \cite{Iosue_2024}]
    Let $\CX_n \subset S'\lr{\R^{n}}$. The measure space $\lr{\CX_n, \Sigma, \mu_{\CX_n}}$ is a \textit{rigged} $t$-design if there are finite constants $\des_{t'} > 0 $, such that
    \begin{equation}
        \int_{\CX_n} d\mu_{\CX_n}\lr{X} \ketbra{X}^{\otimes t'} = \des_{t'} \Pi_{t'} \; \forall t' \leq t.
    \end{equation}
\end{definition}

In this definition the objects $\Pi_t$ appearing on the right are the projectors onto the symmetric subspaces of the Hilbert space $\CH^{\otimes t}$. 
The relaxation of $L^2\lr{\R^n}$ to $S'\lr{\R^n}$ allows $(t\geq 2)$-designs to exist. Despite elements of $S'\lr{\R^n}$ not being normalizable states, this relaxation often does not affect the physicality of the design itself.
The reason is because vectors corresponding to \textit{physical} states are those with finite energy and finite position and momentum moments, and hence belong to $S\lr{\R^n}$. 
Inner products of tempered distributions (elements of $S'\lr{\R^n}$) and Schwartz functions (elements of $S\lr{\R^n}$) are always well-defined.
It follows that even though a rigged $t$-design must consist of tempered distributions, measuring from the POVM that it defines is perfectly well-defined, as expectation values of physical observables over tempered distributions are well-defined.
This ultimately allows shadow tomography with rigged $t$-designs to be a valid protocol \cite{Iosue_2024}.

However, we may also encounter situations where we want the elements of our designs to correspond to normalizable states. To this end, Ref.~\cite{Iosue_2024} defined regularizers $R_{\beta}: S'\lr{\R^{n}} \rightarrow L^2\lr{\R^{n}}$ for $\beta > 0$, where $\beta$ is some parameter such that $R_{0}=I$, as well as the regularized projectors $\Pi_t^{(R_{\beta})}= R_{\beta}^{\otimes t} \Pi_t R_{\beta}^{\otimes t}$

\begin{definition}[Regularized rigged designs \cite{Iosue_2024}]
    An $R_{\beta}$-regularized (rigged) $t$-design is a measure space $\lr{\CY, \Sigma_Y, \mu_{\CY}}$ with $\CY \subset L^2\lr{\R^{n}}$, such that
    \begin{equation}
        \int_{\CY} d\mu_{\CY}\lr{Y} \ketbra{Y}^{\otimes t'} = \des_{t'} \Pi^{(R_{\beta})}_{t'}/\Tr\lrq{\Pi^{(R_{\beta})}_{t'}} \; \forall t' \leq t.
    \end{equation}
\end{definition}

\subsection{Warm up: a CV $1$-design from random GKP states}

In this first application we use the mean value formula to show that that GKP states provided by random lattices $\Lambda \in Y_n$ form a CV $1$-design. Note that, due to its over-completeness, the set of coherent states $\ketbra{\bs{\alpha}}$ also possesses this property. In contrast to the coherent states, labeled by $\bs{\alpha}\in \R^{2n}$, however, the parametrizing space $Y_n$ has more interesting geometry and remains measurable. The main goal of this example is to provide a first demonstration of how to use the mean value formula for $Y_n$.
Let us begin by defining the lattice state
\begin{equation}
    \rho_{\Lambda}=\sum_{\bs{\lambda}\in \Lambda} \ketbra{\bs{\lambda}}=\sum_{\lambda\in \Lambda} D\lr{\bs{\lambda}}\ketbra{0}D^{\dagger}\lr{\bs{\lambda}},\label{eq:rhoLambda}
\end{equation}
which distinguishes itself from $\ketbra{\Lambda}$ in that its Wigner function is described by a fully classical probability distribution. The state $\rho_{\Lambda}$ can be obtained by performing a heterodyne POVM on $\Lambda$ as
\begin{align}
    \int_{\R^{2n}} d\bs{\alpha}\, \ket{\bs{\alpha}}\!\braket{\bs{\alpha}|\Lambda}\!\braket{\Lambda|\bs{\alpha}}\!\bra{\bs{\alpha}} & = \int_{\R^{2n}} d\bs{\alpha}\, \sum_{\bs{\lambda}, \bs{\lambda}'\in \Lambda}
    \ket{\bs{\alpha}}\!\braket{\bs{\alpha}|\bs{\lambda}}\!\braket{\bs{\lambda}'|\bs{\alpha}}\!\bra{\bs{\alpha}}
    \nonumber                                                                                                                                                                                  \\  & =\int_{\R^{2n}} d\bs{\alpha}\, \sum_{\bs{\lambda}, \bs{\lambda}'\in \Lambda} e^{-\frac{\pi}{2}\lr{\|\bs{\alpha}-\bs{\lambda}\|^2+\|\bs{\alpha}-\bs{\lambda}'\|^2}-i\pi \lr{\bs{\lambda}-\bs{\lambda}'}^TJ\bs{\alpha} } \ketbra{\bs{\alpha}}.
\end{align}
In the limit of large shortest lattice vector $\lambda_1\lr{\Lambda}\rightarrow \infty$, the sum is essentially supported on parts where $\bs{\lambda}=\bs{\lambda}'$, such that the state becomes
\begin{equation}
    \widetilde{\rho_{\Lambda}}=\int_{\R^{2n}} d\bs{\alpha}\, \sum_{\bs{\lambda}\in \Lambda} e^{-\pi\|\bs{\alpha}-\bs{\lambda}\|^2} \ketbra{\bs{\alpha}},
\end{equation}
which approximates $\rho_{\Lambda}$ up to an increased uncertainty around each lattice point.

Using the mean value formula, we obtain the following lemma.
\begin{lem}
    The set of formal states
    \begin{equation}
        \rho_{\Lambda}^{\times} =\rho_{\Lambda}-\ketbra{0}= \sum_{\bs{\lambda}\in\Lambda-\lrc{0}} \ketbra{\bs{\lambda}}
    \end{equation}
    for all $\Lambda \in Y_n$ forms a CV $1$-design. That is, we have that
    \begin{align}
        \int_{Y_n} d\mu\lr{\Lambda}  \rho_{\Lambda}^{\times}  = I.
    \end{align}
\end{lem}
\proof
We use the fact that coherent sates resolve the identity, and apply the mean-value formula
\begin{align}
    \int_{Y_n} d\mu\lr{\Lambda}  \rho_{\Lambda}^{\times} & = \int_{\R^{2n}} d\bs{\alpha}d\bs{\beta}\,\lrq{ \int_{Y_n} d\mu\lr{\Lambda} \sum_{\bs{\lambda}\in \Lambda-\lrc{0}}\braket{\bs{\alpha} | \bs{\lambda}}\braket{\bs{\lambda}|\bs{\beta}}}\ket{\bs{\alpha}}\!\bra{\bs{\beta}} \nonumber \\
     & =\int_{\R^{2n}} d\bs{\alpha}d\bs{\beta}\,\lrq{ \int_{\mathbb{R}^{2n}} d\bs{\lambda}\,\braket{\bs{\alpha} | \bs{\lambda}}\braket{\bs{\lambda}|\bs{\beta}}}\ket{\bs{\alpha}}\!\bra{\bs{\beta}}                    \nonumber \\
     & =\int_{\R^{2n}} d\bs{\alpha}d\bs{\beta}\,\braket{\bs{\alpha} | \bs{\beta}}\ket{\bs{\alpha}}\!\bra{\bs{\beta}}                                                                                                             \nonumber \\
     & = I.
\end{align}
where we have used the fact that $\braket{\bs{\alpha} | \bs{\lambda}}\braket{\bs{\lambda}|\bs{\beta}}$ is an integrable function of $\bs{\lambda}$.
\endproof

The states $\rho^{\times}$ are GKP states that have decohered due to a heterodyne measurement, making for a poor ``classical'' approximation to the corresponding GKP states. It is hence reasonable to expect the original family of GKP states to possess an even richer set of properties.

\section{State $2$-designs from displaced lattices}\label{sec:designs}

Displacement operators can be used to expand not just trace-class operators, but also some unitary operators. Applying this expansion to the parity operator, whose characteristic function is $2^{-n}$, and plugging in the mean-value formula yields the following useful fact.
\begin{lem}
    It holds that
    \begin{equation}
        \int_{Y_n} d\mu\lr{\Lambda} \sum_{\bs{\lambda}\in \Lambda - \lrc{0}} D\lr{\bs{\lambda}} = 2^n e^{i\pi \hat{N}}.
    \end{equation}
\end{lem}
\proof
We have that for every trace class operator $O$ that preserves Schwartz space, $\Tr\lrq{D\lr{\bs{\lambda}}O}$ is an integrable function of $\bs{\lambda}\in \R^{2n}$.
Hence, we can apply the mean value formula to obtain for every $O$
\begin{equation}
    \int_{Y_n} d\mu\lr{\Lambda} \sum_{\bs{\lambda}\in \Lambda - \lrc{0}} \Tr\lrq{D\lr{\bs{\lambda}}O} = \int_{\mathbb{R}^{2n}} d\bs{\lambda}\Tr\lrq{D\lr{\bs{\lambda}}O} = 2^n \Tr\lrq{e^{i\pi\hat{N}}O}.
\end{equation}
Since displacement operators $D\lr{\bs{\alpha}}$ form a complete operator basis, the above equality implies that more generally
\begin{equation}
    \int_{Y_n} d\mu\lr{\Lambda} \sum_{\bs{\lambda}\in \Lambda - \lrc{0}} D\lr{\bs{\lambda}} = 2^n e^{i\pi \hat{N}}.
\end{equation}
Note that the above integrals and sums are well-defined in their action on the space of physical states, such that all physical expectation values and inner products involving these  operator sums and -integrals are well-defined and consistent with the above equality.
\endproof

\subsection{A rigged $2$-design from displaced GKP states}
Let $\CP\lr{\Lambda}$ denote a fundamental domain for the lattice $\Lambda$. In the following, we prove a second moment formula.

\begin{them}[GKP states form a rigged $2$-design]\label{them:2design}

The ensemble $\CX_n$, w.r.t.~the Haar measure $\mu\lr{\Lambda}$ over the space of lattices $Y_n$ and the uniform measure over the fundamental domain $\CP\lr{\Lambda}$, forms a rigged $2$-design. In particular, it holds that 

\begin{equation}
    \int_{Y_n}d\mu\lr{\Lambda}\int_{\CP\lr{\Lambda}} d\bs{\alpha}\, D\lr{\bs{\alpha}} \Pi_{\Lambda} D^{\dagger }\lr{\bs{\alpha}} =  I.
\end{equation}
and

    \begin{equation}
        \int_{Y_n} d\mu\lr{\Lambda}\int_{\CP\lr{\Lambda}} d\bs{\alpha}\, D^{\otimes 2}\lr{\bs{\alpha}} \Pi_{\Lambda}^{\otimes 2} D^{\dagger \, \otimes 2}\lr{\bs{\alpha}} = I\otimes I + \widehat{E},
    \end{equation}
    where
    \begin{equation}
        \widehat{E}=e^{-i\frac{\pi}{2} \sum_j \lr{\hat{a}_j-\hat{b}_j}^{\dagger}\lr{\hat{a}_j-\hat{b}_j}}
    \end{equation}
    is the bosonic exchange, or SWAP, operator between $n$ pairs of modes $\{\hat a_j,\hat b_j\}$.
\end{them}

\proof

We first show that the given set of states yields a rigged $1$-design. This can in fact be shown by only the properties of the displacements in $\CP\lr{\Lambda}$. 
We have
\begin{align}
    \int_{Y_n}d\mu\lr{\Lambda}\int_{\CP\lr{\Lambda}} d\bs{\alpha}\, D\lr{\bs{\alpha}} \Pi_{\Lambda} D^{\dagger }\lr{\bs{\alpha}} & =
    \int_{Y_n}d\mu\lr{\Lambda}\sum_{\bs{\lambda}\in \Lambda} e^{i\Phi \lr{\bs{\lambda}}} D\lr{\bs{\lambda}} \int_{\CP\lr{\Lambda}} d\bs{\alpha}\,e^{-i2\pi \bs{\alpha}^TJ\bs{\lambda}}.\label{eq:rigged_onedesign}
\end{align}
To evaluate this integral, first notice that
\begin{align}
    \delta\lr{\bs{x}}=\int_{\R^{2n}} d\bs{\alpha}\,
    e^{-i2\pi \bs{\alpha}^TJ\bs{x}}=\sum_{\bs{\lambda}\in \Lambda}\int_{\CP\lr{\Lambda}} d\bs{\alpha}\,
    e^{-i2\pi \lr{\bs{\alpha}+\bs{\lambda}}^TJ\bs{x}}.
\end{align}
Using Poisson summation, and assuming $\Lambda=\Lambda^{\perp}$,
it holds that this expression equals 
\begin{align}
    \sum_{\bs{\lambda}\in \Lambda} \int_{\CP\lr{\Lambda}} d\bs{\alpha} \int_{\R^{2n}}d\bs{y}\,
    e^{-i2\pi \lr{\bs{\alpha}+\bs{y}}^TJ\bs{x}}e^{i2\pi \bs{y}^TJ\bs{\lambda}} & =
    \sum_{\bs{\lambda}\in \Lambda} \int_{\CP\lr{\Lambda}} d\bs{\alpha} \int_{\R^{2n}}d\bs{y}\,
e^{-i2\pi\bs{\alpha}^TJ\bs{x}} e^{-i2\pi \bs{y}^TJ\lr{\bs{x}-\bs{\lambda}}}                                                                                                                                          \nonumber\\
   & =\sum_{\bs{\lambda}\in \Lambda} \int_{\CP\lr{\Lambda}} d\bs{\alpha} e^{-i2\pi\bs{\alpha}^TJ\bs{\lambda}}\delta\lr{\bs{x}-\bs{\lambda}},
\end{align}
such that it has to hold that
\begin{equation}
    \int_{\CP\lr{\Lambda}} d\bs{\alpha} e^{-i2\pi\bs{\alpha}^TJ\bs{\lambda}} = \delta_{\bs{0}, \bs{\lambda}},\;\bs{\lambda}\in \Lambda.\label{eq:int_PLambda}
\end{equation}
Inserting this expression in Eq.~\eqref{eq:rigged_onedesign} yields the $1$-design property

\begin{equation}
    \int_{Y_n}d\mu\lr{\Lambda}\int_{\CP\lr{\Lambda}} d\bs{\alpha}\, D\lr{\bs{\alpha}} \Pi_{\Lambda} D^{\dagger }\lr{\bs{\alpha}} =  I.
\end{equation}

We proceed to show the design property also for the second moment. To this end compute
\begin{align}
    \widehat{M}_{\Lambda}\coloneqq\int_{\CP\lr{\Lambda}} d\bs{\alpha}\, D^{\otimes 2}\lr{\bs{\alpha}} \Pi_{\Lambda}^{\otimes 2} D^{\dagger \, \otimes 2}\lr{\bs{\alpha}} & = \sum_{\bs{\lambda}, \bs{\lambda}'\in \Lambda} e^{i\lrq{\Phi \lr{\bs{\lambda}}+\Phi \lr{\bs{\lambda}'}}} D\lr{\bs{\alpha}}D\lr{\bs{\lambda}}D^{\dagger}\lr{\bs{\alpha}}\otimes D\lr{\bs{\alpha}}D\lr{\bs{\lambda}'}D^{\dagger}\lr{\bs{\alpha}} \\
     & =\sum_{\bs{\lambda}, \bs{\lambda}'\in \Lambda} e^{i\lrq{\Phi \lr{\bs{\lambda}}+\Phi \lr{\bs{\lambda}'}}}\lrc{\int_{\CP\lr{\Lambda}} d\bs{\alpha}\,
        e^{-i2\pi \bs{\alpha}^TJ\lr{\bs{\lambda}+\bs{\lambda}'}} }D\lr{\bs{\lambda}}\otimes D\lr{\bs{\lambda}'}. \label{eq:second_1}
\end{align}

To evaluate the integral in the curly brackets, we again use the expression derived in Eq.~\eqref{eq:int_PLambda} to obtain
\begin{align}
    \widehat{M}_{\Lambda}=\int_{\CP\lr{\Lambda}} d\bs{\alpha}\, D^{\otimes 2}\lr{\bs{\alpha}} \Pi_{\Lambda}^{\otimes 2} D^{\dagger \, \otimes 2}\lr{\bs{\alpha}}=\sum_{\bs{\lambda}\in \Lambda} D\lr{\bs{\lambda}}\otimes D\lr{\bs{-\lambda}},
\end{align}
where we have also used that $\Phi\lr{\bs{\lambda}}=\Phi\lr{-\bs{\lambda}}$ is a symmetric function and an integer multiple of $\pi$~\cite{Conrad_2022} such that the total phase in each part of the sum has now become trivial.

Following Ref. \cite{Blume_Kohout_2014}, we now express each pair of modes in terms of a center-of-mass mode and a relative difference mode.
There exists a Gaussian unitary (a beam splitter transformation) $U$ inducing a symplectic transformation $S\in \Sp_{4n}\lr{\R}$ such that
\begin{equation}
    S\begin{pmatrix} \bs{\lambda} \\ -\bs{\lambda} \end{pmatrix} =\begin{pmatrix} \sqrt{2}\bs{\lambda} \\ \bs{0} \end{pmatrix},
\end{equation}
and in particular
\begin{equation}
    U\widehat{M}_{\Lambda} U^{\dagger} =\sum_{\bs{\lambda}\in \Lambda} D\lr{\sqrt{2}\bs{\lambda}}\otimes I~.
\end{equation}

We can now average this expression over the space of all lattices in $Y_n$, again using the fact that displacement operators form a complete operator basis and are operators with compact characteristic function so that the mean value formula becomes applicable. Note that the following sums and integrals converge in terms of well-defined operators on Schwartz space. This yields
\begin{align}
    U \widehat{M} U^{\dagger} & \coloneqq \int_{Y_n}d\mu\lr{\Lambda} \sum_{\bs{\lambda}\in \Lambda} D\lr{\sqrt{2}\bs{\lambda}}\otimes I \\
  & =I\otimes I + \int_{\R^{2n}} d\bs{\lambda} D\lr{\sqrt{2}\bs{\lambda}}\otimes I                                    \\
  & =I\otimes I + 2^{-n}\lrq{\int_{\R^{2n}} d\bs{\lambda} D\lr{\bs{\lambda}}}\otimes I                                \\
  & = I\otimes I + e^{i\pi\hat{N}}\otimes I.
\end{align}
Finally, let us introduce generalized annihilation operators $\bs{\hat{a}}, \bs{\hat{b}}$ on the two tensor factors.
Using the fact that $U^{\dagger}\hat{a}_iU=(\hat{a}_i-\hat{b}_i)/\sqrt{2}$  yields

\begin{equation}
    U\lr{e^{i\pi\hat{N}}\otimes I} U^{\dagger}=e^{-i\frac{\pi}{2} \sum_j \lr{\hat{a}_j-\hat{b}_j}^{\dagger}\lr{\hat{a}_j-\hat{b}_j}}\eqqcolon \widehat{E},
\end{equation}
with $\widehat{E}$ the bosonic swap operator.
\endproof

\subsection{Regularized $2$-designs from finitely squeezed GKP states}
There is a fundamental problem with the states $\ket{\Lambda}$ defined above: they are unphysical (i.e., are not in Schwartz space). This is most easily recognized by computing their Wigner function $W_{\Pi_{\Lambda}}\lr{\bs{x}}=2^n \Tr\lrq{\Omega \lr{\bs{x}}\Pi_{\Lambda}}$, which can be verified to constitute a non-compact function on phase space and fails to be normalizable as $\Tr\lrq{\Pi_{\Lambda}}$ formally diverges. 
The average photon number $\langle\hat{N}\rangle$ of such a state diverges as well.

The typical way to deal with this problem within analyses of GKP error correction is to replace $\ket{\Lambda}$ by an approximate state \cite{GKP,Menicucci_2014}
\begin{equation}
    \ket{\Lambda_{\beta}}= N_{\beta} e^{-\beta \hat{N}} \ket{\Lambda}~,
\end{equation}
where the envelope operator $R_{\beta}=e^{-\beta \hat{N}}$ implements an isotropic smooth cutoff to the state and converts the unphysical, formal, state into one that is producible in experiments \cite{GKP, Terhal_2020, Tzitrin_2020, Matsuura_2020, sivak2023real, Konno_2024, Campagne_Ibarcq_2020, Fluehmann_2019}. Here, $N_{\beta}$ is the relevant normalization constant (see Ref.~\cite{Terhal_2020} for an explicit expression).

The finite squeezing error admits a decompositon into displacement operators given by \cite{CahillGlauber, Noh_2020}
\begin{equation}
    e^{-\beta \hat{N}}=\lr{1-e^{-\beta}}^{-n} \int_{\R^{2n}} d\bs{x}\, e^{-\frac{\pi \|\bs{x}\|^2}{\Delta^2}}D\lr{\bs{x}},\label{eq:Rbeta_disp}
\end{equation}
where $\Delta^2=2\tanh\lr{\beta /2} = \beta + O\lr{\beta^2}$ is typically regarded as the square of the \textit{squeezing parameter} as it corresponds to the variance of the width of each Gaussian peak of $\ket{\Lambda_{\beta}}$ in phase space.
Furthermore, $\Delta^{-2}$ quantifies the variance of an enveloping Gaussian on the state that renders it normalizable (see Fig.~\ref{fig:GKPOverlap} for an illustration of the structure of such states).

A simple strategy to obtain a physically valid set of states for Thm.~\ref{them:2design} is to replace every state $\ket{X}$ in the ensemble by states of the form $N_{\beta} R_{\beta} \ket{X}$ with $N_{\beta}=\braket{X|R_{2\beta}|X}$ the normalization factor. By linearity of the theorem above, we immediately obtain the following implication.

\begin{cor}\label{cor:rigged2design}
    For lattices $\Lambda \in Y_n$ and $\bs{\alpha} \in \CP\lr{\Lambda}$ vectors in the fundamental domains of each lattice $\Lambda$, let $\beta>0$ be a small parameter. The set of states
    \begin{equation}
        R_{\beta} \lr{D\lr{\bs{\alpha}} \Pi_{\Lambda} D^{\dagger }\lr{\bs{\alpha}}}R_{\beta}^{\dagger}.
    \end{equation}
    forms a regularized $2$-design.
\end{cor}
\proof

By linearity of the integral, we have analogous to the previous proof
\begin{align}
    \int_{Y_n} d\mu\lr{\Lambda}\int_{\CP\lr{\Lambda}} d\bs{\alpha}\, R_{\beta}^{ \otimes 2}D^{\otimes 2}\lr{\bs{\alpha}} \Pi_{\Lambda}^{\otimes 2} D^{\dagger \, \otimes 2}\lr{\bs{\alpha}}R_{\beta}^{\dagger\,\otimes 2}=R_{\beta}^{\otimes 2}\lr{I\otimes I + \widehat{E}}R_{\beta}^{\dagger\,\otimes 2},
\end{align}
with $\widehat{E}$ as in Thm.~\ref{them:2design}.
\endproof

\subsection{Continuous variable frame potentials}\label{sec:frame}

Building on Ref.~\cite{Iosue_2024}, we examine the ensemble of GKP states defined in the main text through the lens of continuous variable \textit{frame potentials}. Frame potentials are an important tool in the analysis of evenly distributed states and we discuss their interpretation further below. To maintain well-defined expressions throughout, we need to work with appropriately regularized quantities, regularized through a Hermitian regularization operator $R$.
The end of this section contains a general discussion on the frame potential and its interpretation.

Let $\Pi_t$ be the projector onto the symmetric subspace of $L^2(\R^{n})^{\otimes t} $, and for a positive definite Hermitian trace-class operator $ R $, define $ \Pi_t^{(R)} = R^{\otimes t} \Pi_t R^{\otimes t} $.
For any measure space $ \CX_n $ over $ S'(\R^n) $, define
\begin{align}
    W_t^{(R)}(\CX_n)        & \coloneqq \int_{\CX_n} d\mu_{\CX_n}\lr{X} d\mu_{\CX_n}\lr{Y}\, \abs{\braket{X|R|Y}}^{2t} \nonumber \\
    X_t^{(R)}(\CX_n)        & \coloneqq \int_{\CX_n} d\mu_{\CX_n}\lr{X} \braket{X|R^2|X}^{t}\nonumber                          \\
    \CF_t^{(R)}(\CX_n, \des_t) & \coloneqq W_t^{(R)}(\CX_n) - 2 \des_t X_t^{(R)}(\CX_n) + \des_t^2 \Tr\lrq{\Pi_t^{(R)}}.
\end{align}
The following proposition is a modification of \cite[Prop.~6]{Iosue_2024}.

\begin{prop}[\cite{Iosue_2024}]\label{prop:t_design}
    For any ensemble $ \CX_n $ and any positive real number $ \des_t $, $\CF_t^{(R)}(\CX_n, \des_t) \geq 0 $. Furthermore, there exists positive real numbers $ \des_{t'} $ such that $ \CF_{t'}^{(R)}(\CX_n, \des_{t'}) = 0 $ for each $ t'=1,2,\dots,t $ if and only if $ \CX_n $ is a rigged $ t $-design.
\end{prop}

There are a few easy facts that we need:
\begin{align}
    \Tr\lrq{\Pi_1^{(R)}} & = \Tr\lrq{R^2}\nonumber                                                                     \\
    \Tr\lrq{\Pi_2^{(R)}} & = \frac{1}{2}\lr{\Tr \lrq{R^2}^2 + \Tr\lrq{R^4}}\nonumber                                   \\
    \Tr\lrq{\Pi_3^{(R)}} & = \frac{1}{6}\lr{\Tr\lrq{R^2}^3 + 3 \lr{\Tr\lrq{ R^4}}\lr{\Tr\lrq{R^2}} + 2 \Tr\lrq{R^6}} .
\end{align}

Finally,
for the operator $R$, let $c_R\lr{\bs{\alpha}} \coloneqq \Tr\lrq{D^{\dagger}\lr{\bs{\alpha}}R} $ denote its characteristic function. We have
\begin{align}
    c_{R^2}(0) & = \Tr\lrq{R^2} = \int_{\R^{2n}} d\bs{\alpha}\,\abs{c_R(\alpha)}^2\nonumber                                                                                                                                                                                                                                     \\
    c_{R^4}(0) & = \Tr\lrq{R^4} = \int_{\R^{2n}} d\bs{\alpha}_1d\bs{\alpha}_2d\bs{\alpha}_3\,c_R\lr{\bs{\alpha}_1}c_R\lr{\bs{\alpha}_2}c_R\lr{\bs{\alpha}_3}c_R^*\lr{\bs{\alpha}_1+\bs{\alpha}_2+\bs{\alpha}_3} e^{-i\pi \lr{\bs{\alpha}_1^T J \bs{\alpha}_2+\bs{\alpha}_1^T J \bs{\alpha}_3+\bs{\alpha}_2^T J \bs{\alpha}_3}}.
\end{align}

In a very long calculation we will now proceed to show the following theorem, using the proposition above. This calculation serves as independent verification that the ensemble $\CX_n$ of GKP states discussed in the main text indeed yields a rigged $2$-design.

\begin{them}
    The ensemble of GKP states $\CX_n \subset \CS'\lr{\R^n}$ defined in the main text forms a rigged $2$-design with $\des_1 = 1$ and $\des_2=2$.
\end{them}

\proof
Using the frame potential, we can check if our ensemble $\CX_n$ defined in the main text is a $t=1,2$-design.
Specifically, we need to express $X_t^{(R)}(\CG)$ and $W_t^{(R)}(\CG)$ for each $t$ in terms of $\Tr\lrq{R^2}$ and  $\Tr\lrq{R^4}$.

Recall that elements of $\CX_n$ are of the form $D\lr{\bs{\alpha}}\ket{\Lambda}$, where $\ketbra{\Lambda}= \sum_{\bs{\lambda} \in \Lambda} e^{i \Phi\lr{\bs{\lambda}}} D\lr{\bs{\lambda}}$.
In general, our computations will include expressions of the form
\begin{align}
    \braket{\Lambda |D^{\dagger}\lr{\bs{\alpha}} R D\lr{\bs{\alpha}'}|\Lambda'}
     & = \sum_{\bs{\lambda}\in\Lambda} \int_{\R^{2n}} d\bs{\gamma}\, c_{R}\lr{\bs{\gamma}}e^{i \Phi\lr{\bs{\lambda}}} \Tr\lrq{D\lr{\bs{\lambda}}D^{\dagger}\lr{\bs{\alpha}}D\lr{\bs{\gamma}}D\lr{\bs{\alpha}'} } \\
     & = \sum_{\bs{\lambda}\in\Lambda} \int_{\R^{2n}} d\bs{\gamma}\, c_{R}\lr{\bs{\gamma}}e^{i \Phi\lr{\bs{\lambda}}} e^{i\pi\bs{\lambda}^TJ\bs{\alpha}-i\pi\bs{\gamma}^TJ\bs{\alpha}' } \underbrace{\Tr\lrq{D\lr{\bs{\lambda}-\bs{\alpha}} D\lr{\bs{\gamma}+\bs{\alpha}'} }}_{\delta\lr{\bs{\gamma}-\bs{\alpha}+\bs{\alpha}'+\bs{\lambda}}} \\
     & =\sum_{\bs{\lambda}\in\Lambda} \, e^{i \Phi\lr{\bs{\lambda}}} c_{R}\lr{\bs{\alpha}-\bs{\alpha}'-\bs{\lambda}} e^{i\pi\bs{\lambda}^TJ\bs{\alpha}-i\pi\lr{\bs{\alpha}-\bs{\alpha}'-\bs{\lambda}}^TJ\bs{\alpha}' }.
\end{align}

If $\bs{\alpha}=\bs{\alpha}'$ this is
\begin{align}
    \braket{\Lambda |D^{\dagger}\lr{\bs{\alpha}} R D\lr{\bs{\alpha}}|\Lambda} = \sum_{\bs{\lambda}\in\Lambda} \, e^{i \Phi\lr{\bs{\lambda}}} c_{R}\lr{-\bs{\lambda}} e^{i2\pi\bs{\lambda}^TJ\bs{\alpha} }.
\end{align}

We begin with $t=1$. We compute

\begin{align}
    X_1^{(R)}\lr{\CX_n} & = \int_{Y_n} d\mu\lr{\Lambda} \int_{\CP\lr{\Lambda}}d\bs{\alpha} \braket{\Lambda |D^{\dagger}\lr{\bs{\alpha}} R^2 D\lr{\bs{\alpha}}|\Lambda}                                                                                                              \\
                      & =\int_{Y_n} d\mu\lr{\Lambda} \int_{\CP\lr{\Lambda}}d\bs{\alpha} \int_{\R^{2n}} d\bs{\gamma}\,\sum_{\bs{\lambda} \in\Lambda}  e^{i \Phi\lr{\bs{\lambda}}} e^{-i2\pi \bs{\alpha}^TJ\bs{\lambda}} c_{R^2}\lr{\bs{\gamma}}\Tr\lrq{D\lr{\bs{\lambda}}D\lr{\bs{\gamma}} } \\
                      & =c_{R^2}\lr{0}=\Tr\lrq{R^2}.
\end{align}

Similarly, we have
\begin{align}
    W_1^{(R)}(\CX_n) & = \int_{Y_n} d\mu\lr{\Lambda} d\mu\lr{\Lambda'}  \int_{\CP\lr{\Lambda}}d\bs{\alpha}\int_{\CP\lr{\Lambda}'}d\bs{\alpha}'\, \braket{\Lambda|D^{\dagger}\lr{\bs{\alpha}}RD\lr{\bs{\alpha}'}|\Lambda'}\braket{\Lambda'|D^{\dagger}\lr{\bs{\alpha}'}RD\lr{\bs{\alpha}}|\Lambda}                                                                                                 \\
                   & = \int_{Y_n} d\mu\lr{\Lambda} d\mu\lr{\Lambda'} \sum_{\bs{\lambda}\in \Lambda}\sum_{\bs{\lambda'}\in \Lambda'}  \int_{\CP\lr{\Lambda}}d\bs{\alpha}\int_{\CP\lr{\Lambda}'}d\bs{\alpha}'\, e^{i\Phi\lr{\bs{\lambda}}}e^{i\Phi\lr{\bs{\lambda}'}} e^{-i2\pi \bs{\alpha}^TJ\bs{\lambda}-i2\pi \bs{\alpha}'^TJ\bs{\lambda}'} \Tr\lrq{D\lr{\bs{\lambda}} RD\lr{\bs{\lambda}'}R } \\
                   & =\Tr\lrq{R^2},
\end{align}
where we have again used Eq.~\eqref{eq:int_PLambda}.
Altogether, this yields
\begin{equation}
    \CF_1^{(R)}(\CX_n,\des_1) =\Tr\lrq{R^2}-2\des_1\Tr\lrq{R^2}+\des_1^2\Tr\lrq{R^2}
\end{equation}
We see that, with $\des_1=1$, we obtain $\CF_1^{(R)}(\CX_n,\des_1=1)=0$.

Now, on to \textbf{$t=2$}. Here we have
\begin{align}
    X_2^{(R)}\lr{\CX_n} & = \int_{Y_n} d\mu\lr{\Lambda} \int_{\CP\lr{\Lambda}}d\bs{\alpha} \braket{\Lambda |D^{\dagger}\lr{\bs{\alpha}} R^2 D\lr{\bs{\alpha}}|\Lambda}^2                                                                                                                                                                                            \\
                      & = \int_{Y_n} d\mu\lr{\Lambda}\,  \sum_{\bs{\lambda}, \bs{\lambda}' \in\Lambda}  e^{i \Phi\lr{\bs{\lambda}}-i \Phi\lr{\bs{\lambda}'}}  c_{R^2}\lr{-\bs{\lambda}} c_{R^2}\lr{\bs{\lambda}'}\underbrace{\int_{\CP\lr{\Lambda}}d\bs{\alpha}\, e^{i2\pi\lr{\bs{\lambda}-\bs{\lambda}'}^TJ\bs{\alpha} }}_{\delta_{\bs{\lambda}, \bs{\lambda}'}} \\
                      & =\int_{Y_n} d\mu\lr{\Lambda}\sum_{\bs{\lambda} \in\Lambda}\, \abs{c_{R^2}\lr{\bs{\lambda}}}^2                                                                                                                                                                                                                                             \\
                      & =\abs{c_{R^2}\lr{0}}^2 + \int_{\R^{2n}} d\bs{\lambda}\, \abs{c_{R^2}\lr{\bs{\lambda}}}^2                                                                                                                                                                                                                                                            \\
                      & = \Tr\lrq{R^2}^2+\Tr\lrq{R^4}.
\end{align}

We also need to compute
\begin{align}
    W_2^{(R)}(\CX_n) & = \int_{Y_n} d\mu\lr{\Lambda} d\mu\lr{\Lambda'}  \int_{\CP\lr{\Lambda}}d\bs{\alpha}\int_{\CP\lr{\Lambda}'}d\bs{\alpha}'\, \abs{\braket{\Lambda|D^{\dagger}\lr{\bs{\alpha}}RD\lr{\bs{\alpha}'}|\Lambda'}}^4 \nonumber                                                                                                                                                                                     \\
                   & = \hdots \braket{\Lambda|D^{\dagger}\lr{\bs{\alpha}}RD\lr{\bs{\alpha}'}|\Lambda'}\braket{\Lambda'|D^{\dagger}\lr{\bs{\alpha}'}RD\lr{\bs{\alpha}}|\Lambda}\braket{\Lambda|D^{\dagger}\lr{\bs{\alpha}}RD\lr{\bs{\alpha}'}|\Lambda'}\braket{\Lambda'|D^{\dagger}\lr{\bs{\alpha}'}RD\lr{\bs{\alpha}}|\Lambda} \nonumber                                                                                      \\
                   & = \int_{Y_n} d\mu\lr{\Lambda} d\mu\lr{\Lambda'} \int_{\R^{2n}} d\bs{\gamma}_1d\bs{\gamma}_2d\bs{\gamma}_3d\bs{\gamma}_4\,c_R\lr{\bs{\gamma}_1}c_R\lr{\bs{\gamma}_2}c_R\lr{\bs{\gamma}_3}c_R\lr{\bs{\gamma}_4} \sum_{\bs{\lambda}_1, \bs{\lambda}_2 \in \Lambda; \bs{\lambda}'_1, \bs{\lambda}'_2 \in \Lambda'}  \nonumber                                                                                          \\
                   & \hspace{1.5cm} \int_{\CP\lr{\Lambda}}d\bs{\alpha}\int_{\CP\lr{\Lambda}'}d\bs{\alpha}' e^{-i2\pi \bs{\alpha}'^TJ\lr{\bs{\lambda}'_1+\bs{\lambda}'_2}-i2\pi \bs{\alpha}^TJ\lr{\bs{\lambda}_1+\bs{\lambda}_2}}
    \\ & \hspace{1.5cm}   \Tr\lrq{D\lr{\bs{\gamma}_1}D\lr{\bs{\lambda}'_1}D\lr{\bs{\gamma}_2} D\lr{\bs{\lambda}_1}D\lr{\bs{\gamma}_3} D\lr{\bs{\lambda}'_2} D\lr{\bs{\gamma}_4} D\lr{\bs{\lambda}_2}} \nonumber\\
                   & = \int_{Y_n} d\mu\lr{\Lambda} d\mu\lr{\Lambda'} \int_{\R^{2n}} d\bs{\gamma}_1d\bs{\gamma}_2d\bs{\gamma}_3d\bs{\gamma}_4\,c_R\lr{\bs{\gamma}_1}c_R\lr{\bs{\gamma}_2}c_R\lr{\bs{\gamma}_3}c_R\lr{\bs{\gamma}_4} \sum_{\bs{\lambda} \in \Lambda; \bs{\lambda}' \in \Lambda'}  \nonumber                                                                                                                               \\
                   & \hspace{1.5cm} \Tr\lrq{D\lr{\bs{\gamma}_1}D\lr{\bs{\lambda}'}D\lr{\bs{\gamma}_2} D\lr{\bs{\lambda}}D\lr{\bs{\gamma}_3} D^{\dagger}\lr{\bs{\lambda}'} D\lr{\bs{\gamma}_4} D^{\dagger}\lr{\bs{\lambda}}}                                                                                                                                                                                                   \\
                   & = \int_{Y_n} d\mu\lr{\Lambda} d\mu\lr{\Lambda'} \int_{\R^{2n}} d\bs{\gamma}_1d\bs{\gamma}_2d\bs{\gamma}_3d\bs{\gamma}_4\,c_R\lr{\bs{\gamma}_1}c_R\lr{\bs{\gamma}_2}c_R\lr{\bs{\gamma}_3}c_R\lr{\bs{\gamma}_4} \sum_{\bs{\lambda} \in \Lambda; \bs{\lambda}' \in \Lambda'}  \nonumber                                                                                                                               \\
                   & \hspace{1.5cm} \underbrace{\Tr\lrq{D\lr{\bs{\gamma}_1}D\lr{\bs{\gamma}_2} D\lr{\bs{\gamma}_3}  D\lr{\bs{\gamma}_4}}}_{e^{-i\pi\bs{\gamma}_1^TJ\bs{\gamma}_2}e^{-i\pi\bs{\gamma}_3^TJ\bs{\gamma}_4}\delta\lr{\bs{\gamma}_1+\bs{\gamma}_2+\bs{\gamma}_3+\bs{\gamma}_4}} e^{-i2\pi\bs{\lambda}^TJ\lr{\bs{\gamma}_3+\bs{\gamma}_4-\bs{\lambda}'}}e^{-i2\pi \bs{\lambda}'^TJ\lr{\bs{\gamma}_2+\bs{\gamma}_3}} \\
                   & = \int_{Y_n} d\mu\lr{\Lambda} d\mu\lr{\Lambda'} \int_{\R^{2n}} d\bs{\gamma}_1d\bs{\gamma}_2d\bs{\gamma}_3\,c_R\lr{\bs{\gamma}_1}c_R\lr{\bs{\gamma}_2}c_R\lr{\bs{\gamma}_3}c_R^*\lr{\bs{\gamma}_1+\bs{\gamma}_2+\bs{\gamma}_3} \sum_{\bs{\lambda} \in \Lambda; \bs{\lambda}' \in \Lambda'}  \nonumber                                                                                                               \\
                   & \hspace{1.5cm} e^{-i\pi\bs{\gamma}_1^TJ\bs{\gamma}_2}e^{-i\pi\bs{\gamma}_3^TJ\lr{-\bs{\gamma}_1-\bs{\gamma}_2}} e^{i2\pi\bs{\lambda}^TJ\lr{\bs{\gamma}_1+\bs{\gamma}_2+\bs{\lambda}'}}e^{-i2\pi \bs{\lambda}'^TJ\lr{\bs{\gamma}_2+\bs{\gamma}_3}}
\end{align}

We have \footnote{Note that although $f(\bs{\lambda})=e^{-i2\pi \bs{\lambda}^TJ\bs{x}}$ is strictly speaking not an integrable function, the mean value formula may be applied to a regularized form $\tilde{f}_{\beta}(\bs{\lambda})=(2\pi/\beta)^{-1/2}e^{-\frac{\beta}{2}\|\bs{\lambda}\|^2-i2\pi \bs{\lambda}^TJ\bs{x}}$ and the result is obtained in the limit $\beta \to 0$.},

\begin{equation}
    \int_{Y_n} d\mu\lr{\Lambda} \sum_{\bs{\lambda}\in \Lambda} e^{-i2\pi \bs{\lambda}^T J \bs{x}} = 1+ \delta\lr{\bs{x}},
\end{equation}

such that we can further simplify

\begin{align}
    \hdots & =\int_{Y_n} d\mu\lr{\Lambda'} \int_{\R^{2n}} d\bs{\gamma}_1d\bs{\gamma}_2d\bs{\gamma}_3\,c_R\lr{\bs{\gamma}_1}c_R\lr{\bs{\gamma}_2}c_R\lr{\bs{\gamma}_3}c_R^*\lr{\bs{\gamma}_1+\bs{\gamma}_2+\bs{\gamma}_3}   \nonumber                                                             \\
           & \hspace{1.5cm}\sum_{\bs{\lambda}' \in \Lambda'} e^{-i\pi\bs{\gamma}_1^TJ\bs{\gamma}_2}e^{-i\pi\bs{\gamma}_3^TJ\lr{-\bs{\gamma}_1-\bs{\gamma}_2}} \lrc{1+\delta\lr{\bs{\gamma}_1+\bs{\gamma}_2+\bs{\lambda}'}} e^{-i2\pi \bs{\lambda}'^TJ\lr{\bs{\gamma}_2+\bs{\gamma}_3}} \\
           & = \int_{\R^{2n}} d\bs{\gamma}_1d\bs{\gamma}_2d\bs{\gamma}_3\,c_R\lr{\bs{\gamma}_1}c_R\lr{\bs{\gamma}_2}c_R\lr{\bs{\gamma}_3}c_R^*\lr{\bs{\gamma}_1+\bs{\gamma}_2+\bs{\gamma}_3}   \nonumber                                                                                         \\
           & \hspace{1.5cm} e^{-i\pi\bs{\gamma}_1^TJ\bs{\gamma}_2}e^{-i\pi\bs{\gamma}_3^TJ\lr{-\bs{\gamma}_1-\bs{\gamma}_2}} \lrc{1+\delta\lr{\bs{\gamma}_1+\bs{\gamma}_2}} \nonumber                                                                                                  \\
           & +  \int_{\R^{2n}} d\bs{\gamma}_1d\bs{\gamma}_2d\bs{\gamma}_3\,c_R\lr{\bs{\gamma}_1}c_R\lr{\bs{\gamma}_2}c_R\lr{\bs{\gamma}_3}c_R^*\lr{\bs{\gamma}_1+\bs{\gamma}_2+\bs{\gamma}_3} \nonumber                                                                                          \\
           & e^{-i\pi\bs{\gamma}_1^TJ\bs{\gamma}_2}e^{-i\pi\bs{\gamma}_3^TJ\lr{-\bs{\gamma}_1-\bs{\gamma}_2}}\int_{\R^{2n}} d\bs{\lambda}'  \lrc{1+\delta\lr{\bs{\gamma}_1+\bs{\gamma}_2 + \bs{\lambda}'}}e^{-i2\pi \bs{\lambda}'^TJ\lr{\bs{\gamma}_2+\bs{\gamma}_3}}                            \\
           & = \int_{\R^{2n}} d\bs{\gamma}_1d\bs{\gamma}_2d\bs{\gamma}_3\,c_R\lr{\bs{\gamma}_1}c_R\lr{\bs{\gamma}_2}c_R\lr{\bs{\gamma}_3}c_R^*\lr{\bs{\gamma}_1+\bs{\gamma}_2+\bs{\gamma}_3}   \nonumber                                                                                         \\
           & \hspace{1.5cm} e^{-i\pi\bs{\gamma}_1^TJ\bs{\gamma}_2}e^{-i\pi\bs{\gamma}_3^TJ\lr{-\bs{\gamma}_1-\bs{\gamma}_2}} \lrc{1+\delta\lr{\bs{\gamma}_1+\bs{\gamma}_2}} \nonumber                                                                                                  \\
           & +  \int_{\R^{2n}} d\bs{\gamma}_1d\bs{\gamma}_2d\bs{\gamma}_3\,c_R\lr{\bs{\gamma}_1}c_R\lr{\bs{\gamma}_2}c_R\lr{\bs{\gamma}_3}c_R^*\lr{\bs{\gamma}_1+\bs{\gamma}_2+\bs{\gamma}_3} \nonumber                                                                                          \\
           & e^{-i\pi\bs{\gamma}_1^TJ\bs{\gamma}_2}e^{i\pi\bs{\gamma}_3^TJ\lr{\bs{\gamma}_1+\bs{\gamma}_2}}  \lrc{\delta\lr{\bs{\gamma}_2+\bs{\gamma}_3}+e^{i2\pi \lr{\bs{\gamma}_1+\bs{\gamma}_2}^TJ\lr{\bs{\gamma}_2+\bs{\gamma}_3}} }
\end{align}

This final sum has $4$ parts, the first equals $\Tr\lrq{R^4}$. The second is
\begin{align}
    \int_{\R^{2n}} d\bs{\gamma}_1d\bs{\gamma}_3\,|c_R\lr{\bs{\gamma}_1}|^2 |c_R\lr{\bs{\gamma}_3} |^2 = \Tr\lrq{R^2}^2,
\end{align}
same as the third
\begin{align}
    \int_{\R^{2n}}
    d\bs{\gamma}_1d\bs{\gamma}_2 |c_R\lr{\bs{\gamma}_1}|^2 |c_R\lr{\bs{\gamma}_2}|^2  =\Tr\lrq{R^2}^2.
\end{align}

Finally, the last term is computed as
\begin{align}
     & \int_{\R^{2n}} d\bs{\gamma}_1d\bs{\gamma}_2d\bs{\gamma}_3\,c_R\lr{\bs{\gamma}_1}c_R\lr{\bs{\gamma}_2}c_R\lr{\bs{\gamma}_3}c_R^*\lr{\bs{\gamma}_1+\bs{\gamma}_2+\bs{\gamma}_3}e^{-i\pi\bs{\gamma}_1^TJ\bs{\gamma}_2}e^{i\pi\bs{\gamma}_3^TJ\lr{\bs{\gamma}_1+\bs{\gamma}_2}} e^{i2\pi \lr{\bs{\gamma}_1+\bs{\gamma}_2}^TJ\lr{\bs{\gamma}_2+\bs{\gamma}_3}} \\
     & =\int_{\R^{2n}} d\bs{\gamma}_1d\bs{\gamma}_2d\bs{\gamma}_4\,c_R\lr{\bs{\gamma}_1}c_R\lr{\bs{\gamma}_2}c_R^*\lr{\bs{\gamma}_1+\bs{\gamma}_2+\bs{\gamma}_4}c_R\lr{\bs{\gamma}_4}e^{-i\pi\bs{\gamma}_1^TJ\bs{\gamma}_2}e^{-i\pi\bs{\gamma}_4^T\lr{\bs{\gamma}_1+\bs{\gamma}_2}} =\Tr\lrq{R^4}.
\end{align}

Thus in total, we have
\begin{equation}
    W_2^{(R)}(\CX_n)=2\Tr\lrq{R^4}+2\Tr\lrq{R^2}^2
\end{equation}
and

\begin{equation}
    \CF_2^{(R)}(\CX_n, \des_2)=2\Tr\lrq{R^4}+2\Tr\lrq{R^2}^2 - 2 \des_2 \lrc{\Tr\lrq{R^2}^2+\Tr\lrq{R^4}} +  \frac{\des_t^2}{2}\lrc{\Tr \lrq{R^2}^2 + \Tr\lrq{R^4}}.
\end{equation}
From this equation, we can observe that we have, with $\des_2=2$, $\CF_2^{(R)}(\CX_n, \des_2=2)=0$. Under the premise of prop.~\ref{prop:t_design} we hence have verified that $\CX_n$ is a rigged $2$-design.

\endproof

\subsubsection*{The meaning of the frame potential}

We now give a basic interpretation of the frame potential in \textit{finite-dimensional} systems.
The frame potential has its roots in Welch bounds \cite{welch1974lower-bounds-on,datta2009geometry-of-the}.
There are many great expositions of the frame potential, its relation to designs, mutually unbiased bases, and scrambling and chaos \cite{klappenecker2004constructions-o,klappenecker2005mutually-unbias,scott2008optimizing-quan,belovs2008a-criterion-for,roberts2017chaos-and-compl}.
Here, we give a slightly different argument for its applicability.

Ultimately, we want to characterize some ensemble $\CX_n$ of states via a number $F(\CX_n) \in \R$.
Because states are elements of projective space, this function $F$ must be a function of projectors rather than states. Additionally, there is no ordering on $\CX_n$, meaning $F(\CX_n)$ should be independent of reordering or relabeling the elements of $\CX_n$.
We are thus led to
\begin{equation}
    F_t^{(k)}(\CX_n) = \Tr \E_{\psi_1, \dots,\psi_k\in \CX_n} (\ket{\psi_1}\bra{\psi_1})^{\otimes t}\dots (\ket{\psi_k}\bra{\psi_k})^{\otimes t} = \E_{\psi_1, \dots,\psi_k\in \CX_n}\braket{\psi_1\vert\psi_2}^t \braket{\psi_2\vert\psi_3}^t \dots \braket{\psi_k\vert\psi_1}^t,
\end{equation}
where we can tune $t$ to probe different moments of $\CX_n$.

We then notice that $F_t^{(1)}(\CX_n) = 1$ for any $\CX_n$ (because in finite-dimensions we can always assume the states are unit-normalized), and so is trivial.
The first nontrivial function that can characterize the $t^{\rm th}$ moment of $\CX_n$ is thus
$F_t^{(2)}(\CX_n)$.
$F_t^{(2)}(\CX_n)$ is in fact the standard frame potential for finite-dimensional systems, and as such we will notate $F_t(\CX_n) \equiv F_t^{(2)}(\CX_n)$.

If we are interested in measuring how distributed an ensemble of states is (\textit{i.e.}~how much it ``covers'' the full space of states and how spread it is over the space), we want the average magnitude of the overlap between the states in the ensemble to be small. Therefore, the $F_t(\CX_n)$ being small is a measure of how evenly spread $\CX_n$ is. This then gives some intuition as to why the frame potential shows up as a measure of scrambling and chaos.

When thinking about $t$-designs, we need ensembles $\CX_n$ that maximally cover the whole space of states in order for them to mimic the whole space itself.
Therefore, $\CX_n$ is a $t$-design if and only if $F_t(\CX_n)$ is minimized (the proof of this is analogous to the proof of Prop.~\ref{prop:t_design}).
When $\CX_n$ does mimic the $t^{\rm th}$ moment of the whole space of states, then clearly $F_t^{(k)}(\CX_n)$ is determined for all $k$.
Thus, we have found that even though $F_t^{(k)}(\CX_n)$ in principle contains more information about the $t^{\rm th}$ moment of $\CX_n$ than just $F_t(\CX_n) = F_t^{(2)}(\CX_n)$, $F_t(\CX_n)$ alone already characterizes how far $\CX_n$ is from being sufficiently spread over the whole space of states.
In particular, via some continuity analysis, $F_t^{(k)}(\CX_n)$ can be bounded in terms of how far $F_t(\CX_n)$ is from its minimum, and this bound gets tighter as $F_t(\CX_n)$ approaches its minimum.

In the CV case, $W_t^{(R)}(\CX_n)$ is the analogue of $F_t(\CX_n)$.
The function $X_t^{(R)}(\CX_n)$ is an artifact of the fact that the states are not normalized.
In the case of regularized rigged designs, $X_t^{(R)}(\CX_n)$ can in fact be gotten rid of \cite[Prop.~6]{Iosue_2024}, making the analogy between the finite- and infinte-dimensional frame potentials more direct.

\subsection{Diagonal coherent state basis}\label{sec:coherent}

In this appendix, we provide another proof that the displaced GKP ensemble forms a rigged $2$-design by working in the coherent state basis (a.k.a. the $P$-representation). We then perform the analogous $3$-design calculation, but are unfortunately unable to complete it.

The key point is that the diagonal elements of an operator in the coherent state basis fully determine the operator \cite{CahillGlauber, Gerry2004}.
Hence, in order to check if an ensemble $\CX_n$ is a rigged $t$-design, we only need to check that
\begin{equation}
    \int_{\CX_n} dX \prod_{i=1}^t\abs{\braket{\bs{\alpha}_i \vert X}}^2 = \des_t \braket{\bs{\alpha}_1,\dots,\bs{\alpha}_t|\Pi_t|\bs{\alpha}_1,\dots,\bs{\alpha}_t}\label{eq:coherent_check}
\end{equation}
for some $ \des_t > 0 $.
For $\CX_n $ the ensemble of GKP codes defined in the main text, this amounts to showing that
\begin{equation}
    g(\bs{\alpha}_1,\dots,\bs{\alpha}_t) = \des_t \bra{\bs{\alpha}_1,\dots,\bs{\alpha}_t}\Pi_t \ket{\bs{\alpha}_1,\dots,\bs{\alpha}_t},
\end{equation}
where
\begin{align}
    g(\bs{\alpha}_1,\dots,\bs{\alpha}_t)
     & \coloneqq \int_{Y_n}d{\mu(\Lambda)} \int_{\CP(\Lambda)}d{\bs\beta} \prod_{i=1}^t \Tr\left[ D(\bs{\beta})\ket\Lambda \bra\Lambda D(-\bs{\beta}) D(\bs{\alpha}_i)\ket 0 \bra 0 D(-\bs{\alpha}_i) \right]                                \nonumber \\
     & = \int_{Y_n}d{\mu(\Lambda)} \int_{\CP(\Lambda)}d{\bs\beta} \prod_{i=1}^t \sum_{\bs\lambda_i\in \Lambda} e^{i \Phi(\bs\lambda_i)} \Tr\left[ D(\bs\beta)D(\bs\lambda_i) D(-\bs\beta) D(\bs{\alpha}_i)\ket 0 \bra 0 D(-\bs{\alpha}_i) \right] \nonumber\\
     & = \int_{Y_n}d{\mu(\Lambda)} \int_{\CP(\Lambda)}d{\bs\beta} \prod_{i=1}^t \sum_{\bs\lambda_i\in \Lambda} e^{i \Phi(\bs\lambda_i)} \bra 0 D(-\bs{\alpha}_i) D(\bs\beta) D(\bs\lambda_i) D(-\bs\beta) D(\bs{\alpha}_i)\ket 0                 \nonumber \\
     & = \int_{Y_n}d{\mu(\Lambda)} \int_{\CP(\Lambda)}d{\bs\beta} \prod_{i=1}^t \sum_{\bs\lambda_i\in \Lambda} e^{i \Phi(\bs\lambda_i)} e^{2\pi i \bs\lambda_i^T J (\bs\beta-\bs{\alpha}_i)} \bra 0 D(\bs\lambda_i) \ket 0                    \nonumber  \\
    &\stackrel{{\text{Eq.~\eqref{eq:int_PLambda}}}}{=} \int_{Y_n}d{\mu(\Lambda)}\sum_{\substack{\bs\lambda_1,\dots,\bs\lambda_t \in \Lambda                                                                                                                                                 \nonumber  \\ \text{s.t.~} \bs\lambda_1+\dots+\bs\lambda_t = 0}} e^{i \sum_{j=1}^t \Phi(\bs\lambda_j)} e^{2\pi i \sum_{j=1}^t \bs{\alpha}_j^T J \bs\lambda_j} \prod_{j=1}^t \braket{0\vert \bs\lambda_j} \nonumber \\
     & = \int_{Y_n}d{\mu(\Lambda)}\sum_{\substack{\bs\lambda_1,\dots,\bs\lambda_t \in \Lambda                                                                                                                                                   \\ \text{s.t.~}\bs\lambda_1+\dots+\bs\lambda_t = 0}} e^{i \sum_{j=1}^t \Phi(\bs\lambda_j)} e^{2\pi i \sum_{j=1}^t \bs{\alpha}_j^T J \bs\lambda_j} e^{-\frac{1}{2}\sum_{j=1}^t \norm{\bs\lambda_j}^2}   .
\end{align}

The RHS of Eq.~\eqref{eq:coherent_check} yields
\begin{align}
     & \Pi_1 = I && \implies \bra{\bs{\alpha}}\Pi_1 \ket{\bs{\alpha}}= 1  \nonumber \\
     & \Pi_2 = \frac{1}{2}(I\otimes I + \text{SWAP})
     && \implies \bra{\bs{\alpha}_1,\bs{\alpha}_2}\Pi_2 \ket{\bs{\alpha}_1,\bs{\alpha}_2}= \frac{1}{2} + \frac{1}{2}\abs{\braket{\bs{\alpha}_1\vert\bs{\alpha}_2}}^2  .
\end{align}
Furthermore, using $\Pi_3=\frac{1}{6}\sum_{\sigma\in S_3}W_\sigma $, where $W_\sigma$ permutes the tensor factors, we also have
\begin{equation}
    \label{eq:t3-sym-projector}
    \bra{\bs{\alpha}_1,\bs{\alpha}_2,\bs{\alpha}_3}\Pi_3 \ket{\bs{\alpha}_1,\bs{\alpha}_2, \bs{\alpha}_3}
    = \frac{1}{6}\left[ 1 + \abs{\braket{\bs{\alpha}_1\vert\bs{\alpha}_2}}^2+\abs{\braket{\bs{\alpha}_1\vert\bs{\alpha}_3}}^2 + \abs{\braket{\bs{\alpha}_2\vert\bs{\alpha}_3}}^2 + 2\Re\left(\braket{\bs{\alpha}_1\vert\bs{\alpha}_2}\braket{\bs{\alpha}_2\vert\bs{\alpha}_3}\braket{\bs{\alpha}_3\vert\bs{\alpha}_1}\right) \right].
\end{equation}

For $t=1$, we have
\begin{align}
    g(\bs{\alpha})
     & = \int_{Y_n}d{\mu(\Lambda)}\sum_{\substack{\bs\lambda \in \Lambda \\ \text{s.t.~}\bs\lambda = 0}} e^{i \Phi(\bs\lambda)} e^{2\pi i \bs{\alpha}^T J \bs\lambda} e^{-\norm{\bs\lambda}^2/2} \nonumber\\
     & = 1                                                             \nonumber  \\
     & = \bra{\bs{\alpha}}\Pi_1 \ket{\bs{\alpha}},
\end{align}
therefore showing that the displaced GKP ensemble forms a $1$-design with $\des_1 = 1$.

For $t=2$, we have
\begin{align}
    g(\bs{\alpha}_1,\bs{\alpha}_2)
                               & =  \int_{Y_n}d{\mu(\Lambda)}\sum_{\substack{\bs\lambda_1,\bs\lambda_2 \in \Lambda                                                                     \nonumber \\ \text{s.t.~}\bs\lambda_1+\bs\lambda_2 = 0}} e^{i \sum_{j=1}^2 \Phi(\bs\lambda_j)} e^{2\pi i \sum_{j=1}^2 \bs{\alpha}_j^T J \bs\lambda_j} e^{-\frac{1}{2}\sum_{j=1}^2 \norm{\bs\lambda_j}^2} \\
                               & =  \int_{Y_n}d{\mu(\Lambda)}\sum_{\bs\lambda\in \Lambda} e^{2\pi i \bs\lambda^T J (\bs{\alpha}_2-\bs{\alpha}_1)} e^{-\norm{\bs\lambda}^2}                 \nonumber \\
                               & = 1+ \int_{Y_n}d{\mu(\Lambda)}\sum_{\bs\lambda\in \Lambda\setminus \{0\}} e^{2\pi i \bs\lambda^T J (\bs{\alpha}_2-\bs{\alpha}_1)} e^{-\norm{\bs\lambda}^2} \nonumber\\
                               & \stackrel{\text{lem.~\ref{lem:mean-value}}}{=} 1+ \int_{\R^{2n}} d \bs{\lambda} ~ e^{2\pi i \bs\lambda^T J (\bs{\alpha}_2-\bs{\alpha}_1)} e^{-\norm{\bs\lambda}^2}                                               \nonumber \\
     & = 1 + e^{-\norm{\bs{\alpha}_2-\bs{\alpha}_1}^2}                                                                                                           \nonumber \\
                               & = 1+ \abs{\braket{\bs{\alpha}_1\vert\bs{\alpha}_2}}^2                                                                                                   \nonumber   \\
                               & = 2\bra{\bs{\alpha}_1,\bs{\alpha}_2}\Pi_2 \ket{\bs{\alpha}_1,\bs{\alpha}_2}.
\end{align}
This therefore shows that the displaced GKP ensemble forms a rigged $2$-design with $\des_2 = 2$.

The analogous calculation for $t=3$ requires a second moment analogue of Lem.~\ref{lem:mean-value} beyond the extension provided in Ref.~\cite{Kelmer_2019}.
For $t=3$, we have
\begin{align}
    g(\bs{\alpha}_1,\bs{\alpha}_2,\bs{\alpha}_3)
     & = \int_{Y_n}d{\mu(\Lambda)}\sum_{\substack{\bs\lambda_1,\bs\lambda_2,\bs\lambda_3 \in \Lambda                                                                                                                                                                                                          \nonumber  \\ \text{s.t.~}\bs\lambda_1+\bs\lambda_2+\bs\lambda_t = 0}} e^{i \sum_{j=1}^t \Phi(\bs\lambda_j)} e^{2\pi i \sum_{j=1}^t \bs{\alpha}_j^T J \bs\lambda_j} e^{-\frac{1}{2}\sum_{j=1}^3 \norm{\bs\lambda_j}^2}. 
%     %
%      & = \int_{Y_n}d{\mu(\Lambda)}\sum_{\bs\lambda_1,\bs\lambda_2 \in \Lambda } e^{2\pi i \left(  \bs{\alpha}_1^T J \bs\lambda_1 + \bs{\alpha}_2^T J \bs\lambda_2 - \bs{\alpha}_3^T J (\bs\lambda_1+\bs\lambda_2)  \right)} e^{-\norm{\bs\lambda_1}^2/2-\norm{\bs\lambda_2}^2/2-\norm{\bs\lambda_1+\bs\lambda_2}^2/2}\nonumber \\
%     %
%      & = \int_{Y_n}d{\mu(\Lambda)}\sum_{\bs\lambda_1,\bs\lambda_2 \in \Lambda } e^{2\pi i \left(  \bs\lambda_1^T J (\bs{\alpha}_3-\bs{\alpha}_1) + \bs\lambda_2^T J (\bs{\alpha}_2-\bs{\alpha}_1)  \right)} e^{-\norm{\bs\lambda_1}^2/2-\norm{\bs\lambda_2}^2/2-\norm{\bs\lambda_1+\bs\lambda_2}^2/2}  .
\end{align}
To complete this calculation, we need an expression for the second moment integral over $Y_n$. Once this is known, we would need to compare the result to Eq.~\eqref{eq:t3-sym-projector}.

\section{Design-based CV shadow tomography with GKP states}\label{sec:CVshadows}

It was shown in Ref.~\cite{Iosue_2024} that a rigged $2$-design immediately gives rise to a classical shadow tomography protocol for states on the CV Hilbert space. A distinctive feature of the classical shadow tomography protocol is that it constructs a set of POVM elements, whose measurement statistics suffice to reconstruct expectation values of many observables with specific properties while utilizing only a relatively small number of samples \cite{Huang_2020}.

\subsection{Shadow tomography from $2$-designs}

We recall the CV classical shadow procedure proposed in Ref.~\cite{Iosue_2024}. Let $\lr{\CX_n, \mu_{\CX_n}}$ form a set of (improper) states $\ketbra{X},\; \ket{X}\in \CX_n$ together with a normalized measure $\mu_{\CX_{n}}: \CX_n\to \R$, such that it forms a rigged $2$-design. Then measuring a state $\rho$ with respect to $\ketbra{X}$ yields the POVM
\begin{align}
\CM_{\CX_n}\lr{\rho} & =\int_{\CX_n}d\mu_{\CX_n}\lr{X} \braket{X|\rho|X} \ketbra{X}                             \nonumber \\
   & =  \int_{\CX_n}d\mu_{\CX_n}\lr{X} \Tr_1\lrq{\lr{\rho \otimes I}  \ketbra{X}^{\otimes 2}}  \nonumber\\
   & =  \Tr_1\lrq{\lr{\rho \otimes I}  \lrc{I\otimes I + \widehat{E}} }                    \nonumber \\
   & =  I+\rho, \label{eq:CV_shadow_M}
\end{align}
where we inserted the $2$-design property in the third line.
Note, that in this expression it is not actually necessary to think about the  $\ketbra{X}$ as physical states, instead they can be regarded as the classical pointer of the associated measurement outcome that was obtained with probability distribution $\Tr\lrq{\ketbra{X}\rho}=\braket{X|\rho|X}$. Inverting this expression as a function of $\rho$, we find that expectation values of observables $O$ are obtained as
\begin{align}
\braket{O} & = \Tr\lrq{ \lr{\int_{\CX_n}d\mu_{\CX_n}\lr{X} \braket{X|\rho|X} \ketbra{X} - I}O }            \nonumber     \\
   & = \int_{\CX_n}d\mu_{\CX_n}\lr{X} \braket{X|\rho|X}\braket{X|O|X} - \Tr\lrq{O}.\label{eq:estimator_app}
\end{align}

In this recipe, for each sampled set of measurements $X_i, \, i=1,2,\cdots, N$, the ensemble 
of reconstructed ``states''
\begin{equation}
\lrc{\tilde{\rho}_i=\ketbra{X}_i - I}_{i=1}^N
\end{equation}
is regarded as the classical shadow of the input quantum state. Again, here we notice that it is in fact not relevant whether one can regard the expression $\ketbra{X}$ as physical states. The only elements necessary for the classical shadow tomography protocol to function are that they form rigged $2$-designs, that a quantum POVM exists with measurement probability distribution $P\lr{X|\rho}=\braket{X|\rho|X}$ exists for input states $\rho$, and that the expression $\braket{X|O|X} $ is sufficiently well behaved to be evaluated under the above integral.

\subsection{Local shadows vs. global shadows}
In Clifford-based classical shadow tomography for discrete variable systems \cite{Huang_2020} the ensemble of states used to build the shadow protocol is typically chosen to be a random set of stabilizer states obtained from Clifford-twirling a Pauli measurement channel. In this construction, the Clifford-twirled measurement channel is projected onto a depolarizing channel such that the outcome statistics over the stabilizer-state-like pointer stats reflects that of the depolarized input state. This depolarizing channel shortens the length of the Bloch vector of the input state by a factor of $f=(2^n+1)^{-1}$ \footnote{Or $f=(d^{n}+1)^{-1}$ for multi-qudit systems.} and to reproduce the expectation values of the input state faithfully, in classical post-processing of the output stabilizer states, one needs to reamplify by this factor. While this re-amplification process is necessary to obtain the correct expectation values, it also amplifies the variance of the estimator to scale exponentially in the system (or subsystem) size $n$. It is for this reason that a distinction is typically being made between \textit{local} and \textit{global} classical shadows and, for practical purposes, the tomography protocol is typically restricted to be only applied locally. Interestingly, when the POVM used is a Pauli-$Z$ measurement on each of the qubits, it is known that this depolarizing channel with probability $p=1/3$ that is projected onto is \textit{entanglement breaking} despite remaining classically invertible as a matrix. 
Via the classical-post processing protocol of classical shadow tomography, this then allows to estimate the expectation values of non-local observables of entangled states by passing through their depolarized, unentangled shadows \cite{Huang_2020}. The price to pay for this trick, however, is an exponential overhead in the the degree of locality $k$ of the observable, which enters the variance upper bounds with a factor of $f^{-k}$, where $f=1/3$ in the case of qubit systems.

From Eq.~\eqref{eq:CV_shadow_M} we recognize that the tensor power of the random measurement channel over a local ensemble $\CX_1$ randomly deletes the supported mode or preserves it. Similar to the strategy in Ref.~\cite{Huang_2020}, it is possible to define a local shadow protocol by reconstructing the shadow state
\begin{equation}
    \bigotimes_{i=1}^n \lr{\ketbra{X_i} - I}
\end{equation}
upon sampling outcome $X_i$ on each mode $i=1,...,n$. As, together with the inversion, each local channel $\CM_{\CX_1}^{-1} \ketbra{X} \cdot \ketbra{X}$ averages to the identity channel, we see that again the full expectation value of the state can be reconstructed. 

An interesting observation that can be made in Eq.~\eqref{eq:CV_shadow_M} is that, in our GKP-state based shadow protocol, the corresponding factor is $f=1$. As the tomographic data obtained from measuring relative to the ensemble of GKP states $\CX_n$ does not correspond to a depolarized version of the input state but reproduces it exactly on average up to the constant identity offset. Note that the output states of the channel in Eq.~\eqref{eq:CV_shadow_M} are not physical, such that the POVM is generally not a trace preserving channel. We have shown that the interpretation of the classical shadow as an actual quantum state in our protocol is not necessary to reproduce meaningful expectation values for the given observables, which ultimately circumvents the appearance of a factor $f<1$ as discussed above. The implication is interesting: Estimating non-local observables using local ($n=1$) shadows as per our protocol is a-priori not accompanied by a variance-, and sample overhead -blow up in the locality of the observable originating from the need to reamplify the result. Yet, in the bounds we will be able to provide, we still observe a non-trivial scaling of the variance in the number of modes $n$. We investigate this further as follows.

\subsection{Global and local design-based shadow tomography for $3$-designs}\label{sec:design_based_shadows}

In the previous sections, we have leveraged the rigged $2$-design property of our ensemble $\CX_n$ together with knowledge about the specific structure of the ensemble and/or additional physical assumptions in order to bound the variance scaling. While we have not been able to show whether a stronger $3$-design property also holds for our ensemble, we explore what this property would imply. The following discussion applies to design-based classical shadow tomography protocols for either finite or inifinite systems (via the definition of rigged designs).

Throughout this section, we will always assume that $O$ is traceless in order to simplify calculations. Assume that we have a measure space $\lr{\CX, \mu}$ that is a $3$-design. In other words, for all $t \leq 3$, $\int_{\CX} X^{\otimes t} d\mu(X) = \des_t \Pi_{t}$, where each $X\in \CX$ is a positive operator, $\des_1=1$, and $\des_t > 0$.
A rigged $3$-design in infinite-dimensions and a quantum state $3$-design in finite dimensions satisfies exactly this criteria.
In particular, in $d$ dimensions for $d$ finite,
if we choose $\des_t = d/\Tr(\Pi_t)$ (ie $\des_2 = 2/\lr{d+1}$ and $\des_3 = 6/(d+1)(d+2)$) and require each $X$ to be rank 1, then we get the standard definition of a quantum state $3$-design. 
In infinite-dimensions, if we allow $\des_2,\des_3$ to be arbitrary constants and require each $X$ to be rank 1, then we get the definition of a rigged $3$-design.

\subsubsection{Global shadows}
Define
\begin{equation}
M(\rho) 
= \int_{\CX} \Tr[\rho X] X ~ \mathrm d\mu(X) 
= \frac{\des_2}{2}(I + \rho) .
\end{equation}

Suppose that we can sample from the probability distribution $dP(X,\rho) = \Tr(\rho X) ~\mathrm d \mu(X)$. Given a sample $X$ from this distribution, we have the estimator $o(X) = \frac{2}{\des_2}\Tr[X O]$. The expectation value of our estimator is 
\begin{align}
\mathbb E_{X \sim dP}[o(X)] 
&= \int_{\CX} o(X) ~dP(X) \nonumber\\
%
%&= \frac{2}{a_2}\int_{\CX} \Tr[O E] ~dP(E) \\
%
&= \frac{2}{\des_2}\int_{\CX} \Tr[OX] \Tr[\rho X] ~ d\mu(X) \nonumber\\
&= 2\Tr[\Pi_2 (\rho \otimes O)] \nonumber\\
%
%&= \Tr[\rho O] + \Tr[O]\\
%
&= \Tr[\rho O] 
\end{align}
The variance of our estimator is 
\begin{align}
\mathrm{Var}_{X \sim dP}[o(X)]
&= \int_{\CX} o(X)^2 ~dP(X)  \; - \Tr[\rho O]^2\nonumber\\
&= \frac{4}{\des_2^2}\int_{\CX} \Tr[X O]^2 \Tr[\rho X] ~ d\mu(X) \; -  \Tr[\rho O]^2\nonumber \\
&= \frac{4\des_3}{\des_2^2}  \Tr[\Pi_3(O \otimes O \otimes \rho)]  - \Tr[\rho O]^2\nonumber\\
&\leq \frac{2\des_3}{\des_2^2} \Tr[O^2],
\end{align}
where the last step is executed using Eq.~(E9) in Ref.~\cite{Iosue_2024}.
We see that the efficiency of a global shadow tomography protocol with a $3$-design depends on $2\des_3 / \des_2^2$. 
In finite dimension $d$, when using quantum state $3$-designs, the scaling dependencies of this fraction on $d$ cancel and it follows that the variance of the corresponding estimator is $\sim \Tr[O^2]$.
In infinite dimensions, a rigged $3$-design satisfies $\des_2, \des_3 \sim 1$, and thus again we obtain a variance scaling of $\sim \Tr[O^2]$.

We highlight that the factor $\des_2$ alone is not important, but rather the ratio $2\des_3/\des_2^2$.
In general, even in the displaced GKP case (where we are unsure if the ensemble forms a rigged $3$-design),
the fact that $\des_2 \sim 1$ (as opposed to $\des_2 \sim 2^{-n}$ as is typically the case in qubit protocols) is not the key difference dictating the variance scaling of the corresponding shadow tomography protocol.
Ultimately, what matters is the ratio $2\des_3/\des_2^2$. For finite $d$ and utilizing quantum state $3$-designs, we have
\begin{equation}
    \des_2 = \frac{2}{d+1} \quad \text{and}\quad \des_3 = \frac{6}{(d+1)(d+2)}, \label{eq:alpha_23}
\end{equation} 
yielding $\frac{2\des_3}{\des_2^2} \xrightarrow{d=2^n\to \infty} 3$, which is precisely the scaling observed in Ref.~\cite{Huang_2020}.

\subsubsection{Local shadows}

In this section, we will assume that our observable $O$ is of form $O = O_1 \otimes \dots \otimes O_n$, and we assume that $\Tr[O] = 0$ (for example, in finite dimensions, $O$ can be a Pauli observable). We expand our state as $\rho = \sum_\alpha c_\alpha \rho^{(\alpha)}_1 \otimes \dots \otimes \rho^{(\alpha)}_n$, with $\Tr[\rho^{(\alpha)}_i] = 1$ for each $i$ and $\alpha$.
We sample from the ensemble $\CX=\CX_1^{\otimes n}$ with measure $dP(X_1 \otimes \dots \otimes X_n)$ defined by 
\begin{equation}
dP(X_1 \otimes \dots \otimes X_n) = 
\Tr\!\left[\rho(X_1 \otimes  \dots \otimes  X_n)\right] ~d\mu(X_1) \dots d\mu(X_n).
\end{equation}
Our estimator is 
\begin{equation}
o(X_1 \otimes \dots \otimes X_n) = \left(\frac{2}{\des_2} \right)^n  \Tr[X_1 O_1] \dots \Tr[X_n O_n].
\end{equation}

The expectation value of our estimator is then 
\begin{align}
\mathbb E_{X_1\otimes \dots \otimes X_n \sim dP}[o(X_1\otimes \dots \otimes X_n)] 
&= \int_{\CX_1^{\otimes n}} o(X_1\otimes \dots \otimes X_n) ~dP(X_1\otimes \dots \otimes X_n) \nonumber \\
&= \left(\frac{2}{\des_2} \right)^n\sum_\alpha c_\alpha \prod_{i=1}^n \int_{\CX_1} \Tr[X_i O_i]\Tr[X_i \rho_i^{(\alpha)}]~d\mu(X_i) \nonumber\\
&= \sum_\alpha c_\alpha \prod_{i=1}^n \Tr[O_i \rho_i^{(\alpha)}] \nonumber \\
&= \Tr[\rho O] 
\end{align}
The variance of our estimator is
\begin{align}
\mathrm{Var}_{X_1\otimes \dots \otimes X_n \sim dP}[o(X_1\otimes \dots \otimes X_n)]
&\leq \int_{\CX_1^{\otimes n}} o(X_1\otimes \dots \otimes X_n)^2 ~dP(X_1\otimes \dots \otimes X_n) \nonumber \\
&= \sum_\alpha c_\alpha \prod_{i=1}^n \frac{4}{\des_2^2}\int_{\CX_1} \Tr[X_i O_i]^2\Tr[X_i \rho_i^{(\alpha)}]~d\mu(X_i) \nonumber \\
&= \sum_\alpha c_\alpha \prod_{i=1}^n \frac{2\des_3}{\des_2^2}  \Tr[O_i^2] \nonumber \\
&= \left(\frac{2\des_3}{\des_2^2}\right)^n  \prod_{i=1}^n \Tr[O_i^2] \nonumber \\
&= \left(\frac{2\des_3}{\des_2^2}\right)^n \Tr[O^2] 
\end{align}
Again we see that the relevant factor is $\frac{2\des_3}{\des_2^2}$. 
For fixed $d=2$, using Eq.~\eqref{eq:alpha_23}, this value is $2\des_3/ \des_2^2=9/4<3$.

If the design was such that $2\des_3\leq \des_2^2$, this design-based shadow tomography protocol is expected attain an asymptotic variance of $0$ for \textit{any} observable. This suggests that in general a parameter tradeoff $2\des_3\geq \des_2^2$ must hold and the optimal ensemble is the one that minimizes this inequality. Observe that here we have used the \textit{local shadow tomography} protocol to derive a statement about the global shadow as well as general parameter tradeoffs between the factors $\des_t$ of any $3$-design.

\section{Continuous variable shadow tomography using GKP states}\label{sec:GKPShadows}

\subsection{Bounding the estimator variance}
When estimating the expectation value of a set of observables $O_i,\; i=1,2,\cdots, M$, using estimators given by
\begin{equation}
    \tilde{o}_i=\braket{X|\rho|X}\braket{X|O_i|X} - \Tr\lrq{O_i}
\end{equation}
as in Eq.~\eqref{eq:estimator_app}, we need to bound its variance as it dictates the necessary sample overhead to obtain a good estimate. The variance ${\rm Var}\lrq{\tilde{o}}=\braket{\tilde{o}^2}-\braket{\tilde{o}}^2\leq \braket{\tilde{o}^2}$ is naturally upper bounded by the second moment, so we focus on bounding that.

To bound the estimator second moments without additional assumptions, we employ existing bound for second moments of functions over symplectic lattices derived in Ref.~\cite{Kelmer_2019}. %
Let $f\in L^p\lr{\R^{2n}}$, the $p$-norm is given by
\begin{equation}
    \|f\|_p=\lr{\int_{\R^{2n}}d\bs{x}\, f\lr{x}^p}^{\frac{1}{p}}.
\end{equation}
We also make use of Riemann zeta function

\begin{equation}
    \zeta\lr{z}=\sum_{k\in \N}k^{-z}.
\end{equation}

The first statement is a second moment bound for functions of symplectic lattices, which we adapt from Refs.~\cite{das2024centrallimittheoremslattice,ChasingShadows}.

\begin{lem}[symplectic second moments \cite{das2024centrallimittheoremslattice,ChasingShadows}]\label{lem:symplectic_second_moments}

    Let $f\in L^1\lr{\R^{2n}}\cap L^2\lr{\R^{2n}}$ be a non-negative function and $\Lambda \in Y_n$, with
    \begin{equation}
    F_f\lr{\Lambda}=\sum_{\bs{x}\in \Lambda-\lrc{0}} f\lr{\bs{x}}.
    \end{equation}
    
    For $n=1$, we have that
    \begin{equation}
        \int_{Y_1}d\mu\lr{\Lambda}\, \abs{F_f\lr{\Lambda}}^2 = \|f\|_1^2+\|f\|_2^2,
    \end{equation}

    and for $n>2$ it holds that
    
    \begin{equation}
        \int_{Y_n}d\mu\lr{\Lambda}\, \abs{F_f\lr{\Lambda}}^2 \leq \|f\|_1^2+\frac{4\zeta\lr{n}^2}{\zeta\lr{2n}}\|f\|_2^2.
    \end{equation}
\end{lem}
Note that the restriction to non-negativity of $f$ is purely cosmetic can be easily removed. In the following application of this lemma $f$ will be non-negative by default.
\proof
In the case $n=1$ this statement follows from the fact that $\Sp_{2}\lr{\R}=\SL_{2}\lr{\R}$, such that in particular the spaces $X_2=\SL_{2}\lr{\Z}\backslash \SL_{2}\lr{\R}$ and $Y_1$ are identical. The statement is then provided by Ref.~\citep[Proposition 2.2]{das2024centrallimittheoremslattice}. For $n>1$ this bound was proven in Ref.~\citep[Lemma 2]{ChasingShadows} using the results of Ref.~\cite{Kelmer_2019}. Note that the typically cited requirement of compact support for $f$ can generally be relaxed to any $f$ that decays sufficiently fast such that the series converge and integrals and sums may be exchanged or alternatively, that $f\in L^2\lr{\R^{2n}}$ is bounded and non-negative (see Ref.~\citep[p.3]{Kelmer_2019})
\endproof

Note that while the second statement in this lemma also principally also holds true for $n=1$, as $\lim_{n\to 1}\zeta \lr{n}=\infty$, the bound becomes trivial. As the next step we show that

\begin{lem}[Second moment bound]\label{lem:variance_X_alt}
    Let $\ket{X}\in \CX_n$ for an $n$-mode CV quantum system as defined in Eq.~\eqref{eq:GKP_ensemble} and let $O=O^{\dagger}$ be a trace class observable with characteristic function $c_O$.
    Relative to the uniform measure over $\CX_n$, it holds that
    \begin{equation}
        \int_{\CX_n}d\mu\lr{X} \braket{X|\rho|X} \braket{X|O|X}^2 \leq \abs{\Tr\lrq{O}}^2+2\abs{\Tr\lrq{O}} \|c_O\|_1+\|c_O\|_1^2+f_n\Tr\lrq{O^2}\eqqcolon \widehat{V}^n_O, 
    \end{equation}
    where $f_1=1$ and for $n>1$,
    \begin{equation}
        f_n=\frac{4\zeta\lr{n}^2}{\zeta\lr{2n}}.
    \end{equation}
\end{lem}

\proof
Denote $\ket{X}=\ket{\Lambda, \bs{\alpha}}=D\lr{\bs{\alpha}}\ket{\Lambda}$.
We have
\begin{align}
    M&= \int_{Y_n}d\mu(\Lambda)\int_{P(\Lambda)} d\bs{\alpha}\, \bra{\Lambda,\bs{\alpha}}  \rho \ket{\Lambda,\bs{\alpha}}  \bra{\Lambda,\bs{\alpha}} O \ket{\Lambda,\bs{\alpha}}^2 \nonumber\\
    &= \int_{Y_n}d\mu(\Lambda)\int_{P(\Lambda)} \sum_{\bs{\lambda}_1, \bs{\lambda}_2, \bs{\lambda}_3 \in \Lambda}\, e^{i\Phi\lr{\bs{\lambda}_1}+i\Phi\lr{\bs{\lambda}_2}+i\Phi\lr{\bs{\lambda}_3}+i2\pi\bs{\alpha}^TJ\lr{\bs{\lambda}_1+\bs{\lambda}_2+\bs{\lambda}_3}} c_{\rho}\lr{\bs{\lambda}_1}c_{O}\lr{\bs{\lambda}_2}c_{O}\lr{\bs{\lambda}_3} \nonumber\\ 
    &=\int_{Y_n}d\mu\lr{\Lambda} \sum_{\bs{\lambda}_2, \bs{\lambda}_3 \in \Lambda}\,e^{i\Phi\lr{\bs{\lambda}_2+\bs{\lambda}_3}+i\Phi\lr{\bs{\lambda}_2}+i\Phi\lr{\bs{\lambda}_3}} c^*_{\rho}\lr{\bs{\lambda}_2+\bs{\lambda}_3}c_O\lr{\bs{\lambda_2}}c_O\lr{\bs{\lambda_3}}.
\end{align}

Since $\abs{c_{\rho}\lr{\bs{\gamma}}}=\abs{\Tr\lrq{D^{\dagger}\lr{\bs{\gamma}}\rho}}\leq 1$ by the unitarity of $ D^{\dagger}\lr{\bs{\gamma}}$, we can bound

\begin{align}
    \abs{M}&\leq \int_{Y_n}d\mu\lr{\Lambda} \sum_{\bs{\lambda}_2, \bs{\lambda}_3 \in \Lambda} \abs{c_O\lr{\bs{\lambda_2}}}\abs{c_O\lr{\bs{\lambda_3}}}\nonumber\\ &= \int_{Y_n}d\mu\lr{\Lambda} \lrc{\abs{c_O\lr{0}}+F_{\abs{c_O}}\lr{\Lambda}}^2\nonumber\\
    &= \int_{Y_n}d\mu\lr{\Lambda} \abs{c_O\lr{0}}^2+2\abs{c_O\lr{0}}F_{\abs{c_O}}\lr{\Lambda}+F_{\abs{c_O}}\lr{\Lambda}^2
\end{align}
Using the mean value formula, lemma \ref{lem:mean-value}, together with lemma~\ref{lem:symplectic_second_moments}, we obtain for $n=1$

\begin{equation}
    |M|\leq \abs{c_O\lr{0}}^2+2\abs{c_O\lr{0}}\|c_O\|_1+\|c_O\|_1^2+\|c_O\|_2^2,
\end{equation}

and for  $n>1$,
\begin{equation}
 |M|\leq   \abs{c_O\lr{0}}^2+2\abs{c_O\lr{0}}\|c_O\|_1 +\|c_O\|_1^2+\frac{4\zeta \lr{n}^2}{\zeta\lr{2n}}\|c_O\|_2^2.
\end{equation}

Finally, we use that $c_O\lr{0}=\Tr\lrq{O}$ and $\|c_O\|_2^2=\Tr\lrq{O^2}$.
\endproof

\begin{figure}
    \centering
    \includegraphics[width=.4\linewidth]{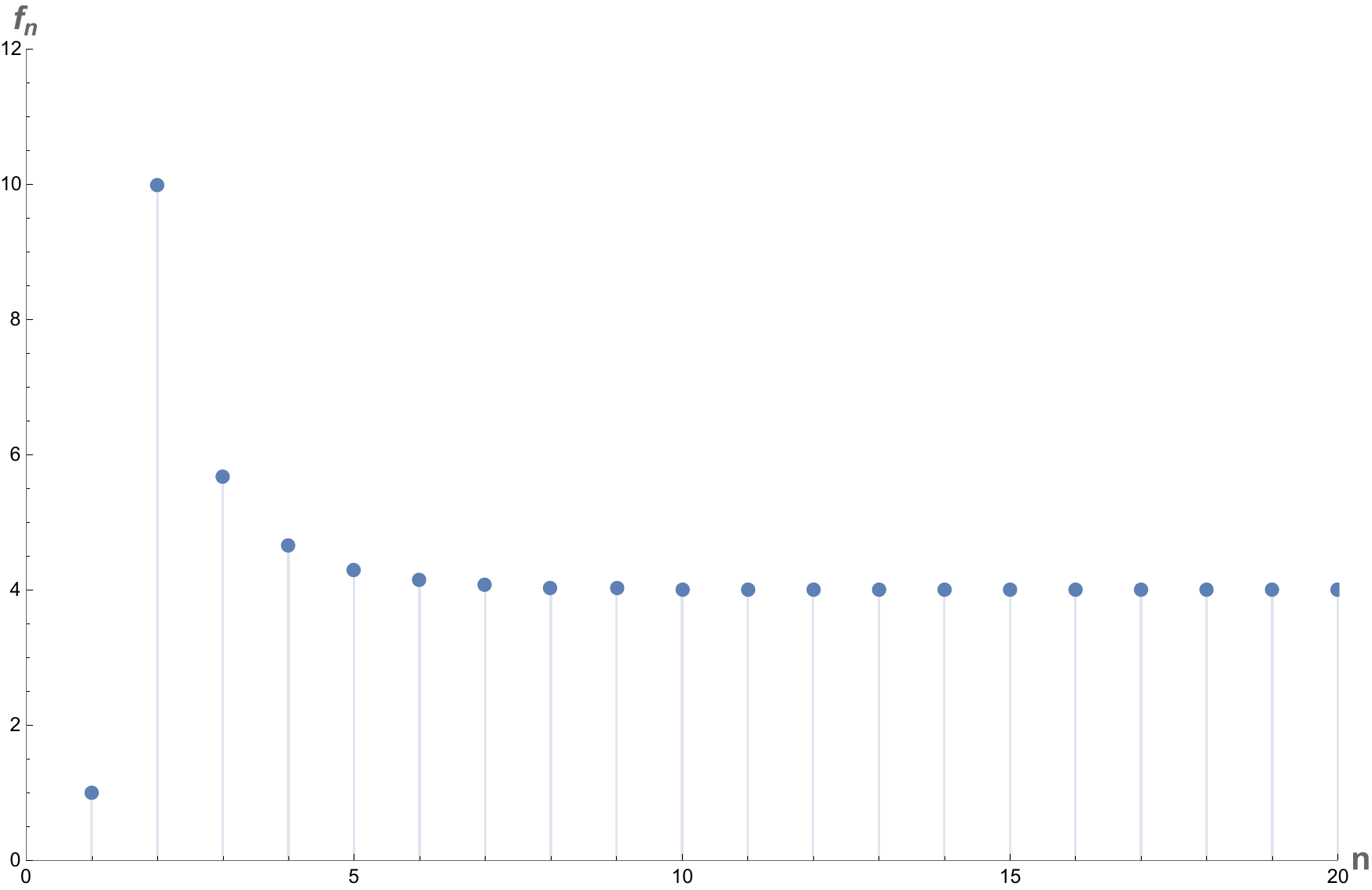}
    \caption{The sequence $\lrc{f_n}_{n=1}^{\infty}$ that parametrizes the variance upper bound for the GKP shadow tomography estimators. Note that the sequence converges to $f_n \to 4$ for $n\to \infty$.}
    \label{fig:fn_plot}
\end{figure}

We illustrate the sequence $f_n$ in Fig.~\ref{fig:fn_plot}. We observe that the sequence is bounded from above by $f_n\leq f_2=10$ and, as $n$ grows, we have
\begin{equation}
    \lim_{n\to \infty} f_n =4.
\end{equation}

The boundedness of the factor can also be verified by the fact that for $n\geq 2$, $\zeta(n)\leq \zeta(2n)$, such that $\zeta(n)^2/\zeta(2n)\leq\zeta(n)\leq \zeta(2)$. The actual bound observed indeed is a little bit stronger, and stems from that observation that for $n\geq 2$, $f_n$ is monotonically decreasing such that the maximum is attained at $f_2$.

\subsection{Physical state constraints}

The previous statements have been derived entirely without external assumptions on the input state or strong constraints on the locality structure of the ensemble $\CX$. In the following sections, we investigate how the incorporation of such constraints manifests in the sample complexity of the shadow tomography protocol. 
We define the generalized effective squeezing parameter via \cite{Duivenvoorden_Sensor} 

\begin{equation}
\Delta^2_{\bs\lambda}\lr{\rho}= -\frac{2}{\pi}\log\lr{\abs{c_{\rho}\lr{\bs\lambda}}}.
\end{equation}

This parameter captures \textit{how translationally} invariant a state is relative to the lattice vector $\bs{\lambda}\in \Lambda$. Equivalently, it parametrizes how strongly the state is squeezed along the phase space hyperplane $\bs{\lambda}^{\perp}\subset \R^{2n}$. For $n=1$, it holds that approximatively $\Delta^2 \geq \frac{4}{\Tr\lrq{\rho \hat{n}}}$ \cite{Duivenvoorden_Sensor}.

We make the following assumption on physical states.

\begin{ass}[structure of physical state]\label{ass:physical}
    Let $M=\lr{\bs{\xi}_1,\hdots,\bs{\xi}_{2n}}^T$ be a basis for the lattice $\Lambda$. Then it holds that
    for any physical state $\rho$ and $\bs{n}\in \Z^{2n}$

    \begin{equation}
    \abs{c_{\rho}\lr{\sum_{i=1}^{2n} n_i\bs{\xi}_i}} \leq \prod_{i=1}^{2n} |c_{\rho}\lr{\bs{\xi}_i}|^{\abs{n_i}}.
    \end{equation}
    
\end{ass}

Note that this assumption is well motivated, as the characteristic function of any physical state is Schwartz, i.e. it decays super-polynomially in phase space. As a random lattice will typically have a shortest vector length $\lambda_1\lr{\Lambda} \sim \sqrt{n}$, we expect all vectors $\bs{\xi}_i$ in the selected basis to follow at least this length scale.

Using this assumption, we have

\begin{prop}[physicality]\label{prop:physicality_0}
    For any physical state $\rho$, and any $\ket{X}\in \CX_n$ it holds that 

\begin{equation}
    \braket{X|\rho | X} \leq \lr{2\lr{1-e^{-\frac{\pi}{2}\tilde{\Delta}^{2}\lr{\rho}}}^{-1}-1}^{2n},
\end{equation}
where
\begin{equation}
\tilde{\Delta}\lr{\rho}\coloneqq \min_{M \in \Sp_{2n}\lr{\R}} \min_{i} \Delta_{\bs\xi_i}\lr{\rho}
\end{equation}
with $M=(\bs\xi_1,.., \bs\xi_{2n})^T$. 

\end{prop}
\proof
Let $M=\lr{\bs{\xi}_1,\hdots, \bs{\xi}_{2n}}^T$ be a minimal basis for $\Lambda$. It holds that, using assumption \ref{ass:physical},
\begin{align}
        \abs{\braket{X|\rho | X}} &\leq \sum_{\bs{\lambda}\in \Lambda} |c_{\rho}\lr{\bs{\lambda}}| =\sum_{\bs{\lambda}\in \Lambda} e^{-\frac{\pi}{2}\Delta^2_{\bs\lambda}\lr{\rho} } 
        \nonumber\\
        &\leq\prod_{i=1}^{2n} \lr{\abs{c_{\rho}\lr{0}}+2\sum_{n_i\in \N} |c_{\rho}\lr{\bs{\xi}_i}|^{n_i}} \nonumber\\
        &=\prod_{i=1}^{2n} \lr{1+2\sum_{n_i\in \N} e^{-\frac{\pi}{2}n_i\Delta^2_{\bs\xi_i}\lr{\rho}}} \nonumber\\
        &=\prod_{i=1}^{2n} \lr{2\lr{1-e^{-\frac{\pi}{2}\Delta^2_{\bs\xi_i}\lr{\rho}}}^{-1}-1}. 
\end{align}

Let $\tilde{\Delta}_{M}\lr{\rho}=\min_{i} \Delta_{\bs\xi_i}\lr{\rho}$, where $M=\lr{\bs{\xi}_1,\hdots, \bs{\xi}_{2n}}^T$.

We can thus bound
\begin{equation}
    \abs{\braket{X|\rho | X}} \leq \lr{2\lr{1-e^{-\frac{\pi}{2}\tilde{\Delta}^{2}_{M}\lr{\rho}}}^{-1}-1}^{2n}.
\end{equation}

Finally, minimizing over the possible lattice choices and all possible choices of bases yields the result.
\endproof

Following the arguments in Refs.~\cite{Terhal_2016, Duivenvoorden_Sensor}, we further estimate that a general lower bound for the effective squeezing is set by the average photon number of the state, $\tilde{\Delta}\lr{\rho}\geq \frac{4}{\Tr\lrq{\rho \hat{N}}}$. For any fixed lattice $\Lambda$, let $M=S^T=\lr{\bs{\xi}_1,\hdots, \bs{\xi}_{2n}}^T$ denote a choice of shortest basis. It is reasonable to expect that any state that delivered a small value of effective squeezing $\tilde{\Delta}_{M}$ for the lattice basis $M$ should also have a high degree of approximative symmetry under the corresponding group of GKP stabilizer operators.
The state $\ket{\Lambda}$ may be obtained as $\ket{\Lambda}=U_S\ket{\Z^{2n}}=U_S\ket{\Z^{2}}^{\otimes n}$ for $U_S$ a Gaussian unitary transform \cite{Conrad_2022}. To obtain the approximate version, we replace each individual state $\ket{\Z^2}$ with an approximate state $\ket{\widetilde{\Z^2}}\approx e^{-\Delta^2 \hat{n}}\ket{\Z^2}/\sqrt{\braket{\Z^2|e^{-2\Delta^2 \hat{n}}|\Z^2}}$ with effective squeezing parameter $\Delta^2\geq 4/\overline{n}$, where $\overline{n}=\braket{\widetilde{\Z^2}|\hat{n} | \widetilde{\Z^2}}$ is the average photon number of the individual state \cite{Duivenvoorden_Sensor, surfGKP}. Let $I=\lr{\bs{e}_1,\hdots, \bs{e}_{2n}}^T$ be a basis that stabilizes the state $\ket{\Z^2}$ and $S^T=I  S^T$. We have, by the properties of the trace,
\begin{equation}
    \Tr\lrq{D\lr{S\bs{e}_i} U_S\ketbra{\widetilde{\Z^{2n}}} U_S^{\dagger}}= \Tr\lrq{D\lr{\bs{e}_i} \ketbra{\widetilde{\Z^{2n}}} }.
\end{equation}
In particular, the finite squeezing parameters satisfy 
\begin{align}
    \Delta_{S^T}^2\lr{U_S\ketbra{\widetilde{\Z^{2n}}} U_S^{\dagger}} 
    &=\Delta_{I}^2\lr{\ketbra{\widetilde{\Z^{2n}}}}\\
    & \geq 4/\overline{N},
\end{align}
where the last inequality stems from the fact that, optimizing over all possible states, a minimal value for $\Delta_{I}^2=\Delta^2$ could be obtained by considering states of the form $\ket{0}^{\otimes (n-1)} \otimes \ket{\widetilde{\Z^2}}$, where $\ket{\widetilde{\Z^2}}$ attaims the minimal $\Delta^2$ the total average photon number allows for. We hence have motivated the following conjecture.
\begin{conj}[Lower bound for the effective squeezing parameter]\label{conj:Delta_lower}
    Given a physical state $\rho$ on an $n$-mode quantum system with average photon number $\overline{N}=\Tr\lrq{\rho\hat{N}}$, it holds with $M=\lr{\bs{\xi}_1,\hdots, \bs{\xi}_{2n}}^T$, that
    \begin{equation}
    \tilde{\Delta}^2\lr{\rho}\coloneqq \min_{M \in \Sp_{2n}\lr{\R}} \min_{i} \Delta^2_{\bs\xi_i}\lr{\rho} \geq \frac{4}{\overline{N}}.
    \end{equation}
\end{conj}

Using conjecture~\ref{conj:Delta_lower}, for any state with average photon number at most $\overline{N}$ the factor under the exponent can be bounded by 

\begin{align}
    \gamma &=2\lr{1-e^{-\frac{\pi}{2}\tilde{\Delta}^2}}^{-1}-1  \nonumber \\
    &\leq 2\lr{1-e^{-2\pi /\overline{N}}}^{-1}-1 \nonumber\\
    &\approx \overline{N}/\pi \label{eq:gamma_est}
\end{align}
in the limit $\overline{N}\gg 1$.

We note that, following a strategy proposed in Ref.~\cite{Duivenvoorden_Sensor}, a possible proof to this the conjecture may be obtained by viewing $\rho$ as a ``sensor state" that is employed to estimate a displacement $D\lr{\bs{x}}$ by executing measurements on copies of $\rho_{\bs{x}}=D\lr{\bs{x}}\rho D^{\dagger}\lr{\bs{x}}$. We expect that a Cramer-Rao bound for this estimation problem would imply the result.

\subsection{Variance bounds with physicality assumptions}

We have also attempted to derive a similar bound for the local ensemble $\CX_1^{\otimes n}$, but have not been able to distill a sufficiently nice and interpretable expression. Instead, to derive a corresponding bound for the ensemble $\CX_1^{\otimes n}$, we invoke the physicality bound from assumption \ref{ass:physical} and proposition \ref{prop:physicality_0}.

\begin{lem}[Second moment of global ensemble, physical]\label{lem:variance_X_phys}
    Let $\CX_n$ be the ensemble of GKP states and let $\rho$ be any physical state, for which assumption \ref{ass:physical} holds with average photon number at most $\overline{N}$.
    We have that for any hermitian trace class observable $O$,
    \begin{equation}
        \int_{\CX_1^{\otimes n}}d\mu\lr{X} \,\braket{X|  \rho |X}\braket{X | O | X}^2 \leq  \lr{\overline{N}/\pi}^{2n}\lr{\Tr\lrq{O}^2+\Tr\lrq{O^2}} \eqqcolon \lr{\overline{N}/\pi}^{2n}\tilde{V}_O.
\end{equation}
\end{lem}
\proof
 We have that for any hermitian trace class observable $O$
    \begin{align}
        \int_{\CX_1^{\otimes n}}d\mu\lr{X} \,\braket{X|  \rho |X}\braket{X | O | X}^2 &\leq   \lr{\overline{N}/\pi}^{2n}\int_{\CX_n}d\mu\lr{X}\, \Tr\lrq{\lr{O\otimes O} \ketbra{X}^{\otimes 2}}
        \nonumber \\ &= \lr{\overline{N}/\pi}^{2n} \lr{\Tr\lrq{O}^2+\Tr\lrq{O^2}} \nonumber \\
     &\eqqcolon \lr{\overline{N}/\pi}^{2n}\tilde{V}^{n}_O,
\end{align}
where we have used bound in prop.~\ref{prop:physicality_0} the $2$-design property, Thm.~\ref{them:2design} and the bound in Eq.~\ref{eq:gamma_est}.
\endproof

\begin{lem}[Second moment of local ensemble, physical]\label{lem:variance_X_loc_phys}
    Let $\CX_1^{\otimes n}$ be the $n$-fold ensemble of single-mode GKP states and let $\rho$ be any physical state, for which assumption \ref{ass:physical} holds with average photon number at most $\overline{N}$.
    We have that for any hermitian trace class observable $O$,
    \begin{equation}
        \int_{\CX_1^{\otimes n}}d\mu\lr{X} \,\braket{X|  \rho |X}\braket{X | O | X}^2 \leq  \lr{\overline{N}/\pi}^{2n}\sum_{\bs{k} \subseteq \bs{n}} \Tr_{\bs{n}-\bs{k}}\lrq{\Tr_{\bs{k}}\lrq{O}^2} \eqqcolon \lr{\overline{N}/\pi}^{2n}\tilde{V}^{\rm loc}_O,
\end{equation}
where  $\sum_{\bs{k} \subseteq \bs{n}}$ denotes the sum over all elements $\bs{k}$ of the power set  of $\bs{n}=\lrc{1,\hdots,n}$ and the outer trace ranges over the complement of $\bs{k}$.
\end{lem}

\proof
Let $\Lambda_{\oplus}^n=\Lambda_1\oplus \hdots \oplus \Lambda_n$ denote a direct sum composite lattice and let $\bs{\alpha}_{\oplus}^n=\bs{\alpha}_1\oplus \hdots \oplus \bs{\alpha}_n$ denote a composite vector, which is such that $\bs{\alpha}_{\oplus}^n\in \CP\lr{\Lambda_{\oplus}^n}$. Note that $\Lambda_{\oplus}^n$ is symplectic if its components $\Lambda_i$ are. Each local shadow sample is represented by a composite state

\begin{equation}
    \bigotimes_{i=1}^n\ket{X_i}= \ket{\Lambda_{\oplus}^n; \bs{\alpha}_{\oplus}^n}
\end{equation}
such that we have using prop.~\ref{prop:physicality_0}

\begin{align}
    M&=\int_{\CX_1^{\otimes n}}d\mu\lr{X}\, \braket{X | \rho | X}\braket{X | O | X}^2 \nonumber\\
    &\leq \lr{\overline{N}/\pi}^{2n}\int_{\CX_1^{\otimes n}}\braket{X | O | X}^2 \nonumber\\
&=\lr{\overline{N}/\pi}^{2n}\int_{Y_1^{\oplus n}}d\mu\lr{\Lambda_{\oplus}^n} \,\sum_{ \bs{\lambda}_{\oplus, 1}^n, \bs{\lambda}_{\oplus, 2}^n\in \Lambda_{\oplus}^n} e^{i\Phi\lr{\bs{\lambda}_{\oplus, 1}^n+\bs{\lambda}_{\oplus,2}^n}}\nonumber\\
&\hspace{2cm}
\int_{\CP\lr{\Lambda_{\oplus}^n}}d\bs{\alpha}_{\oplus}^n\, e^{-i2\pi {\bs{\alpha}_{\oplus}^n}^T J \lr{\bs{\lambda}_{\oplus, 1}^n+\bs{\lambda}_{\oplus, 2}^n}}
c_O\lr{\bs{\lambda}_{\oplus, 1}^n}c_O\lr{\bs{\lambda}_{\oplus, 2}^n}\nonumber\\
&=\lr{\overline{N}/\pi}^{2n}\int_{Y_1^{\oplus n}}d\mu\lr{\Lambda_{\oplus}^n} \,\sum_{ \bs{\lambda}_{\oplus}^n\in \Lambda_{\oplus}^n} \abs{c_O\lr{\bs{\lambda}_{\oplus}^n}}^2.
\end{align}
We split the lattice into partitions $\bs{k}\subseteq \bs{n}=\lrc{1,\hdots, n}$, where the multi-index $\bs{k}$ indicates the indices where the lattice vector has non-zero components,
\begin{equation}    
\Lambda_{\oplus}^n=\bigcup_{\bs{k}\subseteq \bs{n}}  \lr{\Lambda-\lrc{0}}^{\bs{k}}\cup \lrc{0}^{\bs{n}-\bs{k}}.
\end{equation}
Using the fact that
\begin{equation}
    c_O\lr{ \bs{\lambda}^{\bs{k}} \oplus\lrc{0}^{\bs{n-}\bs{k}}} = c_{\Tr_{\bs{n-}\bs{k}}\lrq{O}}\lr{\bs{\lambda}},
\end{equation}
we rewrite the sum above as
\begin{align}
    M&\leq \lr{\overline{N}/\pi}^{2n}\sum_{\bs{k}\subseteq \bs{n}}\int_{Y_1^{\oplus n}}d\mu\lr{\Lambda_{\oplus}^n} \,  \sum_{ \bs{\lambda}^{\bs{k}}\in \lr{\Lambda-\lrc{0}}^{\bs{k}}} \abs{c_{\Tr_{\bs{n}-\bs{k}}\lrq{O}}\lr{\bs{\lambda}^{\bs{k}}}}^2 \nonumber\\
&=\lr{\overline{N}/\pi}^{2n}\sum_{\bs{k}\subseteq \bs{n}} \Tr_{\bs{k}}\lrq{\Tr_{\bs{n}-\bs{k}}\lrq{O}^2}.
\end{align}

Where, in the last step, we have performed the integral over the independent components in $\bs{k}$ and used that for $O=O^{\dagger}$,
\begin{equation}
    \int_{\R^{2k}} d\bs{x}\, \abs{c_{O}\lr{\bs{x}}}^2=\Tr\lrq{O^2}.
\end{equation}

\endproof

\subsection{Continuous variable medians of means estimation}

We employ median of means (MoM) estimation to set up an algorithm for the estimation of the expectation values of $M$ observables $\lrc{O_i}$ relative to the statistics obtained from measuring an arbitrary (physical) input state relative to a GKP state, such that the measurement probability is given by $\braket{X|\rho|X}, \ket{X}\in \CX$ as outlined before.

\textit{Medians of means} (MoM) estimation provides a beneficial strategy to implement this task, which has the upshot that the necessary sample complexity admits scaling linearly with the variance of the estimator and \textit{logarithmically} in the number of observables one is interested in estimating. Given $N=KB$ samples of a random variable $\tilde{o}$, MoM estimation proceeds by dividing this set into $K$ batches each of size $B$ and outputs the median value of the arithmetic mean of $\tilde{o}$ on each batch. This strategy biases the expected value away from low-probability events at the tail of the distribution and allows to prove exponential concentration of the resulting estimate around the true mean (see Ref.~\cite{Chen_note_MoM} for a pedagogical introduction).

MoM proceeds as follows. Divide the $N=KB$ samples of $\tilde{o}_i$ into $K$ batches $\CB_i^{(j)},\; j=1,2,\cdots, K$ each of size $B$, and
\begin{align}
    {\rm MoM}\lrq{\tilde{o}_i}  = {\rm median}\lrc{ \overline{o}^{(1)}_i,\hdots, \overline{o}^{(K)}_i}, \quad  \overline{o}^{(j)}_i = \frac{1}{B}\sum_{\tilde{o} \in  \CB_i^{(j)}} \tilde{o}
\end{align}
is the medians of means estimator. See Refs.~\cite{Chen_note_MoM, lerasle2019lecturenotesselectedtopics, Huang_2020} for good explanations of this estimator.

\begin{them}[GKP shadow tomography]\label{them:CV_shadow}
    Let $\epsilon, \delta >0$, and let $O_i=O_i^{\dagger}, i=1,2,\cdots, M$ be a set of $M$ observables on an $n$-mode CV quantum system with integrable characteristic functions $c_{O_i}$. Let
    \begin{equation}
    \begin{aligned}\label{eq:global-variance}
        \tilde{V}_{O_i} &=\lr{\|c_{O_i}\|_1+\abs{\Tr\lrq{O_i}}}^2 + f_n\Tr\lrq{O_i^2} \\ \vspace{.5cm}
        \tilde{V}_{O}   & =\max_i \tilde{V}_{O_i},
    \end{aligned}
    \end{equation}
    and  $N=KB$, with
    \begin{equation}
        B= 34 \tilde{V}_O / \epsilon^2 \quad\text{and} \quad K=2\log\lr{2M/\delta}.
    \end{equation}
    Then, $N$ samples from the state relative to the 
    POVMs defined by the ensemble of GKP states $\CX_n$ suffice to approximate the expectation values of each observable $O_i$ with probability
\begin{equation}
        {\rm Pr}\lr{\max_{i=1,\dots,M}\left\lvert {\rm MoM}\lrq{\tilde{o}_i} -  \Tr\lrq{O_i\rho}\right\rvert \geq \epsilon} \leq \delta.
    \end{equation}
\end{them}
\proof
The proof follows the standard proof using the variance bound $\tilde{V}_O$ from lem.~\ref{lem:variance_X_alt} for each observable, as discussed in Ref.~\cite{Huang_2020}.
\endproof

On the technical level, the key ingredient is that for medians of means, the error probability of the estimator is provided by counting the number of means that deviated from the median. This counting function can be expressed with a simple bounded random variable, such that Hoeffding's inequality establishes its anticoncentration. Naturally, if the estimator is bounded by default, one may directly apply Hoeffding's inequality to obtain a tight anticoncentration bound using a simple arithmetic mean estimator. This was the strategy employed in Ref.~\cite{Iosue_2024} and is particularly useful when $\braket{X | O| X}$
can be bounded. A special case where this is possible is outlined in appendix~\ref{sec:thermal} such that one can employ Popoviciu's inequality on the variance of a bounded estimator $|\sigma|\leq M$ to simply replace $\tilde{V}_O=M^2$ in the statement above.

Note that, using the physicality assumption we may replace the variance estimates in the theorem to obtain the following.

\begin{them}[CV shadow tomography, physical]\label{them:CV_shadow_phys}
    Let $\epsilon, \delta >0$, and let $O_i=O_i^{\dagger},\, i=1,2,\cdots, M$ be a set of $M$ observables on a $n$-mode CV quantum system with integrable characteristic functions $c_{O_i}$. Let $\rho$ be any physical state, for which assumption \ref{ass:physical} holds with average photon number at most $\overline{N}$ and let
    
    \begin{align}
        \tilde{V}_{O_i} & =\lr{\overline{N}/\pi}^{2n} \lr{\Tr\lrq{O_i}^2+\Tr\lrq{O_i^2}} \\
        \tilde{V}_{O} & =\max_i \tilde{V}_{O_i},
    \end{align}
    and  $N=cBK$, with
    \begin{equation}
        B=34\tilde{V}_O / \epsilon^2 \quad\text{and} \quad K=2\log\lr{2M/\delta},
    \end{equation}
    and $c$ a numerical constant.
    Then, $N$ samples from the state relative to the POVMs defined by the ensemble of GKP states $\CX_n\ni\ket{X}=D\lr{\bs{\alpha}}\ket{\Lambda}$, where $\ket{\Lambda}$ is the unique GKP state corresponding to $\Lambda=\Lambda^{\perp}\subset \R^{2n}$, and where $\bs{\alpha}\in \CP\lr{\Lambda}$, suffice to approximate the expectation values of each observable $O_i$ with high probability,
   \begin{equation}
        {\rm Pr}\lr{\max_{i=1,\dots,M}\left\lvert {\rm MoM}\lrq{\tilde{o}_i} -  \Tr\lrq{O_i\rho}\right\rvert \geq \epsilon} \leq \delta.
    \end{equation}
\end{them}
\proof
The proof follows the standard proof using the variance bound $\tilde{V}_O$ from lem.~\ref{lem:variance_X_phys} for each observable, as discussed in Ref.~\cite{Huang_2020} together with the estimate from Eq.~\eqref{eq:gamma_est}.
\endproof

To fully bound the performance of the local GKP shadow protocol, we employ the result derived in lemma~\ref{lem:variance_X_loc_phys}.

Note that the partition-based structure of the bound is consistent with the fact that, written in vectorized notation,  $\CM_{\CX_1}^{\otimes n}= \lr{I\otimes I^* +\sketbra{I}}^{\otimes n}$ behaves like a \textit{random} deletion channel.    
For $n=2$, the expression in lemma~\ref{lem:variance_X_loc_phys} yields (up to the factor $f^{2n}$)
\begin{align}
  \sum_{\bs{k} \subseteq \bs{n}} \Tr_{\bs{n}-\bs{k}}\lrq{\Tr_{\bs{k}}\lrq{O}^2}= \Tr\lrq{O^2}+\Tr\lrq{O}^2   +\Tr_2\lrq{\Tr_1\lrq{O}^2}+\Tr_1\lrq{\Tr_2\lrq{O}^2}.
\end{align}

In comparison to the global shadow protocol, we see that we obtain additional terms from all the non-trivial bi-partitions of the $n$-mode system. 
It is notable that, in contrast to e.g. the qubit-based shadow protocol in Ref.~\cite{Huang_2020}, the performance of the local shadow tomography protocol derived here is not limited by the \textit{locality} of the observable but rather, in the opposite, by its \textit{in-separability}. As long as the total purity of the observable across any bi-partition remains small, the local shadow protocol will be efficient.

Equipped with the above derivation, we arrive at following theorem.

\begin{them}[CV shadow tomography, local and physical]\label{them:CV_shadow_loc}
    Let $\epsilon, \delta >0$, and let $O_i=O_i^{\dagger}, i=1,2,\cdots, M$ be a set of $M$ observables on a $n$-mode CV quantum system with integrable characteristic functions $c_{O_i}$. Assume conjecture~\ref{conj:Delta_lower} and that the input state has average photon number at most $\overline{N}\gg 1$ with exponentially decaying characteristic function according to assumption~\ref{ass:physical}. Let
    \begin{align}
        \tilde{V}_{O_i}^{\rm loc} & = \lr{\overline{N}/\pi}^{2n}\sum_{\bs{k} \subseteq \bs{n}} \Tr_{\bs{n}-\bs{k}}\lrq{\Tr_{\bs{k}}\lrq{O_i}^2} \\
        \tilde{V}_{O}^{\rm loc}   & =\max_i \tilde{V}_{O_i}^{\rm loc},
    \end{align}
    and  $N=BK$, with
    \begin{equation}
        B= 34 \tilde{V}_O^{\rm loc} / \epsilon^2 \quad\text{and} \quad K=2\log\lr{2M/\delta}.
    \end{equation}
    
    Then, $N$ samples from the states relative to the POVMs defined by the ensemble of GKP states $\CX_1^{n}=\bigotimes_{i=k}^n D\lr{\bs{\alpha}_k}\ket{\Lambda_k}$, where $\ket{\Lambda_k}$ is the unique GKP state corresponding to $\Lambda_k=\Lambda_k^{\perp}\subset \R^{2}$ and where $\bs{\alpha}_k\in \CP\lr{\Lambda_k}$, suffice to approximate the expectation values of each observable $O_i$ with
   \begin{equation}
        {\rm Pr}\lr{\max_{i=1,\dots,M}\left\lvert {\rm MoM}\lrq{\tilde{o}_i} -  \Tr\lrq{O_i\rho}\right\rvert \geq \epsilon} \leq \delta.
    \end{equation}
    % \textcolor{red}{Joe: This should be 
    % \begin{equation}
    %     {\rm Pr}\lr{\max_{i=1,\dots,M}\left\lvert {\rm MoM}\lrq{\tilde{o}_i} -  \Tr\lrq{O_i\rho}\right\rvert \geq \epsilon} \leq \delta.
    % \end{equation}
    % Or alternatively, we could phrase it as: 
    % With probability at least $1-\delta$, $\left\lvert {\rm MoM}\lrq{\tilde{o}_i} -  \Tr\lrq{O_i\rho}\right\rvert < \epsilon$ for all $i$.
    % This is how they phrase it in Eq (S14) of \url{https://arxiv.org/pdf/2002.08953}
    % }
\end{them}
\proof
The proof follows the standard proof using the variance bound $\tilde{V}_O^{\rm loc}$ from lem.~\ref{lem:variance_X_loc_phys} for each observable, as discussed in Ref.~\cite{Huang_2020} as well as the estimate in Eq.~\eqref{eq:gamma_est}.
\endproof

\subsection{Example: Thermal state observables}\label{sec:thermal-state-obs}

We investigate the derived sample complexities for a simple observable, the $n$-mode (unnormalized) thermal state $O=e^{-\beta\hat{N}}$. To this end, we compute the variances $\widehat{V}_O^n, \tilde{V}_O, \tilde{V}_O^{\rm loc}$ derived in lemmas~\ref{lem:variance_X_alt}, ~\ref{lem:variance_X_phys} and \ref{lem:variance_X_loc_phys}, respectively.

We begin with the variance per lemma~\ref{lem:variance_X_alt}. We have

\begin{equation}
    \Tr\lrq{e^{-\beta\hat{N}}} = \lr{1-e^{-\beta}}^{-n},\quad \Tr\lrq{e^{-2\beta\hat{N}}} = \lr{1-e^{-2\beta}}^{-n}.
\end{equation}

Using \cite{CahillGlauber, surfGKP}
\begin{equation}
    O=e^{-\beta \hat{N}}=\lr{1-e^{-\beta}}^{-n} \int_{\R^{2n}}d\bs{x}\, e^{-\frac{\|\bs{x}\|^2}{\Delta^2}} D\lr{\bs{x}},
\end{equation}
where $\Delta^2=2\tanh\lr{\beta /2}$, we can also obtain the norm 
\begin{align}
    \|c_O\|_1&=\lr{1-e^{-\beta}}^n \int_{\R^{2n}}d\bs{x}\, e^{-\frac{\pi\|\bs{x}\|^2}{\Delta^2}} \nonumber\\
    &=\lr{1-e^{-\beta}}^n  \left|\frac{2\pi}{\lr{2\pi/\Delta^2 }}\right|^n \nonumber\\
    &=\lr{1-e^{-\beta}}^n \lr{2 \frac{1+e^{-\beta}}{1-e^{-\beta}}}^n \nonumber\\
    &=2^n\lr{1+e^{-\beta}}^n \nonumber\\
    &= \lr{4-2\beta}^n +O\lr{\beta^2}
\end{align}

Thus, the bound in lemma~\ref{lem:variance_X_alt} becomes
\begin{align}
\widehat{V}^n_O&=\abs{\Tr\lrq{O}}^2+2\abs{\Tr\lrq{O}} \|c_O\|_1+\|c_O\|_1^2+f_n\Tr\lrq{O^2} \nonumber \\
&=2^{n+1}+2^n\lr{1+e^{-2\beta}}^n  + 2^{2n}\lr{1+e^{-\beta}}^{2n}+f_n \lr{1-e^{-2\beta}}^{-n} \nonumber\\
&=O\lr{\lr{2\lr{1+e^{-\beta}}}^{2n}}.
\end{align}

For the remaining variances, a quick computation shows that
\begin{align}
\tilde{V}_O&=\Tr\lrq{O^2}+\Tr\lrq{O}^2 \nonumber \\
    &=\lr{1-e^{-2\beta}}^{-n}+\lr{1-e^{-\beta}}^{-2n} \nonumber\\
    &=\lr{1+e^{-\beta}}^{-n}\lr{1-e^{-\beta}}^{-n}+\lr{1-e^{-\beta}}^{-2n}
\end{align}
and 
\begin{align}
    \tilde{V}_O^{\rm loc} &= \sum_{\bs{k} \subseteq \bs{n}} \Tr_{\bs{n}-\bs{k}}\lrq{\Tr_{\bs{k}}\lrq{O}^2} \nonumber\\
    &=\sum_{k=0}^n \binom{n}{k}\lr{1-e^{-2\beta}}^{n-k}\lr{1-e^{-\beta}}^{-2k} \nonumber\\
&=\lr{1-e^{-\beta}}^{-2n}\lr{2e^{-3\beta}-2e^{-\beta}-e^{-4\beta}+2}^n.
\end{align}
In the high-temperature regime ($\beta\to 0$), the local shadow and global shadow follow the same variance scaling with $n$. Restoring the prefactor from the energy assumption, we hence obtain
\begin{equation}
    f^{2n}\tilde{V}_{O}, f^{2n}\tilde{V}_{O}^{{\rm loc}}\sim \lr{\frac{\overline{N}}{\pi\beta}}^{2n}~.
\end{equation}
In the reverse limit, $\beta \to \infty$, we also have $\tilde{V}_O\to 2$, $\tilde{V}_O^{\rm loc}\to 1$. 

It is notable that, under the energy constraint, the global- and local shadow attain the same effective scaling. They exhibit different scalings as $n\to \infty$ depending on whether $\overline{N}>\pi \beta $ or $\overline{N}<\pi \beta $. I particular, if the average photon number of the input system can be assumed to be very low $\overline{N}<\pi \beta $ (remember that we are aready assuming a high temperature limit $\beta \ll 1$), the given variance together with the sample complexity of the shadow tomography protocol may scale benignly in the system size.

\section{Sample complexity: comparison to related protocols}

\subsection{Qubit-based classical shadows}

We would like to make a comparison of the thermal-observable calculation in Sec.~\ref{sec:thermal-state-obs} to related qubit-based protocols. To this end consider, analogous to the above, a canonical choice of thermal state 
\begin{equation}
    O'=e^{-\beta \sum_{i=1}^n \hat{\sigma}_z},
\end{equation}
where each $\hat{\sigma}_z$ is the Pauli-Z operator on qubit ``$i$''. We evaluate the scaling bound estimated in Ref.~\cite{Huang_2020},
\begin{equation}
    \Tr\lrq{O'^2}=\lrq{e^{2\beta}-e^{-2\beta}}^n=\lrq{2\cosh\lr{2\beta}}^n =\lr{2+4\beta^2+ O\lr{\beta^4}}^n.
\end{equation}
In the limit of small $\beta \ll 1$, this bound scales as $2^n$ with $n$. Note that this bound differs from the bound derived from lemma ~\ref{lem:variance_X_alt}, $4^n$ (for $\beta \to 0$) only by an exponential factor, although the Hilbert space employed in our setting is of infinite size, rather than $2^n$.

\subsection{Photon number cutoffs}
Building on the intuition gained from the previous section, we further discuss the sample complexity of our GKP-design based shadow tomography scheme. We first investigate the sample complexity under the assumption that the input state is well represented by a state with a total photon number cutoff $N_{\rm max}$, which is a scenario also investigated in Refs.~\cite{gandhari2023precisionboundscontinuousvariablestate, Iosue_2024}.

Naturally, given a total photon number cutoff on the input state $N_{\rm max}$, this implies a trivial upper bound on the average photon number on the input $\overline{N}\leq N_{\rm max}$.

We can also use this derivation to compare to the multimode homodyne shadow protocol discussed in Ref.~\cite{gandhari2023precisionboundscontinuousvariablestate}, where each mode is cutoff at maximum photon number $N_{1, \rm max}$, such that the total photon number admits a cutoff of $N_{\rm max}=n N_{1, \rm max}$. 
Thus, our variance bound and sample complexity scaling of our shadow tomography protocol scale as $O\lr{\lr{nN_{1, \rm max}}^{2n}}$ under a very pessimistic estimate. While this scaling appears more beneficial in the local photon number $N_{1, \rm max}$, the scaling with the mode number $n$ appears worse. We expect this to be mainly an effect of the conservative estimate provided in conjecture \ref{conj:Delta_lower}.

Homodyne tomography can be bounded using the same type of shadow analysis using a combination of Corr.~IV.1.2 and Eq.~(15) from Ref.~\cite{gandhari2023precisionboundscontinuousvariablestate}.
The former yields a shadow sample complexity bound of the same type as our main theorem, while the latter fills in a key variable in said bound.
Keeping track of $N_{1,\text{max}}$ and $n$ yields a homodyne sample complexity of order $O(nC_{1}^{2n}N_{1,\text{max}}^{20n/3}\log2N_{1,\text{max}})$ (with $C_1$ a constant), roughly a quartic deterioration over our protocol in $N_{1,\text{max}}$. %
However, the shadow bounds of Ref.~\cite{gandhari2023precisionboundscontinuousvariablestate} are not tight, and it remains unclear how well they quantify the performance of homodyne tomography.

\subsubsection{Full state tomography}
Using these findings, we further investigate the application of our shadow tomography protocol to full state tomography.
To this end, note that, in Ref.~\cite{mele2024learningquantumstatescontinuous},
the following lower bound on energy constrained CV state tomography has been proven.
\begin{them}[Optimal CV tomography \cite{mele2024learningquantumstatescontinuous}]\label{them:CVlearning}
Let $\rho$ be an unknown quantum state on $n$ bosonic 
modes satisfying an energy
constraint $\Tr\lrq{\hat N\rho}
\leq  \overline{N}$ for some absolute constant $\overline{N}$. Then the number of copies of $\rho$ required to perform quantum state tomography with precision $\varepsilon$ in trace distance has to scale at least as 
$(\Theta (\overline{N}/n\varepsilon))^{2n}$.
\end{them}

This theorem is proven by showing that an energy constrained quantum state $\rho :\, \Tr\lrq{\hat N\rho}
\leq \overline{N} $ with average photon number $\overline{N}$ is approximated by a rank $r=O\lr{\lr{e\overline{N}/n\varepsilon}^n}$ state projected onto the subspace cutoff at total photon number $N_{\rm max} = \ceil{\overline{N}/\varepsilon^2} $ up to an error $\varepsilon$ in trace distance. The corresponding effective subspace has dimension $D=O\lr{\lr{e\overline{N}/n\varepsilon^{2}}^n}$. This approximation of the state by one restricted to the subspace of the CV Hilbert space then allows to leverage the optimal sampling bound of $O\lr{Dr}$ needed to learn a quantum state \cite{wrightHowLearnQuantum}. 

We denote by $\rho_D$ the input state $\rho$ projected onto the $D$-dimensional subspace, which with $D=O\lr{\lr{e\overline{N}/\nu^2\varepsilon^2}^{n}}$ is $\nu\varepsilon$-close in trace distance to the average energy constrained input state $\rho$. To reconstruct this state using our shadow tomography protocol, we compute the allowed estimation error on each element a complete basis of $D^2$ generalized (normalized) Pauli operators $\lrc{P_i}:\; \Tr\lrq{P_i^{\dagger}P_j}=\delta_{ij}$. Assume each Pauli operator is estimated to within additive error $\epsilon_S$. The trace distance between our estimate $\hat{\rho}$ and the finite target state then is
\begin{equation}
    \|\hat{\rho}-\rho_D\|\leq \epsilon_S\sum_{i=0}^{D^2-1} \|P_i\|_1 = \epsilon_S D^{\frac{5}{2}},\label{eq:trace_distance_bound}
\end{equation}
which is bounded from above by $\lr{1-\nu}\varepsilon$ for
\begin{equation}
    \epsilon_S \leq \lr{1-\nu}\varepsilon D^{-\frac{5}{2}}. 
\end{equation}
In total, we hence obtain with Thm.~\ref{them:CV_shadow_phys} an estimation with error
\begin{equation}
    \|\hat{\rho}-\rho_D+\rho_D-\rho\|_1\leq \lr{1-\nu}\varepsilon + \nu\varepsilon=\varepsilon,
\end{equation}
with a sample complexity scaling as
\begin{align}
O\lr{\lr{\frac{\overline{N}}{\pi}}^{2n}\frac{\tilde{V}_O}{\epsilon_S^2} \log\lr{\frac{D^2}{\delta}}}
&=O\lr{\lr{\frac{\overline{N}}{\pi}}^{2n}\frac{1}{\epsilon_S^2} \log\lr{\frac{D}{\delta^{1/2}}}}.
\label{eq:CV_learning_shadow}
\end{align}
Inserting $\epsilon_S\leq \varepsilon D^{-5/2}$ and $D=O\lr{\lr{e\overline{N}/\nu^2\varepsilon^2}^{n}}$ now yields the sample complexity estimate of

\begin{equation}
    O\lr{\lr{\frac{\overline{N}}{\pi}}^{2n}\frac{D^5}{\varepsilon^2} \log\lr{\frac{D}{\delta^{1/2}}}}=    O\lr{\lr{\frac{e^5\overline{N}^7}{\nu^{10}\varepsilon^{10}\pi^2}}^{n} n\log\lr{\frac{e\overline{N}}{\nu^2\varepsilon^2\delta^{1/2}}}}
\end{equation}

In Refs.~\cite{Becker_2024, mele2024learningquantumstatescontinuous} it has been shown that, in addition to the average photon number constraint as in Thm.~\ref{them:CVlearning}, a restriction on the higher moments of the total photon number of the input state
$\Tr\lrq{\hat{N}^k\rho}$ allows for an improved sample complexity in its dependence on $\varepsilon$.

Nevertheless, this scaling is significantly worse as compared to the optimal bound in Thm.~\ref{them:CVlearning}. This can be attributed to an inefficient choice of observable basis to implement the state tomography protocol as well as the fact that classical shadow tomography is not a particularly optimal way of performing this task. 

As full state tomography requires the execution of $D^2$ known POVMs, it is to be expected that a randomized approach, such as classical shadow tomography, necessarily underperforms. Note that even for qubit systems, the naive shadow-tomography based full-tomography routine discussed here similarly underperforms. To obtain a classical shadow that is indistinguishable from the original state up to trace distance $\varepsilon$, using the results of Ref.~\cite{Huang_2020} and $D=2^n$ together with Eq.~\eqref{eq:trace_distance_bound} also here we estimate a sample complexity scaling as
\begin{equation}
    O\lr{\frac{2^{5n}}{\varepsilon^2}\log\lr{\frac{4^n}{\delta}}},
\end{equation}
which is not quite as bad as the scaling in Eq.~\eqref{eq:CV_learning_shadow} since the system is natively finite and doesn't require to truncate the tails of the input state, but still remains in a similar ballpark.
% of terrible scaling.

\subsection{Lattice based CV shadows}
In Ref.~\cite{ChasingShadows}, the first author of the present manuscript discusses a random Wigner tomography scheme that finds an interpretation of shadow tomography  in the style of Ref.~\cite{Huang_2020} via a suitable logical channel twirl of a physical POVM. 
While also randomizing over lattices, however, the protocol utilizes \textit{classical} probability distributions that mimic the structure of random GKP states. 
No design condition is used or required in the protocol.
By contrast, in showing that GKP states form a 2-design, the present work shows that access to true GKP states serves as a powerful resource for state tomography.
This yields a protocol that is more closely aligned with the original design-based shadow tomography scheme.

\section{Implementation of design-based GKP shadow tomography}\label{sec:Implementation}

Given an input state $\rho$, the original shadow strategy to obtain samples from the effective state $\CM_{\CX_n}\lr{\rho}$ proceeds as follows.
\begin{enumerate}
    \item Sample a lattice $\Lambda \in Y_n$ specified by a generator matrix $M=\lr{\bs{\xi}_1,\hdots, \bs{\xi}_{2n}}^T$ \cite{Conrad_2022}.
    \item Measure each displacement operator $\lrc{D\lr{\bs{\xi}_i}}_{i=1}^{2n}$ by performing GKP syndrome extraction.
\end{enumerate}
In this strategy, for each sampled lattice $\Lambda$, the displacement operator measurement returns an outcome determined by a vector $\bs{\alpha}\in \CP\lr{\Lambda}$ with probability
\begin{equation}
    P_{\Lambda}\lr{\bs{\alpha}} = \braket{\Lambda |D^{\dagger}\lr{\bs{\alpha}} \rho D\lr{\bs{\alpha}}|\Lambda},\label{eq:P_alpha}
\end{equation}
which is precisely the expression $\braket{X|\rho|X}$ that determines the distribution over pointers $X=(\Lambda,\bs{\alpha})$ that we are seeking. This strategy is analogous to the strategy employed in \cite{Huang_2020} and can be implemented with standard techniques to implement GKP stabilizer measurements, for which the circuits are depicted in Fig.~\ref{fig:Disp_circ}.

On the other extreme, a full batching of all samples to directly obtain the desired statistics $\braket{X|\rho | X}$ could be obtained using a SWAP test, provided access to GKP states $\ket{X}\in \CX_n$ \cite{Buhrman_2001, Milburn_Fradkin, ding2024quantumcontroloscillatorkerrcat}. While this strategy highlights the power of access to GKP states $\ket{X}\in \CX_n$, such states are unphysical and this is not a practically feasible protocol.

In this section we seek to improve the former strategy while also explaining the mechanics of the circuits in Fig.~\ref{fig:Disp_circ}. In particular, we show how repeated executions of the circuits allow to directly estimate the values $P\lr{\bs{\alpha}}$ from Eq.~\eqref{eq:P_alpha}, effectively already processing the sample statistics to probabilities for $X$ for the experiment outlined above. 

\subsection{Circuit analysis and batched implementation}

In the following it suffices to show how, provided a state $\rho$, we can estimate $P_{\Lambda}\lr{\bs{0}}=\braket{\Lambda|\rho|\Lambda}$, and the outlined strategy can immediately be extended to estimate $P_{\Lambda}\lr{\bs{\alpha}}$ by inserting appropriate phases. This yields an alternative procedure to estimate the values $\braket{X|\rho|X}$ for the set of states constructed in Thm.~\ref{them:2design}, which is expected to simplify the implementation of our protocol in practice.

By the standard structure of group projectors (see e.g. Ref.~\cite{Conrad_2022}) and the uniqueness of the states fixed by symplectic lattices, we have that
\begin{align}
\braket{\Lambda|\rho|\Lambda}= \Tr\lrq{\Pi_{\Lambda }\rho}
     & =\sum_{\bs{\lambda} \in \Lambda} e^{i\Phi\lr{\bs{\lambda}}} \Tr\lrq{D\lr{\bs{\lambda}}\rho}.
\end{align}
This is precisely the expression for the probability of a syndrome measurement on the state $\rho$ relative to the GKP state specified by $\Lambda$ to yield a zero syndrome.  As this is a measure zero set within the possible syndromes, we can't rely on measuring syndromes and counting how often a zero event occurs to estimate this number but need to find a better strategy. 
First, note that the phase $e^{i\Phi\lr{\Lambda}}=\pm 1$ will always be real due to the symplecticity of the lattice~\cite{Conrad_2022}. By symmetry of the lattice, this means that only the real parts $\Re\lrc{\Tr\lrq{D\lr{\bs{\lambda}}\rho}}$ of the displacement expectations on the r.h.s. will contribute to the sum.
The individual parts of the above sum can be evaluated by finding the values of the characteristic function $c_{\rho}\lr{\bs{\lambda}}= \Tr\lrq{D\lr{\bs{\lambda}}\rho}$ of the state $\rho$ over the lattice $\Lambda$.
Although the sum formally necessitates the evaluation of all such values over $\bs{\lambda}\in \Lambda$, we make the further assumption on the input state $\rho$ to be of \textit{limited bandwidth}, i.e. it has a characteristic function with bounded support, such that there exists some $R_{\rho}\in \R_+$, such that $c_{\rho}\lr{\bs{x}}=0\; \forall\, \|\bs{x}\|\geq R_{\rho}$. 

This assumption effectively limits the maximal resolution of the fine-grained structure of the state but can be physically motivated from the bounded behavior of Wigner functions physical states, see also section \ref{sec:toolbox}. By choosing a model for the regularization of the state, e.g. using the regularizer $R_{\beta}=e^{-\beta \hat{N}}$ an estimate for $R_{\rho}$ can be evaluated, which then provides a means to cutoff the infinite sum to the set $\bs{\lambda} \in \Lambda \cap B_{2n}\lr{R_{\rho}}$, where $B_{2n}\lr{R_{\epsilon}}$ is the $2n$-dimensional $R_{\rho}$ ball (see also a discussion in the previous section and in Ref.~\cite{ChasingShadows}).

Given a quantum state $\rho$, we provide two different settings displayed in Fig.~\ref{fig:Disp_circ} to estimate the values $\braket{D\lr{\bs{\lambda}}}=\Tr\lrq{D\lr{\bs{\lambda}}\rho}$.

\begin{figure}
    \centering
    \includegraphics[width=.7\linewidth]{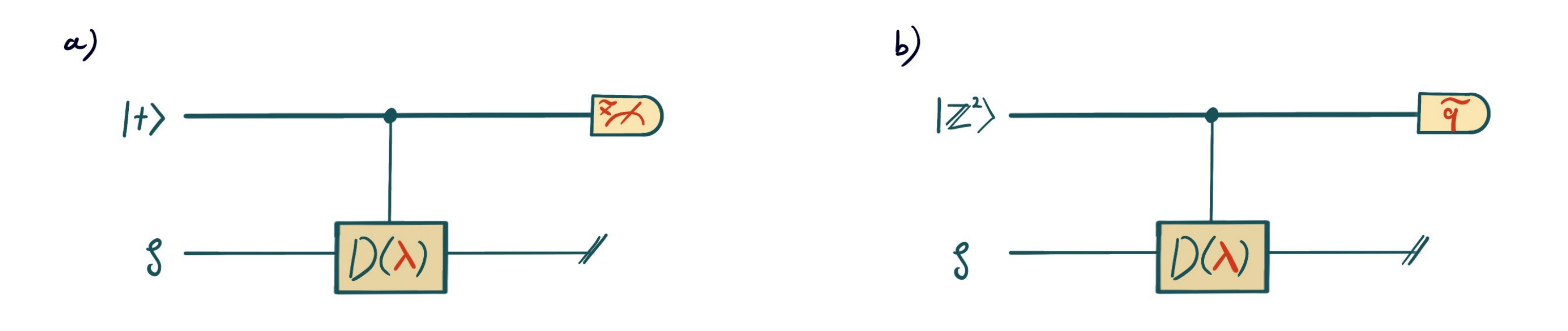}%
    \caption{
    Qubit-assisted and GKP-state-assisted stabilizer measurement circuits that implement the shadow tomography protocol. 
    $a)$ An implementation using a qubit-controlled displacement, where the qubit expectation value, $\langle \tilde X \rangle = P_{+}-P_{-}=\Re\lrc{\Tr\lrq{D\lr{\bs{\lambda}}\rho}}$, encodes the sampling probabilities.  
    $b)$ An auxiliary square-lattice GKP state $|\mathbb{Z}^2\rangle$ together with a mode-controlled displacement operation (a Gaussian unitary gate) is used to sample the probability distribution $P\lr{\tilde{p}}$ of momenta on the auxiliary mode, whose Fourier series coefficients encode all values $\Re\lrc{\Tr\lrq{D\lr{\ell\bs{\lambda}}\rho}}$, $\ell \in \Z$.
    }
    \label{fig:Disp_circ}
\end{figure}

\subsubsection{Estimating the shadow using qubit-controlled displacements}
The first scheme uses an auxiliary qubit initialized in the $\ket{+}=\lr{\ket{0}+\ket{1}}/\sqrt{2}$ that is coupled to the input $n$-mode state via a generalized qubit-controlled displacement operation
\begin{equation}
    D_{\rm qc}\lr{\bs{\lambda}} =\ketbra{0}_a \otimes I + \ketbra{1}_a \otimes D\lr{\bs{\lambda}}
\end{equation}
and finally measured in the Pauli $X$ basis with probabilities to obtain an outcome $\pm 1$  given by
\begin{align}
    P_{\pm} & =\Tr{\lr{\ketbra{\pm}\otimes I} D_{\rm qc}\lr{\bs{\lambda}} \lr{\ketbra{+}_a\otimes \rho}D^{\dagger}_{\rm qc}\lr{\bs{\lambda}}} \nonumber \\
            & =\frac{1}{2}\pm \frac{\Tr\lrq{D\lr{\bs{\lambda}}\rho}}{2},
\end{align}
such that we obtain the desired expectation value as $\braket{X}_a=P_+ - P_- = \Re\lrc{\Tr\lrq{D\lr{\bs{\lambda}}\rho}}$. This procedure corresponds to the established way to implement stabilizer measurements for qubit-controlled GKP quantum error correction.

Qubit-controlled displacement operators of this form, albeit usually only considered to couple to a single mode, are the typical gates considered for implementations of the GKP code, where qubit-like degrees of freedom are natural to the physical architecture.
% and have been used successfully for experimental demonstrations of GKP quantum error correction. 
We refer to Refs.~\cite{Terhal_2015,Terhal_2020, Fluehmann_2019,Campagne_Ibarcq_2020,sivak2023real} for a more in-depth discussion on the implementation and analysis of this scheme.

While the expectation value of this estimator exactly yields the desired outcome, the information gain in each individual execution of the circuit is rather low at a single bit. Using Hoeffdings inequality on the sampled values for a $X$ measurement on the qubit, whose range is bounded to $\{\pm 1\}$, we obtain for $S$ samples that the sample average $\overline{X}^S_a$ approximates the desired value with accuracy
\begin{align}
    P\lr{|\overline{X}^S_a - \Re\lrc{\Tr\lrq{D\lr{\bs{\lambda}}\rho}} | \geq \epsilon} \leq 2e^{-\frac{\epsilon^2}{2}S},
\end{align}
such that, with high probability $1-\delta$ a target accuracy $\epsilon\ll 1$ is achieved with a number of samples scaling as $S=O\lr{\epsilon^{-2} \log\lr{1/\delta}}$.

\subsubsection{Estimating the shadow using auxiliary GKP states}
Alternatively, we may use a GKP encoded auxiliary mode together with a mode-controlled displacement operator (which is a Gaussian unitary operation) to estimate the desired expectation value. 
This procedure (depicted in panel $b$ in Fig.~\ref{fig:Disp_circ}) requires a \textit{fixed} square-lattice GKP resource state and a controlled displacement by primitive vectors of a desired lattice. 

We use an auxiliary GKP state initialized in the single mode GKP ``sensor state'' \cite{Duivenvoorden_Sensor}, which is fixed by displacement in the lattice $\Lambda=\Z^2$, and couple it via the Gaussian unitary operation
\begin{equation}
    D_c\lr{\bs{\lambda}}= e^{-i\hat{q}_a \otimes \bs{\lambda}^TJ\bs{\hat{x}}}
\end{equation}
to the input state. We denote by $\ketbra{\tilde{p}}_p$ the projector onto a fixed momentum $\tilde{p}\in \R$ and by $\rho_{\Z^2}=\ketbra{\Z^2}$ the (approximative) unique GKP state stabilized with lattice $\Z^2$ with wave function in position- or momentum basis $\psi_{\Z^2}\lr{x}= {}_q\!\braket{x|\Z^2}= {}_p\!\braket{x|\Z^2}$. After executing the circuit in Fig.~\ref{fig:Disp_circ}, a measurement of the auxiliary state in the momentum basis produces the distribution%
\begin{align}
    P\lr{\tilde{p}} & =\Tr\lrq{\lr{\ketbra{\tilde{p}} \otimes I}D_c\lr{\bs{\lambda}} \lr{\rho_{\Z^2} \otimes \rho}D_c^{\dagger}\lr{\bs{\lambda}}} \nonumber \\
    & = \int_{\R} dq dq'\, e^{i\tilde{p}\lr{q-q'}} \braket{q'|\rho_{\Z^2}|q} \Tr\lrq{D\lr{\lr{q-q'}\bs{\lambda}/\sqrt{2\pi}} \rho}     \nonumber\\
    & = \int_{\R} dq\, e^{i\tilde{p}q} \braket{\Z^2 |e^{-iq\hat{p}}|\Z^2} \Tr\lrq{D\lr{q\bs{\lambda}/\sqrt{2\pi}}\rho}.\label{eq:P_p}
\end{align}
Using the fact that
$\braket{\Z^2 |e^{-iq\hat{p}}|\Z^2}=\Sha_{\sqrt{2\pi}}\lr{q}=\sum_{n\in \Z} \delta\lr{q-\sqrt{2\pi}n}$,
we find that
\begin{equation}
    P_{\bs{\lambda}}\lr{\tilde{p}}=\sum_{n\in \Z} e^{i\sqrt{2\pi}n\tilde{p} } \Tr\lrq{D\lr{n\bs{\lambda}}\rho}.
\end{equation}

The outcome statistics $P\lr{\tilde{p}}$ is a periodic distribution invariant under $\tilde{p}\mapsto \tilde{p}+\sqrt{2\pi},\, n\in \Z$ and its Fourier coefficients encode the expectation values $\Tr\lrq{D\lr{n\bs{\lambda}}\rho}$ for all $n\in \Z$. Given the GKP state $\ket{\Z^2}$, it hence suffices to estimate the statistics $P_{\bs{\lambda}}\lr{\tilde{p}}$ for \textit{primitive} lattice vectors $\bs{\lambda} \in \Lambda_{\rm pr}$. The set of primitive lattice vectors $\Lambda_{\rm pr}$ is the set of smallest lattice vectors that are not proportional to each other. For example, for $\Z^2$, $\Z^2_{\rm pr}$ contains all vectors $(a, b)^T\in \Z^2$ where the coefficients are coprime ${\rm gcd}\lr{a, b}=1$. Crucially, every lattice vector $\bs{\lambda}\in \Lambda$ can be written as $n\bs{\lambda}_{\rm pr}$ with a primitive lattice vector $\bs{\lambda}_{\rm pr}$ and a unique integer $n$. Using this circuit and again using the finite bandwidth of physical states, we see that the set of vector for which the quantum circuit needs to be evaluated really is $\Lambda_{\rm pr}\cap B_{2n}\lr{R_{\rho}}$.

\begin{figure}
    \center
    \includegraphics[width=.75\textwidth]{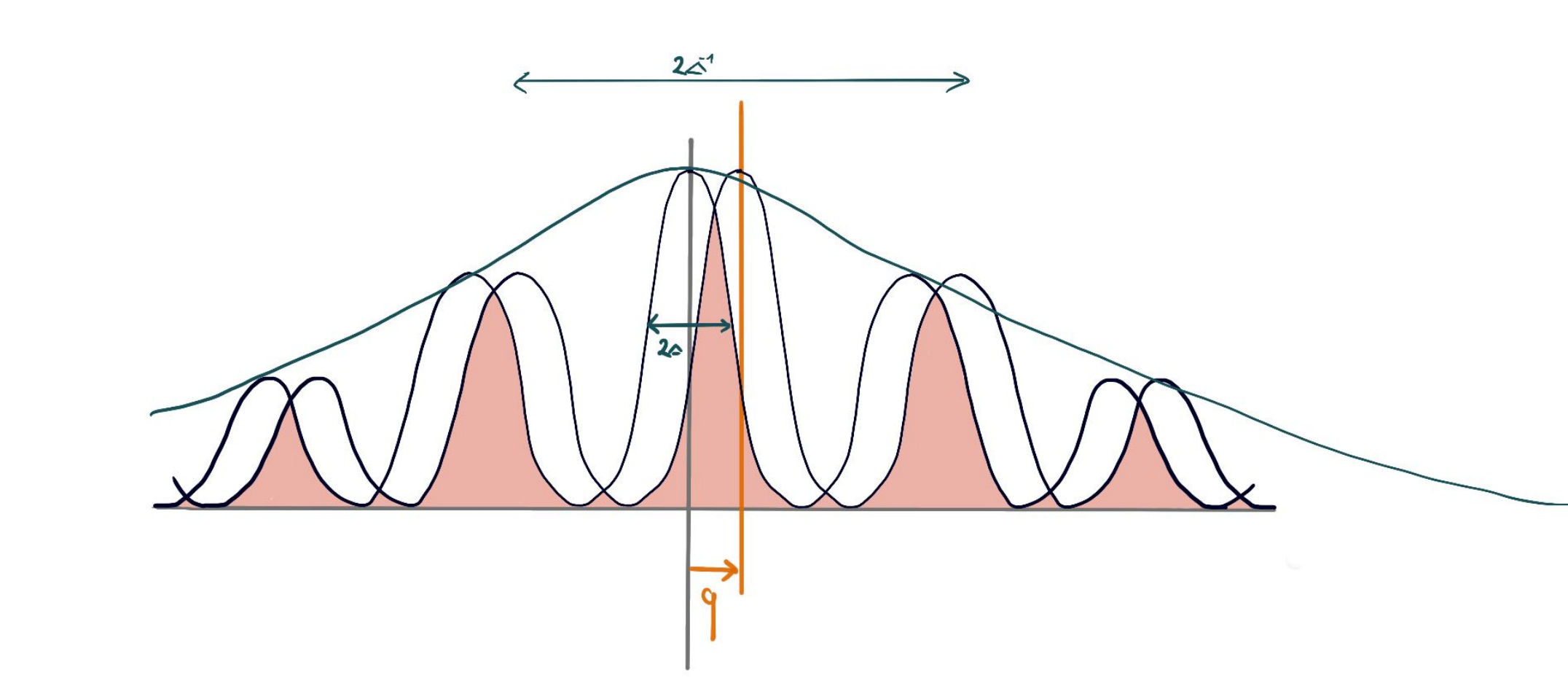}
    \caption{A sketch of the overlap integral of an approximate GKP state with its own displacement. An approximate GKP state approximates a state with periodic structure. For finite bandwidth states the individual peaks are of finite resolution and the state is globally deformed so that it is normalizable. Here, we illustrate the ``standard'' approximation implemented by regularizer $R_{\beta}=e^{-\beta \hat{n}}$, where $\Delta^2=\tanh\lr{\beta /2}$ quantifies the variance of a train of Gaussian peaks regularized by an envelope with variance $\Delta^{-2}$ \cite{GKP, noh2021low} }\label{fig:GKPOverlap}
\end{figure}

In physical implementations, we won't have access to the formal GKP states $\ket{\Z^2}$ but rather states of a regularized form such as $\ket{\Z^2_{\beta}} =N_{\beta} e^{-\beta \hat{N}} $. The effect of using these approximate states can be interpreted using Eq.~\eqref{eq:P_p}. Substituting $\ket{\Z^2}$ with its approximate version changes the factor $\braket{\Z^2 |e^{-iq\hat{p}}|\Z^2} $ into the position basis overlap integral
\begin{eqnarray}
    \braket{\Z_{\beta}^2 |e^{-iq\hat{p}}|\Z_{\beta}^2} = \int_{\R} dq'\, \psi^*_{\Z_{\beta}^2} \lr{q'} \psi_{\Z_{\beta}^2}\lr{q'+q}.
\end{eqnarray}
This hopping-like integral computes ``how periodic'' the approximate state is by computing its overlap with its displaced version. We illustrate the form of this integral in Fig.~\ref{fig:GKPOverlap}. Due to the finite bandwidth of these approximate states, each peak of the distribution now has acquired a finite width $\propto \Delta$, such that values of $q$ close to, but not exactly lying on, integer multiples of $\sqrt{2\pi}$ now can also produce non-zero contributions. Furthermore, due to the presence of an envelope of width $\propto \Delta^{-1}$ on the state (that renders it normalizable), now large values of $q\ll \Delta^{-1}$ can render the corresponding contributions zero, even if $q\in \sqrt{2\pi}\Z$ should be an allowed value by the intended periodicity. 
While this analysis has been provided with the aim of obtaining displacement expectation values, it showcases how the Fourier spectrum of homodyne measurement outcomes under the circuit discussed above contains the information of an infinite number of such displacement operator expectation values. It is to be expected that this information can be of use in the refined characterization of GKP states, e.g. through generalizatons of the effective squeezing parameter \cite{Weigand_2018, Weigand_2020}, and we leave a refined study to future work.

\subsubsection{Typicality}
In the constructions above, we approximate the sum \begin{equation}
    \braket{\Lambda | \rho | \Lambda} = \sum_{\bs{\lambda}\in \Lambda} e^{i\Phi\lr{\bs{\lambda}}} \Tr\lrq{D\lr{\bs{\lambda}} \rho}
\end{equation}
by truncation beyond a radius $R$. Under a randomly chosen lattice $\Lambda\in Y_n$, we wish to estimate how many points there are in our set of displacements to measure. The relevant number of points for our two protocols we need to estimate is hence
\begin{equation}
    |\Lambda \cap B_{2n}\lr{R}| \quad \text{and} \quad |\Lambda_{\rm pr} \cap B_{2n}\lr{R}|.
\end{equation}

In fact, one may apply the mean value formula to obtain the average values of these estimates simply by choosing $f\lr{\lambda}=\chi_{R}\lr{0\neq \|\bs{\lambda}\| \leq R}$ in the mean value formula in Eq.~\eqref{eq:mean_val_app}. We call $\Lambda^{\times}=\Lambda -\lrc{0}$. Applying the mean value formula shows that, that the expected number of (primitive) lattice points of $\Lambda^{\times}$ in $B_{2n}\lr{R}$ is given by the volume of a $2n$-ball $\braket{\Lambda | \rho | \Lambda}$.

\begin{equation}
    \braket{ |\Lambda^{\times} \cap B_{2n}\lr{R}| }_{Y_n}=V_{2n}\lr{R} \quad \text{and} \quad \braket{ |\Lambda_{\rm pr} \cap B_{2n}\lr{R}| }_{Y_n}=V_{2n}\lr{R}/\zeta\lr{n}, \label{eq:mean_count}
\end{equation}
where $\zeta\lr{n}$ is the Riemann zeta function \cite{Kelmer_2019}.

The volume of an $2n$-ball is
\begin{equation}
    V_{2n}\lr{R}= \frac{\pi^n}{n!}R^{2n} \stackrel{\text{Stirling}}{\approx} \frac{1}{\sqrt{2\pi n}}\lr{2\pi e \frac{R^2}{n}}^{n/2} ,
\end{equation}
which vanishes in the limit $n\to \infty$ for $R=o\lr{\sqrt{n}}$ and grows as $R^n$ otherwise, while $\lim_{n\to \infty }\zeta\lr{n}=1$. One observation is hence this: as long as the bandwidth of the state can be assumed to scale as $R=o\lr{\sqrt{n}}$, our protocol to estimate the expectation values $\braket{\Lambda | \rho | \Lambda}$ remains efficient on average.
Ref.~\cite{Kelmer_2019} even derives second moment extensions to the mean value formula over $Y_n$, which yield bounds on the variances of these counters for $n\geq 2$ given by

\begin{align}
    \braket{ \lr{|\Lambda^{\times} \cap B_{2n}\lr{R}| - V_{2n}\lr{R} }^2}_{Y_n} & \leq 4\zeta\lr{n}V_{2n}\lr{R}, \nonumber                      \\
    \braket{ \lr{|\Lambda_{\rm pr} \cap B_{2n}\lr{R}| - V_{2n}\lr{R} }^2}_{Y_n} & \leq 4\frac{V_{2n}\lr{R}}{\zeta\lr{2n}}. \label{eq:var_count}
\end{align}

These bounds tell us that, in the limit of large $n$ and for states of  bandwidth bounded by $R=o\lr{\sqrt{n}}$, the number of displacement operators whose expectation we need to estimate to obtain a good approximation to $\braket{\Lambda | \rho | \Lambda}$ remains small and vanishes asymptotically in $n$. 
In this limit the typical lattice will have shortest vector length scaling with $\sqrt{n}$ and only the trivial point $\lrc{0}$ and a small number of  short vectors close by remain within the ball $B_{2n}\lr{R}$ in phase space.
To contain non-trivial components for typical lattices, the input state is hence required to have bandwidth limit of $R=O\lr{n}$. 

We understand this artifact in analogy to a qubit-based protocol, where the natural analogue of our ensemble $\CX_n$ are the qubit stabilizer states. If the input state is \textit{itself} a stabilizer state, most Pauli expectation values will evaluate to $0$, except on the exponentially small fraction of Pauli operators that exactly align with the stabilizer group of the input state. 
Most information can hence be gained if the input state is instead well-distributed among a large fraction of stabilizer states.

Back to the present situation, a meaningful quantifier for how much a state is overlapping with the various GKP pointer states is the phase-space bandwidth $R$ and lattice instances will yield non-trivial shadows for $R=\Omega\lr{\sqrt{n}}$. 
Provided a fixed length-scale, the number of relevant displacement expectation values is estimated by $ V_{2n}\lr{R}/\zeta\lr{n}+2k \sqrt{V_{2n}\lr{R}/\zeta\lr{2n}}$, where $k=O(1)$ determines the scale up to which we wish to consider fluctuations around the typical instance. When $n$ is fixed, the number of expectation values needed scales efficiently as $R^{2n}$.

\section{Application: variational preparation of thermal states many-body states}\label{sec:thermal}

In this appendix we sketch the application of our protocol to variationally prepare a many-body thermal state based on a perturbation series approach. 
Given a description of a thermal state $\rho_{\beta}\propto e^{-\beta H}$ for an n-mode Hamiltonian $H$, we show for a  GKP state $\ket{\Lambda}$, how to compute the value of $\braket{\Lambda |e^{-\beta H}|\Lambda}$ under a perturbative expansion. 
Having access to these values then one allows to apply the above described protocol to estimate the state overlap $\Tr\lrq{\rho_{\beta}\rho_{\rm model}}$ between the target state $\rho_{\beta}$ and another state $\rho_{\text{model}}$, which in turn can be used to estimate their Hilbert-Schmidt distance $d_{\rm HS}\lr{\rho_{\beta}, \rho_{\rm model}}$. 
By preparing the model state $\rho_{\rm model}=\ketbra{\psi_{\rm model}}$ via a parametrized quantum channel then would allow to approximate the thermal state using well-studied variational techniques. 
The key element in this chain of argumentation is the ability to classically find the values $\braket{\Lambda |e^{-\beta H}|\Lambda}$ needed to implement the classical post-processing in the shadow tomography protocol. Furthermore, as discussed in the previous section, evaluating (or bounding) this value also directly implies a bound on the sample overhead necessary to implement this protocol.

Consider a $n$-mode bosonic system with a Hamiltonian given by
\begin{equation}
    H=\hat{N}+ \epsilon P\lr{\bs{\hat{x}}},
\end{equation}
where $P\lr{\bs{\hat{x}}}$ is a finite polynomial in the quadratures $\hat{q}_1,\hdots, \hat{p}_n$. The thermal state of such a system at inverse temperature $\beta$ is (up to the potentially unknown partition function) proportional to

\begin{equation}
    \rho_{\beta}=e^{-\beta\lr{ \hat{N}+ \epsilon P\lr{\bs{\hat{x}}}}},
\end{equation}
and, in the limit of small $0\leq \epsilon \ll 1$ can be expanded in a perturbative series
\begin{equation}
\rho_{\beta}=\tilde{P}\lr{\bs{\hat{x}}}e^{-\beta \hat{N}},
\end{equation}
where we assume $\tilde{P}\lr{\bs{\hat{x}}}$ to be a polynomial in Weyl-ordered form~\cite{CahillGlauber, Gerry2004}. The existence of this form of expression can be again found using the formula for exponential derivatives~\cite{RossmanLie} to expand $\rho_{\beta}(\epsilon)=\rho_{\beta}(0)+\epsilon \frac{\partial \rho_{\beta}(\epsilon)}{\partial \epsilon}  \vert_{\epsilon =0}$, which yields
\begin{equation}
    \frac{\partial \rho_{\beta}(\epsilon)}{\partial \epsilon}  \vert_{\epsilon =0} = -\beta\int_0^1 dy\, e^{-y\beta \hat{N}} P \lr{\bs{\hat{x}}}e^{y\beta \hat{N}} e^{-\beta \hat{N}}= -\beta\int_0^1 dy\,  P \lr{e^{-y\beta \hat{N}}\bs{\hat{x}}e^{y\beta \hat{N}}} e^{-\beta \hat{N}}. \label{eq:Taylor_rhob}
\end{equation}
It holds for each pair of quadrature and value $r\in \R$ that
\begin{equation}
    e^{-r \hat{n}} \begin{pmatrix}
        \hat{q} \\ \hat{p}
    \end{pmatrix}
    e^{r \hat{n}} = \begin{pmatrix}
        \cosh\lr{r} & i \sinh\lr{r} \\
        -i\sinh\lr{r} & \cosh\lr{r}
    \end{pmatrix}\begin{pmatrix}
        \hat{q} \\ \hat{p}
    \end{pmatrix},
\end{equation}
such that we see that the expression in eq.~\eqref{eq:Taylor_rhob} can be integrated and rewritten in the desired form.

We now leverage the decomposition already provided in Eq.~\eqref{eq:Rbeta_disp},
\begin{equation}
    e^{-\beta \hat{N}}=\lr{1-e^{-\beta}}^{-n} \int_{\R^{2n}} d\bs{\alpha}\, e^{-\pi \|\bs{\alpha}\|^2/2\Delta^2}D\lr{\bs{\alpha}},
\end{equation}
with $\Delta^2=2\tanh\lr{\beta /2}$ to write
\begin{align}
\lr{1-e^{-\beta}}^{n}\tilde{P}\lr{\bs{\hat{x}}}e^{-\beta \hat{N}} & = \int_{\R^{2n}} d\bs{\alpha}\,   e^{-\pi \|\bs{\alpha}\|^2/2\Delta^2} \tilde{P}\lr{\bs{\hat{x}}} D\lr{\bs{\alpha}}                                          \nonumber  \\
& =\int_{\R^{2n}} d\bs{\alpha}\,  e^{-\pi \|\bs{\alpha}\|^2/2\Delta^2} \lrc{\tilde{P}\lr{ \bs{\nabla}_{\bs{\alpha}}} D\lr{\bs{\alpha}} } \nonumber\\
& =\int_{\R^{2n}} d\bs{\alpha}\, \lrc{\tilde{P}\lr{ (1+i \Delta^{-2})\sqrt{\frac{\pi}{2}} J\bs{\alpha} } e^{-\pi \|\bs{\alpha}\|^2/2\Delta^2}}D\lr{\bs{\alpha}} ,
\end{align}
where we defined
\begin{equation}
    \bs{\nabla}_{\bs{\alpha}}\coloneqq
    \frac{-i}{\sqrt{2\pi}}J\bs{\nabla}+\sqrt{\frac{\pi}{2}}\bs{\alpha}
\end{equation}
 and used integration by parts together with the compactness of the Gaussian factor. This expression determines the characteristic function of the perturbative expansion of the thermal state. Using this expression, we can compute for a GKP state $\ket{\Lambda}$ the expectation value
\begin{align}
    \Braket{\Lambda |\tilde{P}\lr{\bs{\hat{x}}}e^{-\beta \hat{N}}|\Lambda} & = \lr{1-e^{-\beta}}^{-n}\sum_{\bs{\lambda} \in \Lambda} e^{i\Phi\lr{\bs{\lambda}}} \tilde{P}\lr{ (1+i \Delta^{-2})\sqrt{\frac{\pi}{2}} J\bs{\lambda} } e^{-\pi \|\bs{\lambda}\|^2/2\Delta^2}. 
\end{align}
This series naturally converges for any finite polynomial $\tilde{P}$. Let $\overline{P}\lr{\bs{x}}$ be a polynomial function of the coefficients of $\bs{x}\in \R^{2n}$ such that \begin{equation}
    \overline{P}\lr{\bs{x}} \geq \left|\tilde{P}\lr{(1+i \Delta^{-2})\sqrt{\frac{\pi}{2}} J\bs{x}}\right|\,\forall \bs{x}\in \R^{2n}.
\end{equation}
The above expression is bounded by
\begin{equation}
    \left|\Braket{\Lambda|\tilde{P}\lr{\bs{\hat{x}}}e^{-\beta\hat{N}}
        |\Lambda}\right|\leq \lr{1-e^{-\beta}}^{n} \sum_{\bs{\lambda} \in \Lambda} \overline{P}\lr{\bs{x}}  e^{-\pi \|\bs{\lambda}\|^2/2\Delta^2}= \Theta_{\Lambda, \overline{P}}\lr{\frac{i}{4\Delta^2}},
\end{equation}
where $\Theta$ are known as a \textit{weighted theta series} in the special case where $\overline{P}$ is a harmonic polymonial
\cite{Elkies_weighted}.

What have we gained? We have shown that the GKP-state expectation value of the perturbative expansion of any (unnormalized) multi-mode thermal state can be evaluated in terms of a Gaussian lattice sum.
We have also bounded this sum by a Gaussian sum which is easy to study and where one can devise tractable approximations. Under the specific assumptions that the perturbative expansion is bounded by a harmonic polynomial, this expression turns into a weighted theta series. 

Weighted theta series are number theoretically interesting objects, encoding information about the distribution of lengths of lattice vectors together with their distribution over spheres. Furthermore, they turn out to be modular forms \cite{Elkies_weighted}, which are functions with a high degree of symmetry that are potentially useful for the derivation of stronger bounds and evaluation tactics for these functions. While we leave a further study of the protocol sketched in this appendix to future work, we have demonstrated that the CV shadow tomography protocol devised in this manuscript can find applications in non-trivial algorithmic tasks, such as estimating the expectation value of multi-mode thermal states relative to a given bosonic state.
This can be leveraged for the variational preparation of such a thermal state.

%\bibliography{Shadow_bib}
\end{document}